\PassOptionsToPackage{figuresright}{rotating}
\documentclass[lettersize,journal]{IEEEtran}
\usepackage{amsmath,amsfonts}
\usepackage{algorithmic}
\usepackage{algorithm}
\usepackage{array}
\usepackage[caption=false,font=normalsize,labelfont=sf,textfont=sf]{subfig}
\usepackage{textcomp}
\usepackage{stfloats}
\usepackage{url}
\usepackage{verbatim}
\usepackage{graphicx}
\usepackage{cite}
\usepackage{caption}
\usepackage{enumitem}
\usepackage{booktabs}
\usepackage{multirow}
\usepackage{listings}
\usepackage{makecell}
\usepackage{rotating}
\usepackage[table]{xcolor}
\definecolor{mycolor1}{HTML}{FF7744}
\definecolor{mycolor2}{HTML}{FFA488}
\definecolor{mycolor3}{HTML}{FFC8B4}
\definecolor{mygreen}{HTML}{098842}

\definecolor{myback1}{HTML}{EAEFF7}
\definecolor{myback2}{HTML}{EBF1E9}

\definecolor{myred}{HTML}{EE204B}
\definecolor{myblue}{HTML}{ccffcc}
\definecolor{mycolor}{HTML}{99ffff}
\usepackage{booktabs} 

\definecolor{darkgreen}{rgb}{0.0,0.5,0.0}
\definecolor{darkred}{rgb}{0.5,0.0,0.0}

\begin{document}

\title{LibEER: A Comprehensive Benchmark and Algorithm Library for EEG-based Emotion Recognition}

\author{Huan Liu, Shusen Yang, Yuzhe Zhang, Mengze Wang, Fanyu Gong, Chengxi Xie, Guanjian Liu, Zejun Liu, Yong-Jin Liu,~\IEEEmembership{Senior Member~IEEE,} Bao-Liang Lu,~\IEEEmembership{Fellow~IEEE,} Dalin Zhang,~\IEEEmembership{Senior Member~IEEE}
\thanks{Huan Liu, Shusen Yang, Yuzhe Zhang, Mengze Wang, Guanjian Liu are with the School of Computer Science and Technology, Xi'an Jiaotong University, Xi'an 710049, China.}
\thanks{Fanyu Gong is with the School of Software Engineering, Xi'an Jiaotong University, Xi'an 710049, China.}
\thanks{Chengxi Xie and Zejun Liu are with the Joint School of Design and Innovation, Xi'an Jiaotong University, Xi'an 710049, China.}
\thanks{Yong-Jin Liu is with the Department of Computer Science and Technology, Tsinghua University, Beijing 100084, China.}
\thanks{Bao-Liang Lu is with the Department of Computer Science and Engineering, Shanghai Jiao Tong University, Shanghai 200240, China.
}
\thanks{Dalin Zhang is with the Space Information Research Institute, Hangzhou Dianzi University, Hangzhou 310018, China.}
\thanks{Corresponding authors: zhangyuzhe@stu.xjtu.edu.cn, dalinzhang@ieee.org}
}

\markboth{Journal of \LaTeX\ Class Files,~Vol.~14, No.~8, August~2021}%
{Shell \MakeLowercase{\textit{et al.}}: A Sample Article Using IEEEtran.cls for IEEE Journals}


\maketitle

\begin{abstract}
EEG-based emotion recognition (EER) has gained significant attention due to its potential for understanding and analyzing human emotions. While recent advancements in deep learning techniques have substantially improved EER, the field lacks a convincing benchmark and comprehensive open-source libraries. This absence complicates fair comparisons between models and creates reproducibility challenges for practitioners, which collectively hinder progress.
To address these issues, we introduce LibEER, a comprehensive benchmark and algorithm library designed to facilitate fair comparisons in EER. LibEER carefully selects popular and powerful baselines, harmonizes key implementation details across methods, and provides a standardized codebase in PyTorch. By offering a consistent evaluation framework with standardized experimental settings, LibEER enables unbiased assessments of seventeen representative deep learning models for EER across the six most widely used datasets. Additionally, we conduct a thorough, reproducible comparison of model performance and efficiency, providing valuable insights to guide researchers in the selection and design of EER models. Moreover, we make observations and in-depth analysis on the experiment results and identify current challenges in this community. 
We hope that our work will not only lower entry barriers for newcomers to EEG-based emotion recognition but also contribute to the standardization of research in this domain, fostering steady development. The library and source code are publicly available at \url{https://github.com/XJTU-EEG/LibEER}.
\end{abstract}


\section{Introduction}
\IEEEPARstart{E}{motion} represents a spectrum of subjective cognitive experiences that underpins interpersonal relationships and human behavior assessment~\cite{emontiondef}. Electroencephalography (EEG), owing to its non-invasive nature, high temporal resolution, portability, and cost-effectiveness, has become a widely used physiological signal in emotion analysis~\cite{ding2023lggnet, khare2020time}. Consequently, EEG-based emotion recognition (EER) has garnered increasing attention across domains such as healthcare~\cite{hasnul2021electrocardiogram, 10697478}, advertising~\cite{gauba2017prediction}, and education~\cite{nandi2021real}. Early EER methods focused on hand-crafted features like power spectral density (PSD)\cite{alsolamy2016emotion} and differential entropy (DE)\cite{DE}, while recent advances in deep learning have enabled automatic extraction of temporal, spatial, and spectral features, leading to state-of-the-art performance~\cite{zhang2022eeg, zhang2024cross, liu2024eeg}.

Despite significant advancements in deep learning model design, the practical implementation of these approaches is often overlooked, making it difficult for newcomers and researchers to keep up with emerging studies. Therefore, this work focuses on the implementation aspects of EER. The specific issues we aim to address are as follows:
\begin{itemize}[leftmargin=*]
    \item \textbf{I1 (Lack of benchmarking in EER):}  
A common practice in many model design studies is to compare new approaches with state-of-the-art (SOTA) models. However, such comparisons are often unreliable due to the absence of comprehensive benchmarking within the EER community. Studies frequently use varying preprocessing techniques, such as artifact removal, and different preprocessing settings, such as varying train/test split ratios. Some studies also adopt questionable practices, such as trial-dependent configurations~\cite{cheng2020emotion, acrnn}, which can introduce data leakage risks. Furthermore, many studies provide ambiguous experimental settings~\cite{HSLT, BiDANN, li2023brain}, which hinder reproducibility. To address these issues, standardizing preprocessing techniques and experimental settings is essential for enabling fair comparisons and identifying genuine progress in the field.

    \item \textbf{I2 (Lack of open-source libraries for deep learning-based EER):} 
    Implementation details, such as training hyperparameters, play a crucial role in model performance. While it is impractical and unnecessary to include every detail in an academic paper, making implementation code publicly available is vital for enabling the reproduction and further development of SOTA research. However, even when code is provided, it is often tied to different platforms, such as PyTorch~\cite{pytorch}, TensorFlow~\cite{tensorflow}, or MATLAB~\cite{matlab}, posing challenges to researchers. Therefore, there is a pressing need for an open-source library based on a standardized platform that provides comprehensive implementation details while achieving performance comparable to the original study.

    \item \textbf{I3 (Lack of in-depth analysis of SOTA EER research):} While deep learning has dominated EER research, many survey papers~\cite{liu2024eeg,alarcao2017emotions} have been published summarizing progress from various perspectives. However, these surveys typically report findings without hands-on experience in reproducing results, which may lead to misdirected future research. This lack of practical insight can obscure real challenges faced in the field. Therefore, an in-depth analysis of SOTA EER research based on reproduction results is essential for uncovering the true challenges and advancing the field.
\end{itemize}

To address these issues, we propose LibEER (\underline{Lib}rary for \underline{E}EG-based \underline{E}motion \underline{R}ecognition), an open-source platform that enables reproduction and analysis of SOTA EER studies. To establish a benchmark (addressing \textbf{I1}), we standardize datasets, evaluation metrics, and experimental settings. Based on a review of recent studies, we include six widely used datasets, SEED~\cite{SEED}, SEED-IV~\cite{SEED-IV}, DEAP~\cite{DEAP}, MAHNOB-HCI~\cite{HCI}, SEED-V~\cite{SEED-V}, and MPED~\cite{MPED}, and support both subject-dependent and cross-subject configurations, using a rigorous train/validation/test split.
To address \textbf{I2}, we implement seventeen representative deep learning-based EER methods within a unified PyTorch training pipeline. LibEER provides diverse EEG preprocessing tools, evaluation protocols, and flexible experimental configurations via user-friendly interfaces.
Finally, to address \textbf{I3}, we conduct extensive reproduction-based comparisons using LibEER and derive four key observations that reveal the strengths and limitations of existing models. While recent efforts like~\cite{zhang2024torcheegemo} have introduced EER toolkits, our work differs by providing more in-depth benchmarking analysis, explicitly focusing on EER-specific models, and systematically reproducing several influential methods that lack open-source implementations. These contributions together offer a more comprehensive and practical foundation for advancing EER research.

In summary, our main contributions are as follows:
\begin{itemize}[leftmargin=*] 
    \item We have pioneered the establishment of the first benchmark specifically designed to enable fair comparisons in EEG-based emotion recognition (EER). This benchmark not only offers an objective reflection of the actual progress in the EER field but also serves as a practical reference for conducting standardized comparative experiments.
    \item We have introduced LibEER as an open-source project on GitHub, offering flexible customization of experimental schemes through user-friendly interfaces. LibEER supports fair and comprehensive evaluation of seventeen popular models across six datasets. Serving both as a toolbox and a platform, it assists practitioners in efficiently utilizing and reproducing EER methods. 
    \item We provide thorough evaluations across various experimental settings, carefully controlling implementation details. These experimental results offer valuable insights into the advancements made and assist researchers in selecting or designing EER models. \end{itemize}

The paper is organized as follows. 
Section \ref{sec:prob} provides the background and formal definition of the EER task.
Section \ref{sec:bench} presents the design of the benchmark.
Section \ref{sec:lib} details the library. 
Section \ref{sec:exp} covers experimental results and in-depth analysis. 
Section \ref{sec:future} identifies future work, and Section \ref{sec:concl} concludes the whole work.

\begin{figure}
    \centering
        \includegraphics[width=1\linewidth]{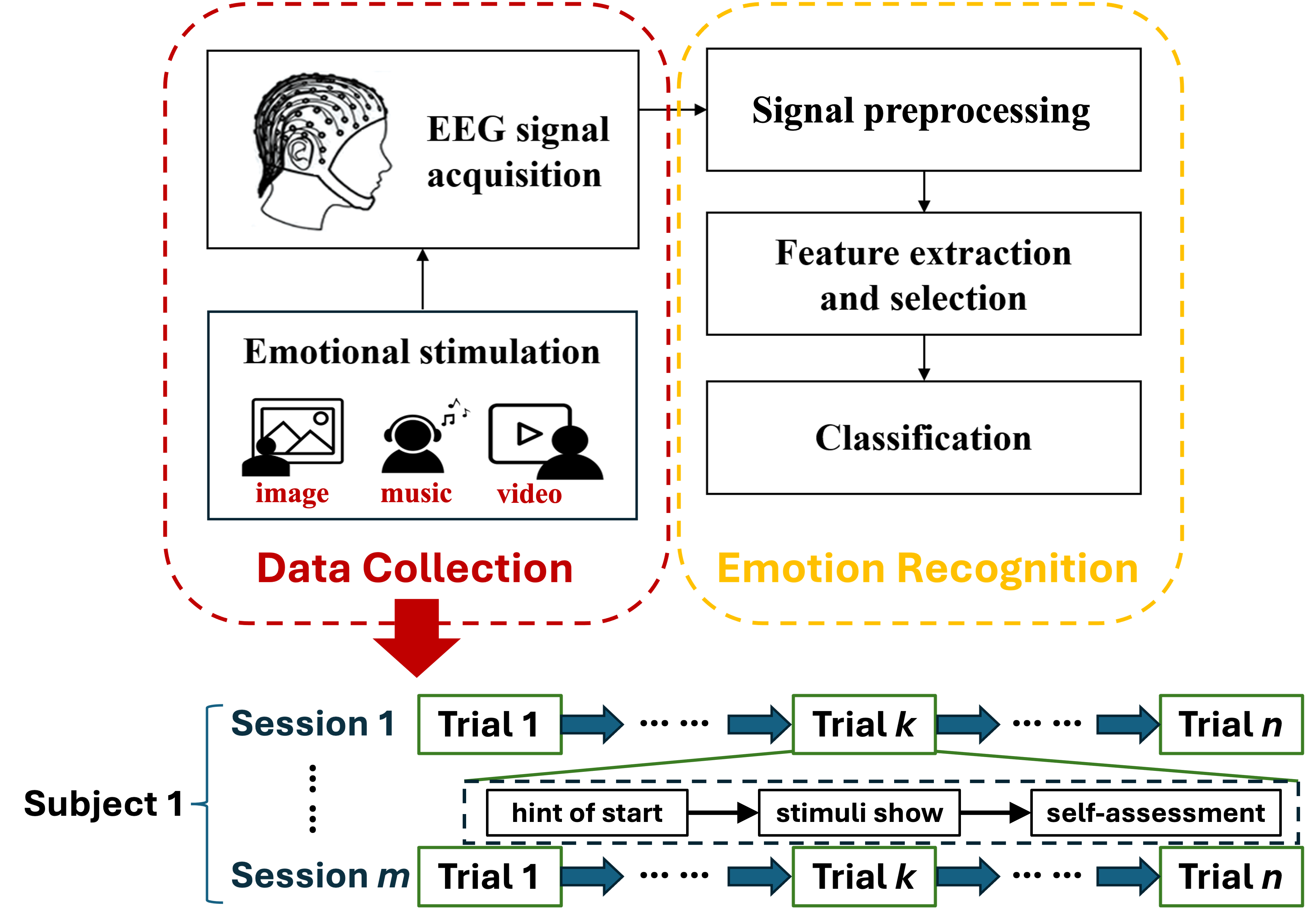}
        \caption{Data collection protocol of EER.}
    \label{workflow}
\end{figure}

\section{Background of EER}\label{sec:prob}
In this section, we present the general background of EEG, including data collection protocol and formal problem definition.

\subsection{Data Collection Protocol}
The general protocol for EER data collection is illustrated in Fig.~\ref{workflow}. Each data collection begins with a start hint shown to the participant. Afterwards, various stimuli, such as images, audio, or video, are presented to evoke specific emotions in the participant, such as happiness, sadness, or fear. After each stimulus, the participant is asked to self-assess their emotional state, providing a subjective emotion score based on their experience during the stimulus. This process is repeated a set number of times with different stimuli to collect enough data.


In this context, the participant exposed to the stimuli, from whom data is subsequently collected, is commonly referred to as a \textit{subject}. A continuous segment of EEG data collected from a subject under a specific stimulus condition is called a \textit{trial}, typically lasting between one and five minutes. Between trials, the subject takes a rest. A series of trials is termed a \textit{session}. Since EEG data from the same subject, even in the same emotional state, may vary significantly over time, datasets generally include several sessions, with an interval of more than 24 hours between consecutive ones.
Given the high cost of EEG data acquisition, a trial is often divided into segments. Two consecutive segments within a trial can be overlapping or non-overlapping. Each segment is called a \textit{sample}, serving as the fundamental unit for EER. Notably, within a single trial, all samples should be designated for either training or test. Splitting samples from the same trial across training and test sets is impractical for real-world applications and considered invalid.

\subsection{Problem Formulation}


Given an EER dataset $D$, a sample is denoted as $x_{i,j,k,q} \in \mathbb{R}^{c \times d}$, where $i$, $j$, $k$, and $q$ refer to the $i$-th subject, $j$-th session, $k$-th trial, and $q$-th sample within the trial, respectively. Here, $c$ represents the number of electrode channels, and $d$ denotes the feature dimension of each sample. The goal of EER is to classify a sample into a designated emotional state, which may be represented dimensionally (e.g., valence, arousal) or categorically (e.g., sadness, joy), via the following mapping:
\begin{equation}
    \mathcal{F}: x \rightarrow \boldsymbol{y},
\end{equation}
where $\boldsymbol{y}$ denotes the emotional state and $\mathcal{F}$ is a mapping function. SOTA map functions are usually neural networks trained on a training set $D_{\text{train}} \subset D$, and evaluated on a test set $D_{\text{test}} \subset D$. Additionally, a validation set $D_{\text{valid}} \subset D$ is often employed to  select the best model. These subsets are required to be mutually exclusive, that is $D_{\text{train}} \cap D_{\text{test}} \cap D_{\text{valid}} = \emptyset$.


\begin{table}[t]
	\caption{Summary of our proposed benchmark.}
	\label{Benchmark}
	\renewcommand{\arraystretch}{1.2}
	\centering

            \begin{tabular}{m{2.5cm}<{\raggedright}m{5.5cm}<{\centering}}
                \noalign{\global\arrayrulewidth=2pt}
                \toprule
                 \textbf{Module}&\textbf{Key Information}\\
                \noalign{\global\arrayrulewidth=0.5pt}			
                
                \midrule
                Data Preprocessing& Bandpass Filtering / Removing Eye Movement  / Extracting DE features / Segmenting data\\
                \midrule
                Data Splitting&Train : Val : Test = 0.6 : 0.2 : 0.2\\
                \midrule
                Experimental Task&Subject-dependent / Cross-subject\\
                \midrule
                Datasets&SEED / SEED-IV / DEAP / MAHNOB-HCI / SEED-V / MPED\\
                \midrule
                Evaluation Metrics& Accuracy / F1-score\\
                
            \noalign{\global\arrayrulewidth=1pt}
            \bottomrule
            \end{tabular}

\end{table}

\begin{figure*}
    \centering
        \includegraphics[width=1\linewidth]{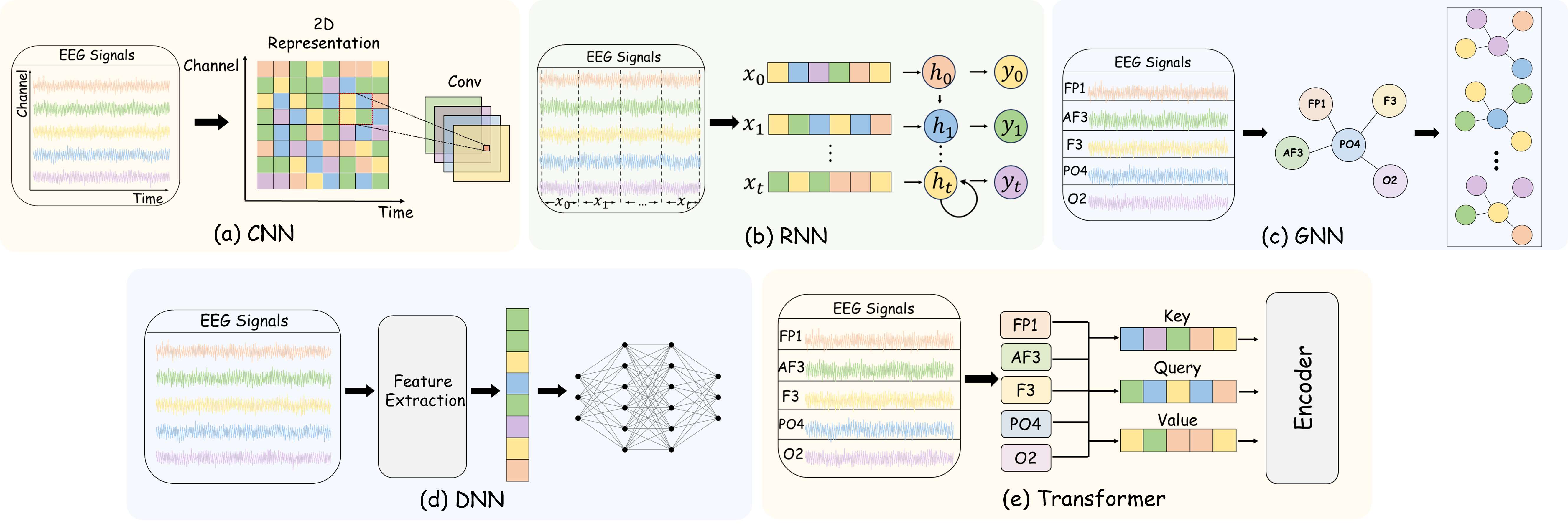}
        \caption{The distinct utilization of EEG signals by different methods.}
    \label{representative EEG methods}
\end{figure*}

\begin{table}[!t]
  \centering
  \caption{Detailed information of selected baseline methods.}
    \renewcommand{\arraystretch}{1}
   {\scriptsize{
    \begin{tabular}{m{2cm}<{\raggedright} m{0.5cm}<{\centering} m{0.8cm}<{\centering} m{1.5cm}<{\centering} m{0.5cm}<{\centering} m{0.9cm}<{\centering}}
      \noalign{\global\arrayrulewidth=1pt}
      \toprule
      \textbf{Model} & \makecell[c]{\textbf{Base.}\\ \textbf{freq.}} & \makecell[c]{\textbf{Citation}\\ \textbf{count}} & \makecell[c]{\textbf{Publication}} & \makecell[c]{\textbf{Publish}\\ \textbf{year}} & \textbf{Type} \\
      \midrule
      EEGNet\cite{eegnet} & 7  & 2979 & JNE  & 2018 & CNN \\
      CDCN\cite{cdcn} & 2  & 142 & TCDS  & 2020 & CNN \\
      Tsception\cite{tsception} & 2  & 136 & TAFFC & 2022 & CNN \\\midrule
      ACRNN\cite{acrnn} & 3  & 338 & TAFFC & 2020 & RNN \\
      BiDANN\cite{BiDANN} & 4 & 250 & TAFFC & 2021 & RNN\\
      R2G-STNN\cite{R2G-STNN} & 8 & 235 & TAFFC & 2022 & RNN \\\midrule
      DGCNN\cite{dgcnn} & 18 & 1105 & TAFFC & 2018 & GNN \\
      GCBNet\cite{gcbnet} & 3  & 302 & TAFFC & 2019 & GNN \\
      GCBNet\_BLS\cite{gcbnet} & 3  & 302 & TAFFC & 2019 & GNN \\
      RGNN\cite{rgnn} & 11 & 545 & TAFFC & 2020 & GNN \\
      NSAL-DGAT~\cite{NSAL-DGAT} &- & 1 &TAFFC &2025 &GNN \\\midrule
      DBN\cite{DBN} & 7  & 1884 & TAMD & 2015 & DNN \\
      DAN~\cite{DAN} &- & 199 & ICONIP &2018 &DNN \\ 
      DANN~\cite{DAN} &- & 199 & ICONIP &2018 &DNN \\ 
      MS-MDA\cite{ms-mda} & 6  & 95 & Front. Neuros. & 2021 & DNN \\
      PR-PL~\cite{PR-PL} & -& 55 & TAFFC& 2024 &DNN\\ \midrule
      HSLT\cite{HSLT} & 2  & 79 & IEEE SENS J & 2022 & Trans. \\
      
      \noalign{\global\arrayrulewidth=1pt}
      \bottomrule
    \end{tabular}
  }}
  \label{Detailed Information Of Selected Baseline Methods}%
\end{table}

%
\section{Benchmark Building}\label{sec:bench}


A comprehensive, well-structured, and transparent benchmark is essential for fair comparisons of research methodologies in the field—a need that remains unmet in the EER domain. In this section, we present our benchmark, covering baseline selection, preprocessing, experimental setups, datasets, and evaluation protocols. For ease of reference, TABLE \ref{Benchmark} provides a summary of the key information in our EER benchmark.

\subsection{Baseline Selection}
We select representative models for the benchmark. 
Specifically, we set up the following criteria:

\begin{itemize}[leftmargin=*]
\item We first collected a total of 60 high-quality papers published in the past year in the field of EEG-based Emotion Recognition (EER) from reputable journals and conferences (such as TAFFC, JBHI, and ICASSP).

\item We then compiled a list of 205 comparative methods cited across these 60 papers.

\item After that, we ranked these 205 methods based on their frequency as comparison methods in the 60 papers and selected the top 20 for replication.

\item These 20 methods, following the common practice in previous studies of categorizing deep learning models by their main architecture~\cite{yang2022survey,alom2019state}, are divided into five categories: CNN, RNN, GNN, DNN, and Transformer. Finally, we selected models with replication results within a 10\% error margin compared to the original reported results, ensuring that each category includes at least one representative model.
\end{itemize}

Based on these steps, we ultimately identified the twelve most representative works~\cite{eegnet, cdcn, tsception, acrnn, dgcnn, rgnn, gcbnet,DBN,ms-mda, HSLT, BiDANN, R2G-STNN} to include in LibEER. In addition, we include two classic transfer learning methods (DAN and DANN~\cite{DAN}) that have demonstrated superior performance in EER, as well as two recent state-of-the-art approaches (NSAL-DGAT~\cite{NSAL-DGAT} and PR-PL~\cite{PR-PL}). These methods are also categorized into the corresponding architecture-based groups according to their backbone model structures. TABLE \ref{Detailed Information Of Selected Baseline Methods} presents detailed information on these seventeen methods. In future work, we will continue to maintain and expand this collection by incorporating more recent and diverse methods. The following subsections will offer detailed descriptions of the five categories, as well as an examination of each selected method.

\subsubsection{\textbf{CNN-based Methods}}
Convolutional Neural Networks (CNNs) are a type of feedforward neural network characterized by their convolutional operations. When applied to EEG signals, CNNs usually treat EEG data as a two-dimensional (2D) representation: one dimension corresponds to the channels, while the other dimension represents either the temporal or spectral features of the EEG data. This structure is illustrated in Fig. \ref{representative EEG methods} (a), which could handle temporal/spectral and spatial information simultaneously. On the other hand, some work relies on 1D CNN to extract temporal and spatial features separately. 
LibEER includes three representative CNN-based methods: EEGNet~\cite{eegnet}, CDCN~\cite{cdcn}, and TSception~\cite{tsception}.



\subsubsection{\textbf{RNN-based Methods}}
Recurrent Neural Networks (RNNs) are inherently designed for sequential data, making them well-suited for extracting temporal features from EEG signals. For spatial feature extraction, RNNs treat channels as sequence nodes, as illustrated in Fig. \ref{representative EEG methods} (b). In practice, Long Short-Term Memory (LSTM) is a widely used variant of RNNs.
Our benchmark includes three RNN-based methods: ACRNN~\cite{acrnn}, BiDANN~\cite{BiDANN}, and R2G-STNN~\cite{R2G-STNN}.

\subsubsection{\textbf{GNN-based Methods}}
Graph Neural Network (GNN) is a type of neural network designed to handle data structured as graphs. A graph consists of nodes and edges, which can represent complex relationships and structures. When processing EEG data, the EEG channels are typically treated as nodes in a graph, as shown in Fig. \ref{representative EEG methods} (c). Since EEG data does not inherently contain information about the edges or the adjacency matrix, using graph neural networks to process EEG data often involves leveraging prior knowledge about the relationships between channels or using learnable adjacency matrices as the graph's adjacency matrix. 
Our benchmark includes five GNN-based methods: DGCNN~\cite{dgcnn}, GCBNet~\cite{gcbnet}, GCBNet\_BLS~\cite{gcbnet}, RGNN~\cite{rgnn}, and NSAL-DGAT~\cite{NSAL-DGAT}. 


\subsubsection{\textbf{DNN-based Methods}}
Deep Neural Network (DNN), in this paper, specifically refers to the feedforward fully connected neural network, also known as Fully Connected Neural Network (FNN) or Multi-Layer Perceptron (MLP). It consists of an input layer, multiple hidden layers composed of fully connected neurons with nonlinear activation functions, and an output layer. Unlike other deep learning architectures such as CNN, RNN, or GNN, DNN focuses on dense layer transformations without convolutional, recurrent, or graph-based structures. In the field of EER, DNN processes EEG data by performing feature transformation and mapping into a high-dimensional feature space, enabling the extraction and utilization of complex patterns within the data, as illustrated in Fig. \ref{representative EEG methods} (d). Our benchmark includes five DNN-based methods: DBN~\cite{DBN}, DAN~\cite{DAN}, DANN~\cite{DAN}, MS-MDA~\cite{ms-mda}, and PR-PL~\cite{PR-PL}.


\subsubsection{\textbf{Transformer-based Methods}}
Transformer uses a self-attention mechanism to capture dependencies between elements in a sequence, effectively handling long-range dependencies. As shown in Fig. \ref{representative EEG methods} (e), the EEG data are typically segmented into multiple patches on channels, regions, or temporal segments when using transformers to process EEG data. These patches are then fed into the transformer model for further analysis. Our benchmark includes one Transformer-based methods: HSLT~\cite{HSLT}. 


\subsection{Data preprocessing and Splitting}
\subsubsection{preprocessing}
A significant challenge in reproducing extant studies on EER is the lack of a standardized data preprocessing protocol. Publicly available EER datasets are not immediately usable and require extensive preprocessing. Moreover, this critical information is rarely detailed in the literature, and many open-source packages do not provide the necessary data scripts. To address this issue, our benchmark proposes a comprehensive data preprocessing framework specifically designed for EER research. The procedure includes: (1) applying bandpass filtering between 0.3 and 50 Hz; (2) eliminating eye movement artifacts using Principal Component Analysis (PCA); (3) extracting DE features across five frequency bands—[0.5, 4], [4, 8], [8, 14], [14, 30], and [30, 50]—followed by processing with a Linear Dynamic System (LDS); and (4) segmenting data using 1-second non-overlapping sliding windows to enhance the dataset.

\subsubsection{Splitting}
Data splitting is a critical process that delineates how a dataset is partitioned into training and test sets, significantly influencing model performance. However, prevailing practices in this domain often exhibit a lack of rationality and the absence of a coherent, standardized approach. For example, many current data-splitting methodologies fail to incorporate a validation set for model selection and do not consistently adhere to cross-trial configurations, resulting in unreliable experimental outcomes.

To mitigate these shortcomings, we propose a systematic and rational data-splitting methodology in our benchmark. Specifically, for tasks that are subject-dependent, we allocate each individual’s data into training, validation, and test sets in a 0.6:0.2:0.2 ratio, adjusting as necessary for datasets that do not divide evenly. The SEED-V dataset contains relatively fewer trials and more classes, so the ratio was adjusted to 1:1:1. In the case of cross-subject tasks, we similarly partition subjects into training, validation, and test sets following the same 0.6:0.2:0.2 ratio. Furthermore, our benchmark rigorously adheres to the cross-trial principle, ensuring that samples from the same trial do not appear in both the training and test sets.

\subsection{Experimental Tasks}
\begin{figure}[t]
    \centering
        \includegraphics[width=1\linewidth]{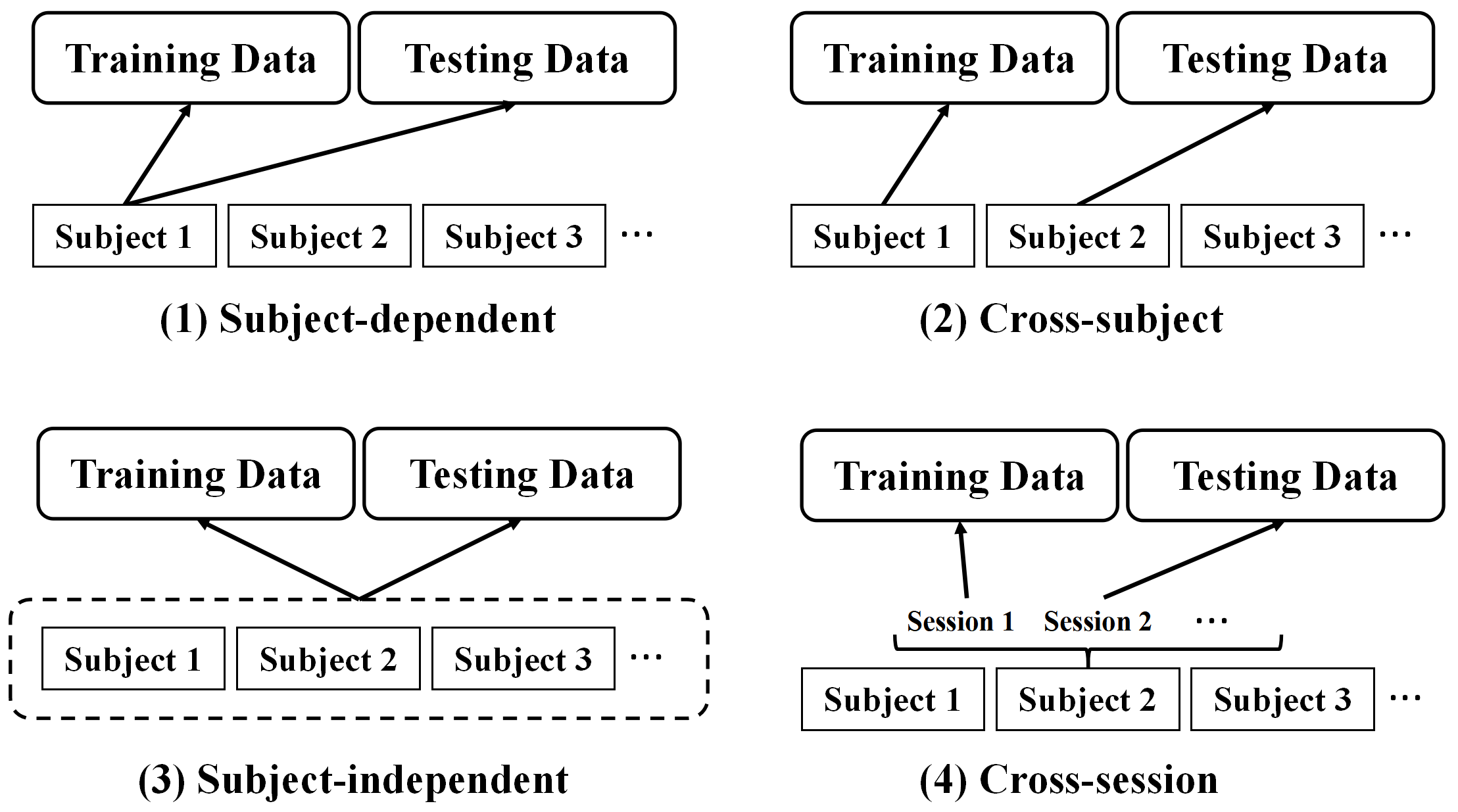}
        \caption{Four different experimental tasks in EER.}
    \label{fourtask}
\end{figure}

The field of EER encompasses various experimental tasks, each designed to address distinct challenges with corresponding models. The primary experimental tasks are categorized as follows: (1) Subject-dependent: This task evaluates model performance for individual subjects, requiring both training and testing data from the same subject. (2) Cross-subject: This task aims to generalize model performance across different subjects, requiring training and testing data from different subjects. (3) Subject-independent: This task assesses model performance without constraining the subject-level split, allowing training and testing data to be randomly divided from mixed samples of multiple subjects. (4) Cross-session: This task examines model performance across different sessions of the same subject, requiring training and testing data from different sessions of the same individual. Figure~\ref{fourtask} provides an intuitive illustration of the differences among the four tasks. 

In previous studies, the definitions of subject-independent and cross-subject tasks have often overlapped, as subject-independent is typically implemented using leave-one-subject-out cross-validation (LOSO-CV)~\cite{dgcnn,rgnn}. To avoid ambiguity, we explicitly adopt cross-subject as one of the core evaluation settings in our benchmark. Cross-subject strictly requires training and testing on entirely different subjects, emphasizing the model’s generalization ability. This task setup is more aligned with real-world applications where models are expected to perform reliably on new, unseen users in large-scale deployment scenarios.
In addition, subject-dependent tasks are included to evaluate a model’s performance under personalized settings, where training and testing are performed on the same subject. This setup is important for scenarios focusing on optimizing accuracy for specific individuals, such as personalized healthcare or user-specific emotion monitoring.
Therefore, our benchmark concentrates on cross-subject and subject-dependent tasks as the most widely adopted and practically meaningful evaluation setups. Additionally, it is important to note that we treat data from different sessions of the same subject as if they originate from distinct subjects. We provide specific definitions for the subject-dependent and cross-subject tasks below.

In \textbf{subject-dependent} tasks, each task is conducted using data from a single subject and the data is split based on individual trials. We denote the training, validation, and test sets for subject \(S_i\) as \(D^{(i)}_{train}\), \(D^{(i)}_{val}\) and \(D^{(i)}_{test}\), respectively. These sets must satisfy two essential constraints:
\begin{equation}
\begin{aligned}
    &D^{(i)}_{train} \cap D^{(i)}_{val} \cap D^{(i)}_{test} = \emptyset, \\
    &D^{(i)}_{train} \cup D^{(i)}_{val} \cup D^{(i)}_{test} = D^{(i)}.
\end{aligned}
\end{equation}
These conditions guarantee that there is no overlap among the trials in the training, validation, and test sets, while ensuring that the union of these sets encompasses all available data from subject $S_i$.
Following the cross-trial principle, the $D^{(i)}_{train}$, $D^{(i)}_{val}$, $D^{(i)}_{test}$ can be defined as:
\begin{equation}
\begin{aligned}
    &D^{(i)}_{train} = \{T^{(i)}_1, T^{(i)}_2, \dots, T^{(i)}_{m_1}\}, \\
    &D^{(i)}_{val} = \{T^{(i)}_{m_1+1}, T^{(i)}_{m_1+2}, \dots, T^{(i)}_{m_2}\}, \\
    &D^{(i)}_{test} = \{T^{(i)}_{m_2+1}, T^{(i)}_{m_2+2}, \dots, T^{(i)}_{m}\}. 
\end{aligned}
\end{equation}
The EEG samples from $D^{(i)}_{train}$ can be represented as $X^{(i)}_{tr} \in \mathbb{R}^{m^{(i)}_{tr} \times c \times d}$, where $m^{(i)}_{tr}$ represents the number of EEG sample of $D^{(i)}_{train}$, $c$ denotes the number of electrode channels and $d$ is the feature dimension. We denote the corresponding labels of $X^{(i)}_{tr}$ as $\boldsymbol{y}^{(i)}_{tr}\in \mathbb{R}^{m^{(i)}_{tr}}$. 
The subject-dependent task involves learning a mapping function $f: X^{(i)}_{tr} \rightarrow \boldsymbol{y}^{(i)}_{tr}$ with cross-entropy loss that accurately predicts the emotion state of \(S_i\).
The validation set $D^{(i)}_{val}$ is used to tune hyperparameters and select the best model, while the test set $D^{(i)}_{test}$ is used to evaluate the final model and estimate its performance on unseen data.

The \textbf{cross-subject} task uses EEG data from all subjects. We denote the 
training, validation, ans test sets as \(D_{train}\), \(D_{val}\), and \(D_{test}\), respectively. These sets also satisfy two constraints:
\begin{equation}
\begin{aligned}
    &D_{train} \cap D_{val} \cap D_{test} = \emptyset, \\
    &D_{train} \cup D_{val} \cup D_{test} = D.
\end{aligned}
\end{equation}
These conditions ensure that there is no overlap subjects between three sets and the union of these sets covers all available data of the dataset. 
Following the cross-subject principle, the \(D_{train}\), \(D_{val}\), and \(D_{test}\) can be defined as:
\begin{equation}
\begin{aligned}
    &D_{train} = \{S_1, S_2, \dots, S_{n_1}\}, \\
    &D_{val} = \{S_{n_1+1}, S_{n_1+2}, \dots, S_{n_2}\}, \\
    &D_{test} = \{S_{n_2+1}, S_{n_2+2}, \dots, S_{n}\}.
\end{aligned}
\end{equation}
Analogously, the EEG samples from $D_{train}$ can be represented as $X_{tr} \in \mathbb{R}^{m_{tr} \times c \times d}$, where $m_{tr}$ represents the number of EEG sample of $D_{train}$. We denote the corresponding labels of $X_{tr}$ as $\boldsymbol{y}_{tr}\in \mathbb{R}^{m_{tr}}$. 
The cross-subject task learns a mapping function $f: X_{tr} \rightarrow \boldsymbol{y}_{tr}$ with the cross-entropy loss.
The validation set $D_{val}$ is used to tune hyperparameters and select the best model, while the test set $D_{test}$ is used to evaluate the final model and estimate its performance on unseen data.

\begin{table*}[t]
	\caption{The statistics of datasets in LibEER.}
	\label{dataset}
	\renewcommand{\arraystretch}{1.2}
	\centering
 {\footnotesize{
            \begin{tabular}{m{2.4cm}<{\raggedright}m{0.8cm}<{\centering}m{0.8cm}<{\centering}m{0.7cm}<{\centering}m{0.8cm}<{\centering}m{1.4cm}<{\centering}m{1.2cm}<{\centering}m{0.5cm}<{\centering}m{1cm}<{\centering}m{4.5cm}<{\centering}}
                \noalign{\global\arrayrulewidth=2pt}
                \toprule
                 \textbf{Dataset}&\textbf{\#Session}&\textbf{\#Subject}&\textbf{\#Trial}&\textbf{\#Channel}&\textbf{Stimulus}&\textbf{Duration}&\textbf{HZ}&\textbf{Features}&\textbf{Labels}\\
                \noalign{\global\arrayrulewidth=0.5pt}			
                \midrule
                
                SEED\cite{SEED}&3&15&15&62&movie clip&185-265s&200&DE&positive/neutral/negative\\
                SEED-IV\cite{SEED-IV}&3&15&24&62&movie clip&43-259s&1000&raw data& happiness/sadness/fear/neutral\\
                DEAP\cite{DEAP}&1&32&40&32&music video&60s&512&raw data&valence/arousal\\
                MAHNOB-HCI\cite{HCI}&1&30&20&32&movie clip&34.9-117s&512&raw data&valence/arousal\\
                SEED-V~\cite{SEED-V} &3&16&15&62&movie clip&120s-240s&1000&raw data&happy/sad/disgust/netural/fear\\
                MPED~\cite{MPED} &1&23&28&62&movie clip&150s-300s&1000&STFT&joy/funny/anger/fear/disgust/sad/neutrality\\
                
            \noalign{\global\arrayrulewidth=1pt}
            \bottomrule
            \end{tabular}
          
            }}
\end{table*}

\subsection{Datasets}
Research in the EER field exhibits variability in dataset selection and often lacks clarity regarding the specific data segments utilized. This study evaluates the performance of popular models using six datasets, which are selected by comprehensively considering factors such as their frequency of use, data scale, and data quality. They will be introduced along with the relevant data segments, as summarized in Table~\ref{dataset}.

\subsubsection{SEED}
The SEED\cite{SEED} dataset consists of 15 subjects (7 males, 8 females), each completing three sessions, with 15 emotion-inducing video trials per session (positive, neutral, or negative, 4 minutes each). EEG data were recorded using a 62-channel cap at 200 Hz, following the 10-20 system. The dataset includes raw data and features such as PSD and DE across five frequency bands. Data from all sessions are used for subject-dependent tasks, while the first session is used for cross-subject tasks.

\subsubsection{SEED-IV}
The SEED-IV\cite{SEED-IV} dataset comprises 15 subjects (7 males, 8 females), each participating in three sessions with 24 trials per session. Each 2-minute trial includes videos evoking happiness, sadness, fear, or neutrality. EEG data were recorded at 1000 Hz using a 62-channel cap. The dataset offers raw data and preprocessed features such as PSD and DE. All sessions' data are used for subject-dependent tasks, while the first session is used for cross-subject tasks.

\subsubsection{DEAP}
The DEAP\cite{DEAP} dataset consists of 32 subjects (16 males, 16 females) who participated in one session containing 40 trials. Each trial involved a 1-minute music video designed to evoke specific emotions, followed by participant ratings for arousal, valence, dominance, and liking on a 1-9 scale. EEG data were recorded using a 32-channel cap at 512 Hz, and the dataset includes both raw and downsampled (128 Hz) versions. All samples are used for both subject-dependent and cross-subject tasks. In the subsequent experiments, we used the DEAP dataset with the valence labels, the DEAP dataset with the arousal labels, and the DEAP dataset with both labels, denoted as DEAP-V, DEAP-A, and DEAP-VA, respectively.

\subsubsection{MAHNOB-HCI}
The MAHNOB-HCI\cite{HCI} dataset includes 30 subjects (13 males, 17 females), though EEG data are completely missing for 2 subjects and partially missing for 3. Each subject participated in a single session with 20 trials, where videos lasting 34.9 to 117 seconds evoked emotions. After each video, participants rated arousal, valence, dominance, and predictability on a 1-9 scale. EEG data were recorded using a 32-channel system at 512 Hz, and the dataset includes both raw and downsampled (128 Hz) versions. Both complete and partial samples are used for subject-dependent and cross-subject tasks. In the subsequent experiments, we used the MAHNOB-HCI dataset with the valence labels, the MAHNOB-HCI dataset with the arousal labels, and the MAHNOB-HCI dataset with both labels, denoted as HCI-V, HCI-A, and HCI-VA, respectively.

\subsubsection{SEED-V}
The SEED-V\cite{SEED-V} dataset contains 16 subjects (6 males, 10 females) who each participated in three experimental sessions. Each session included 15 emotion-inducing movie clips (3 clips per emotion category: happy, sad, neutral, fear, and disgust). EEG signals were recorded using a 62-channel cap following the 10-20 system, accompanied by simultaneous eye movement recordings. The dataset provides differential entropy (DE) features extracted from five frequency bands (delta, theta, alpha, beta, and gamma) for EEG signals. Data from all three sessions are used for subject-dependent and cross-subject tasks.

\subsubsection{MPED}
The MPED\cite{MPED} dataset is a comprehensive multimodal resource for discrete emotion recognition, comprising physiological recordings from 23 Chinese participants (10 males, 13 females). During emotion elicitation experiments using 28 standardized video stimuli representing seven discrete emotions (joy, funny, anger, sadness, disgust, fear, neutrality), 62-channel EEG signals were acquired via an ESI NeuroScan System at 1000 Hz sampling rate following the international 10-20 system.  All data are used for both subject-dependent and cross-subject tasks.

\subsection{Evaluation Methods and Metrics}
The evaluation method determines how the results reported in the paper are calculated. This part of the experimental setup is also where many studies encounter the most significant issues. Currently, many studies report the best results of the model on the test set for each epoch. This evaluation method is seriously flawed as it greatly exaggerates the performance of the model.

Therefore, in our benchmark, our evaluation method reports the performance on the test set of the model that achieved the highest F1 score on the validation set across all epochs. We report both the mean and standard deviation of two metrics, accuracy and F1 score. The former is a key metric for classification tasks, while the latter provides a more reasonable assessment when sample labels are imbalanced. The methods for calculating accuracy and F1 score are presented in Eq.~\ref{acc} and Eq.~\ref{f1}, respectively.
\begin{equation}
\label{acc}
\text{ACC} = \frac{TP + TN}{TP + TN + FP + FN}
\end{equation}
\begin{equation}
\label{f1}
\text{F1 Score} = \frac{2 \times TP}{2 \times TP + FP + FN}
\end{equation}
where TP (True Positives), TN (True Negatives), FP (False Positives), and FN (False Negatives) are the metrics used to calculate accuracy and F1 score. Besides, the random seed is fixed at 2024 to ensure the reproducibility of the results.

\section{LibEER Toolkit Framework}\label{sec:lib}
In this section, we will present the framework of the algorithm library to facilitate researchers' utilization of these models in the domain of EER. Specifically, we will detail the coding procedures associated with the data loader, data split, and model training and evaluation within the LibEER. To facilitate the utilization of the LibEER library in the field of EER, users can easily install via pip, which stremlines the installation process and ensures that all necessary dependencies are managed automatically.
For further details and access to the source code, please visit the official repository at \url{https://github.com/XJTU-EEG/LibEER}.

\begin{figure}[t]
    \centering
    \includegraphics[width=1\linewidth]{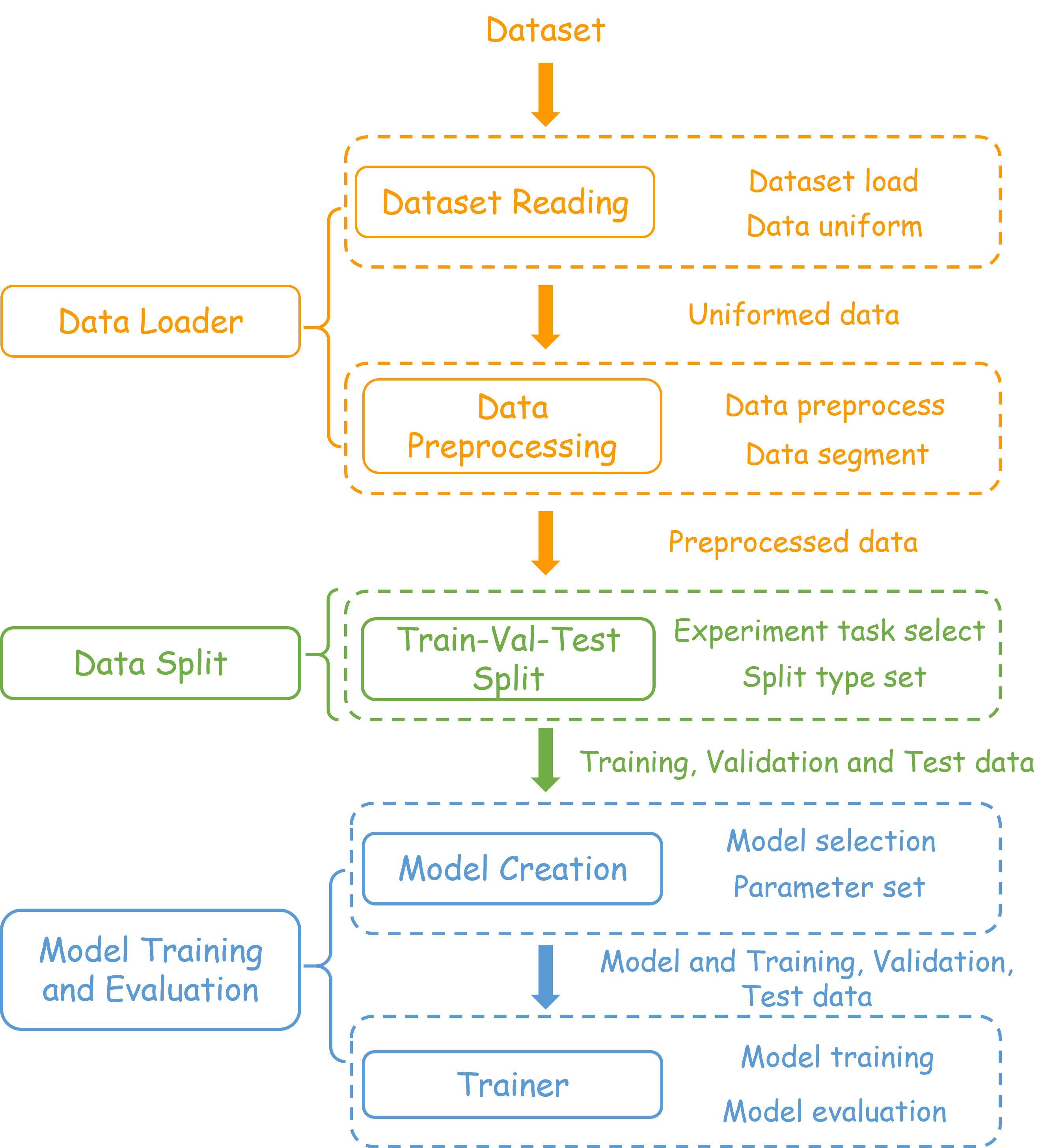}
    \caption{The framework and data pipeline of LibEER.}
    \label{toolkitFramework}
\end{figure}

LibEER offers a user-friendly and equitable platform for EER implemented using PyTorch. It comprises three primary modules: the data loader, data splitting, and model training and evaluation. The data loader module standardizes the data formats across various datasets and provides a range of preprocessing techniques. The data splitting module delivers a cohesive solution applicable to different experimental tasks and partitioning methods. The model training and evaluation module presents a standardized and extensible framework for conducting model training and evaluation. Fig. \ref{toolkitFramework} illustrates the framework of LibEER. Below, we provide detailed explanations for each module.


\subsubsection{Data loader}
The data loader module primarily comprises two components: dataset reading and EEG data preprocessing. Currently, the data formats in EEG emotion recognition datasets lack uniformity, necessitating distinct data loading methods for different datasets, which ultimately reduces research efficiency. To address this challenge, LibEER offers the \textit{get\_uniform\_data()} function, which facilitates the reading of datasets through tailored methods for each dataset, subsequently integrating the data into a standardized format of (session, subject, trial). This approach not only streamlines subsequent preprocessing tasks but also enhances compatibility with various experimental tasks and data-splitting methods.

Additionally, LibEER provides the \textit{preprocess()} function, which allows for the preprocessing of data based on user-defined settings. The \textit{preprocess()} function operates on a trial basis and encompasses three primary steps: noise removal, feature extraction, and sample segmentation. When utilizing raw data directly from the dataset, only sample segmentation is executed. Noise removal aims to eliminate electromyography (EMG) noise, electrooculography (EOG) artifacts, and other forms of interference from the EEG signals. The feature extraction process analyzes the data to derive relevant EEG features, such as DE and PSD features, based on the specified feature types and frequency bands provided by the user. Sample segmentation organizes the data into user-defined sample lengths. Ultimately, the processed data is integrated into the format of (session, subject, trial, sample). An example demonstrating how to use the data loader is provided in our GitHub project.

\subsubsection{Data split}
The data splitting module is primarily responsible for partitioning the dataset into training, testing, and validation sets. LibEER performs this division based on the experimental tasks and splitting methods selected by the user. For the two experimental tasks, the \textit{merge\_to\_part()} function is utilized to integrate the data into a standardized format.

In the context of subject-dependent tasks, the \textit{merge\_to\_part()} function organizes the data into the format of (subject, trial, sample), wherein the data for each subject is treated as an independent sub-task, allowing for subsequent data splitting on a trial basis for each sub-task. Conversely, for cross-subject and cross-session tasks, the \textit{merge\_to\_part()} function consolidates the data into (subject, sample) and (session, sample) formats, with data splitting conducted on a subject or session basis.

To accommodate various splitting methods, the \textit{get\_split\_index()} function further divides the data based on the specified formats and labels. The splitting methods available in LibEER include train-test splitting, which partitions the data into training and testing sets according to a predetermined ratio;train-val-test splitting, which divides the data into training, validation, and testing sets while maintaining balanced label distribution;and cross-validation methods, which segment the data into n folds, ensuring that each fold is utilized as the testing set once. An example illustrating how to perform data splitting with LibEER is provided in our GitHub project.

\begin{table*}[t]
	\caption{Experimental settings and comparison of replication results for reproduced models in LibEER.}
	\label{reproduction}
	\renewcommand{\arraystretch}{1.2}
	
 {\small{{\centering
        \begin{tabular}{m{1.6cm}<{\centering}m{1.5cm}<{\centering}m{2cm}<{\centering}m{1.2cm}<{\centering}m{2cm}<{\centering}m{1.4cm}<{\centering}>{\columncolor{myback1}}m{1.8cm}<{\centering}>{\columncolor{myback1}}m{1.8cm}<{\centering}>{\columncolor{myback1}}m{1.1cm}<{\centering}}
            \noalign{\global\arrayrulewidth=2pt}
            \hline
            \multirow{2}*{Method} & \multicolumn{5}{c}{\textbf{Experimental Settings}} & \multicolumn{3}{>{\columncolor{myback1}}c}{\textbf{Results (Accuracy)}} \\
            \cmidrule(lr){2-6}\cmidrule(lr){7-9} & Dataset&Preprocessing&Task&Splitting \quad(Train : Test)&Evaluation&Reported (\%)&Ours (\%)&Gap (\%)\\
            \noalign{\global\arrayrulewidth=0.5pt}			
            \hline
            EEGNet&SEED& B, R, DE, 1s&dependent &3 : 2 &ACC & —— &68.15$\pm$12.32& —— \\
            \cmidrule(lr){1-9}
            \multirow{3}*{CDCN}&SEED&B, R, DE, 1s &dependent &3 : 2 &ACC&90.63&85.10$\pm$8.80 &\textcolor{mygreen}{\textbf{5.53$\downarrow$}}\\
            \cmidrule(lr){2-9}
            &DEAP-V&B, R, 1s &dependent&  9 : 1&ACC&92.24&92.30$\pm$11.33&\textcolor{red}{\textbf{0.06$\uparrow$}}\\
            \cmidrule(lr){2-9}
            &DEAP-A&B, R, 1s &dependent&9 : 1& ACC&92.92&91.99$\pm$12.20&\textcolor{mygreen}{\textbf{0.93$\downarrow$}}\\
            \cmidrule(lr){1-9}
            \multirow{2}*{TSception}&DEAP-V&B, R, 4s&dependent &9 : 1 & ACC& 61.57$\pm$11.04 &61.89$\pm$12.16 &\textcolor{red}{\textbf{0.32$\uparrow$}}\\
            \cmidrule(lr){2-9}
            &DEAP-A&B, R, 4s &dependent & 9 : 1& ACC&59.14$\pm$7.60  & 60.02$\pm$8.84&\textcolor{red}{\textbf{0.88$\uparrow$}}\\
            \cmidrule(lr){1-9}
            \multirow{2}*{ACRNN}&DEAP-V&B, R, DE, 3s &dependent & 9 : 1&ACC&93.72&86.03$\pm$9.20&\textcolor{mygreen}{\textbf{7.69$\downarrow$}} \\ 
             \cmidrule(lr){2-9}
            &DEAP-A&B, R, DE, 3s &dependent&9 : 1 &ACC
            &93.38 &88.31$\pm$7.77 &\textcolor{mygreen}{\textbf{5.07$\downarrow$}}\\
            \cmidrule(lr){1-9}
            \multirow{2}*{BiDANN}&SEED&B, R, DE, 9s &dependent & 3 : 2& ACC&92.38$\pm$7.04&89.06$\pm$9.42&\textcolor{mygreen}{\textbf{3.32$\downarrow$}}\\
            \cmidrule(lr){2-9}
            &SEED&B, R, DE, 9s &cross &14 : 1 &ACC & 83.28$\pm$9.60 &80.36$\pm$8.25& \textcolor{mygreen}{\textbf{2.92$\downarrow$}}\\
            \cmidrule(lr){1-9}
            \multirow{2}*{R2G-STNN}&SEED&B, R, DE, 9s &dependent & 3 : 2& ACC&93.38$\pm$5.96&88.71$\pm$8.85&\textcolor{mygreen}{\textbf{4.67$\downarrow$}}\\
            \cmidrule(lr){2-9}
            &SEED&B, R, DE, 9s &cross &14 : 1 &ACC &84.16$\pm$7.63  &78.35$\pm$9.46 & \textcolor{mygreen}{\textbf{5.81$\downarrow$}}\\
            \cmidrule(lr){1-9}
            DGCNN&SEED& B, R, DE, 1s &dependent & 3 : 2 & ACC &90.40$\pm$8.49&89.48$\pm$8.49&\textcolor{mygreen}{\textbf{0.92$\downarrow$}}\\
            \cmidrule(lr){1-9}
            GCBNet&SEED&B, R, DE, 1s &dependent & 3 : 2&ACC&92.30$\pm$7.40&89.04$\pm$8.03&\textcolor{mygreen}{\textbf{3.26$\downarrow$}} \\
            \cmidrule(lr){1-9}
            GCBNet\_BLS&SEED&B, R, DE, 1s &dependent & 3 : 2&ACC&94.24$\pm$6.70&88.80$\pm$9.54 &\textcolor{mygreen}{\textbf{5.44$\downarrow$}}\\
            \cmidrule(lr){1-9}
            RGNN&SEED&B, R, DE, 1s &dependent & 3 : 2&ACC&94.24$\pm$5.95&84.66$\pm$10.74&\textcolor{mygreen}{\textbf{9.58$\downarrow$}}\\
            \cmidrule(lr){1-9}
            \multirow{2}*{NSAL-DGAT}& SEED & B, R, DE, 1s & dependent & 3 : 2 &ACC&96.53$\pm$9.10&92.27$\pm$10.09&\textcolor{mygreen}{\textbf{4.26$\downarrow$}} \\
            \cmidrule(lr){2-9}
            &SEED&B, R, DE, 1s &cross & 14 : 1&ACC &95.69$\pm$5.24  & 90.19$\pm$7.32&\textcolor{mygreen}{\textbf{5.50$\downarrow$}}\\
            \cmidrule(lr){1-9}
            DBN&SEED&B, R, DE, 1s &dependent & 3 : 2& ACC&86.91$\pm$7.62&81.18$\pm$8.13&\textcolor{mygreen}{\textbf{5.73$\downarrow$}}\\
            \cmidrule(lr){1-9}
            DAN&SEED& B, R, DE, 1s& cross& 14 : 1&ACC&83.81$\pm$8.56 &76.29$\pm$6.60  &\textcolor{mygreen}{\textbf{7.52$\downarrow$}} \\
            \cmidrule(lr){1-9}
            DANN&SEED& B, R, DE, 1s& cross& 14 : 1&ACC&—— &77.40$\pm$5.86  &——\\
            \cmidrule(lr){1-9}
            MS-MDA &SEED&B, DE, 1s &cross & 14 : 1&ACC&89.63$\pm$6.79&89.65$\pm$7.24&\textcolor{red}{\textbf{0.02$\uparrow$}} \\   
            \cmidrule(lr){1-9}
            \multirow{2}*{PR-PL}&SEED&B, R, DE, 1s & dependent & 3 : 2 & ACC & 94.84$\pm$9.16  & 93.2$\pm$6.66 & \textcolor{mygreen}{\textbf{1.64$\downarrow$}}\\
            \cmidrule(lr){2-9}
            &SEED&B, R, DE, 1s &cross &14 : 1 &ACC & 94.84$\pm$9.16&89.19$\pm$5.34 & \textcolor{mygreen}{\textbf{5.65$\downarrow$}}\\
            \cmidrule(lr){1-9}            
            \multirow{3}*{HSLT}&DEAP-V&B, R, PSD, 6s &cross &14 : 1&ACC, F1 & 66.51 &69.18&\textcolor{red}{\textbf{2.67$\uparrow$}}\\
            \cmidrule(lr){2-9}
            &DEAP-A&B, R, PSD, 6s &cross&31 : 1&ACC, F1 &65.75&68.81&\textcolor{red}{\textbf{3.06$\uparrow$}}\\
            \cmidrule(lr){2-9}
            &DEAP-VA&B, R, PSD, 6s &cross&31 : 1&ACC, F1 &56.93&49.57&\textcolor{mygreen}{\textbf{7.36$\downarrow$}}\\
            \noalign{\global\arrayrulewidth=1pt}
            \hline
        \end{tabular}\\}
        B: Bandpass filtering, R: Removing eye movement, DE: Extracting
DE features, Ns: Segmenting data into N seconds.
        }}
\end{table*}

\subsubsection{Model training and evaluation}
Model training and evaluation are essential processes responsible for training the designated model and assessing its performance. LibEER constructs models based on the selected architecture and parameters, subsequently training them according to the specified training configurations. In scenarios where a validation set is not available, LibEER identifies the optimal performance on the testing set during training as the final outcome for that training iteration. Conversely, when a validation set is utilized, LibEER selects the network weights that yield the best performance on the validation set for evaluation against the testing set, thereby determining the final result.

The ultimate outcomes of the model will be reported according to the performance metrics specified by the user. An example of creating, training, and evaluating a model within LibEER is provided in our GitHub project.

\section{Experimental Results}\label{sec:exp}
In this section, we thoroughly validate the importance of LibEER and our benchmark through extensive experimentation. Initially, we utilize LibEER to replicate the selected models as closely as possible following the experimental configurations detailed in the original publications. Subsequently, we employ LibEER to compare the performance of all models based on the established benchmark. Lastly, we investigate the impact of several key experimental settings on model performance through comparative experiments.

\begin{sidewaystable*}[htp]
	\caption{The mean accuracies and F1 scores (and standard deviations) using the proposed benchmark for subject-dependent EER experiment. The top two methods in each scenario are highlighted using bold and underlined formatting.}
	\label{dependent}
	\renewcommand{\arraystretch}{1}
	\centering
	{\footnotesize{
        \begin{tabular}{m{1cm}<{\centering}m{0.5cm}<{\centering}m{0.8cm}<{\centering}m{0.8cm}<{\centering}m{0.8cm}<{\centering}m{0.8cm}<{\centering}m{0.8cm}<{\centering}m{0.8cm}<{\centering}m{0.8cm}<{\centering}m{0.8cm}<{\centering}m{0.8cm}<{\centering}m{0.8cm}<{\centering}m{0.8cm}<{\centering}m{0.8cm}<{\centering}m{0.8cm}<{\centering}m{0.8cm}<{\centering}m{0.8cm}<{\centering}m{0.8cm}<{\centering}m{0.8cm}<{\centering}}
            \noalign{\global\arrayrulewidth=2pt}
            \hline
            \noalign{\global\arrayrulewidth=0.5pt}	
           \multicolumn{2}{c}{\multirow{2}{*}{Method}}&\multirow{2}{*}{SVM}&
           \multicolumn{3}{c}{CNN}&\multicolumn{3}{c}{RNN}&\multicolumn{5}{c}{GNN}&\multicolumn{4}{c}{DNN}&Trans.\\
           \cmidrule(lr){4-6}\cmidrule(lr){7-9}\cmidrule(lr){10-14}\cmidrule(lr){15-18}\cmidrule(lr){19-19}& &&\scriptsize{EEGNet}&\scriptsize{CDCN}&\scriptsize{TSception}&\scriptsize{ACRNN}&\scriptsize{BiDANN}&\scriptsize{R2G-STNN}&\scriptsize{DGCNN}&\scriptsize{GCBNet}&\scriptsize{GCBNet\quad\_BLS}&\scriptsize{RGNN}&\scriptsize{NSAL-DGAT}&\scriptsize{DBN}&\scriptsize{DAN}&\scriptsize{DANN}&\scriptsize{PR-PL}&\scriptsize{HSLT}\\ 
            \hline
            \multirow{4}*{SEED} 
            &\multirow{2}*{ACC}&\cellcolor{myback1}75.08 &\cellcolor{myback1}58.81 &\cellcolor{myback1}68.23 &\cellcolor{myback1}64.01 &\cellcolor{myback1}49.71 &\cellcolor{myback1}78.37 &\cellcolor{myback1}78.88 &\cellcolor{myback1}\textbf{82.55} &\cellcolor{myback1}\underline{80.56} &\cellcolor{myback1}76.64 &\cellcolor{myback1}76.55 &\cellcolor{myback1}78.62 &\cellcolor{myback1}71.88 &\cellcolor{myback1}60.35 &\cellcolor{myback1}62.00 &\cellcolor{myback1}76.54 &\cellcolor{myback1}64.83\\
             ~&~&\cellcolor{myback1}(19.73) &\cellcolor{myback1}(16.22) &\cellcolor{myback1}(20.35) &\cellcolor{myback1}(16.44) &\cellcolor{myback1}(13.15) &\cellcolor{myback1}(20.36) &\cellcolor{myback1}(17.34) &\cellcolor{myback1}\textbf{(15.61)} &\cellcolor{myback1}\underline{(16.98)} &\cellcolor{myback1}(17.44) &\cellcolor{myback1}(16.92) &\cellcolor{myback1}(17.48) &\cellcolor{myback1}(19.02) &\cellcolor{myback1}(17.03) &\cellcolor{myback1}(21.95) &\cellcolor{myback1}(18.80) &\cellcolor{myback1}(20.47)\\
             \cmidrule(lr){2-19}
            ~&\multirow{2}*{F1} &\cellcolor{myback2}70.82 &\cellcolor{myback2}54.41 &\cellcolor{myback2}63.76 &\cellcolor{myback2}60.53 &\cellcolor{myback2}45.78 &\cellcolor{myback2}75.91 &\cellcolor{myback2}73.72 &\cellcolor{myback2}\textbf{79.89} &\cellcolor{myback2}\underline{77.29} &\cellcolor{myback2}72.52 &\cellcolor{myback2}72.52 &\cellcolor{myback2}74.77 &\cellcolor{myback2}67.39 &\cellcolor{myback2}53.12 &\cellcolor{myback2}53.58 &\cellcolor{myback2}72.19 &\cellcolor{myback2}58.82\\
            ~&~&\cellcolor{myback2}(23.51) &\cellcolor{myback2}(17.59) &\cellcolor{myback2}(24.49) &\cellcolor{myback2}(18.51) &\cellcolor{myback2}(14.18) &\cellcolor{myback2}(24.60) &\cellcolor{myback2}(21.20) &\cellcolor{myback2}\textbf{(18.93)} &\cellcolor{myback2}\underline{(20.92)} &\cellcolor{myback2}(21.20) &\cellcolor{myback2}(20.08) &\cellcolor{myback2}(20.95) &\cellcolor{myback2}(22.81) &\cellcolor{myback2}(19.70) &\cellcolor{myback2}(26.90) &\cellcolor{myback2}(22.85) &\cellcolor{myback2}(23.36)\\
            
            \cmidrule(lr){1-19}         
            \multirow{4}*{SEED-IV} &\multirow{2}*{ACC}  &\cellcolor{myback1}47.80 &\cellcolor{myback1}29.89 &\cellcolor{myback1}52.26 &\cellcolor{myback1}36.06 &\cellcolor{myback1}29.01 &\cellcolor{myback1}51.82 &\cellcolor{myback1}53.04 &\cellcolor{myback1}52.39 &\cellcolor{myback1}53.28 &\cellcolor{myback1}\underline{53.51} &\cellcolor{myback1}45.40 &\cellcolor{myback1}53.48 &\cellcolor{myback1}45.56 &\cellcolor{myback1}44.78 &\cellcolor{myback1}46.49 &\cellcolor{myback1}\textbf{54.59} &\cellcolor{myback1}40.28 \\
            ~&~&\cellcolor{myback1}(23.03) &\cellcolor{myback1}(13.53) &\cellcolor{myback1}(21.97) &\cellcolor{myback1}(15.12) &\cellcolor{myback1}(7.10) &\cellcolor{myback1}(23.71) &\cellcolor{myback1}(23.62) &\cellcolor{myback1}(24.32) &\cellcolor{myback1}(21.05)&\cellcolor{myback1}\underline{(22.45)} &\cellcolor{myback1}(22.90) &\cellcolor{myback1}(21.21) &\cellcolor{myback1}(21.19) &\cellcolor{myback1}(23.84) &\cellcolor{myback1}(24.67) &\cellcolor{myback1}\textbf{(22.13)} &\cellcolor{myback1}(23.80) \\
            \cmidrule(lr){2-19} ~&\multirow{2}*{F1} &\cellcolor{myback2}40.17 &\cellcolor{myback2}26.59 &\cellcolor{myback2}45.26 &\cellcolor{myback2}32.77 &\cellcolor{myback2}19.80 &\cellcolor{myback2}44.65 &\cellcolor{myback2}46.48 &\cellcolor{myback2}45.94 &\cellcolor{myback2}46.26 &\cellcolor{myback2}46.91 &\cellcolor{myback2}38.24 &\cellcolor{myback2}\underline{47.09} &\cellcolor{myback2}37.61 &\cellcolor{myback2}37.18 &\cellcolor{myback2}37.76 &\cellcolor{myback2}\textbf{48.24} &\cellcolor{myback2}30.92 \\
            ~&~&\cellcolor{myback2}(21.68) &\cellcolor{myback2}(13.58) &\cellcolor{myback2}(23.00) &\cellcolor{myback2}(15.08) &\cellcolor{myback2}(5.42) &\cellcolor{myback2}(23.72) &\cellcolor{myback2}(24.76) &\cellcolor{myback2}(24.17) &\cellcolor{myback2}(22.27) &\cellcolor{myback2}(22.46) &\cellcolor{myback2}(23.09) &\cellcolor{myback2}\underline{(20.92)} &\cellcolor{myback2}(20.68) &\cellcolor{myback2}(22.60) &\cellcolor{myback2}(23.89) &\cellcolor{myback2}\textbf{(22.32)} &\cellcolor{myback2}(24.47) \\
            
            \cmidrule(lr){1-19} \multirow{4}*{HCI-V} &\multirow{2}*{ACC}  &\cellcolor{myback1}64.83 &\cellcolor{myback1}61.15 &\cellcolor{myback1}60.48 &\cellcolor{myback1}61.12 &\cellcolor{myback1}60.51 &\cellcolor{myback1}\underline{68.50} &\cellcolor{myback1}63.10 &\cellcolor{myback1}67.83 &\cellcolor{myback1}66.84 &\cellcolor{myback1}\textbf{70.60} &\cellcolor{myback1}64.86 &\cellcolor{myback1}65.17 &\cellcolor{myback1}62.03 &\cellcolor{myback1}63.55 &\cellcolor{myback1}58.94 &\cellcolor{myback1}61.18 &\cellcolor{myback1}64.00 \\
            ~&~&\cellcolor{myback1}(25.95) &\cellcolor{myback1}(16.76) &\cellcolor{myback1}(21.90) &\cellcolor{myback1}(15.52) &\cellcolor{myback1}(16.89) &\cellcolor{myback1}\underline{(24.18)} &\cellcolor{myback1}(23.41) &\cellcolor{myback1}(22.40) &\cellcolor{myback1}(21.42) &\cellcolor{myback1}\textbf{(21.05)} &\cellcolor{myback1}(17.36) &\cellcolor{myback1}(20.52) &\cellcolor{myback1}(24.90) &\cellcolor{myback1}(22.19) &\cellcolor{myback1}(19.16) &\cellcolor{myback1}(12.19) &\cellcolor{myback1}(11.40) \\
            \cmidrule(lr){2-19} ~&\multirow{2}*{F1} &\cellcolor{myback2}55.99 &\cellcolor{myback2}50.35 &\cellcolor{myback2}51.73 &\cellcolor{myback2}50.51 &\cellcolor{myback2}49.39 &\cellcolor{myback2}58.74 &\cellcolor{myback2}53.30 &\cellcolor{myback2}54.78 &\cellcolor{myback2}54.61 &\cellcolor{myback2}\textbf{58.96} &\cellcolor{myback2}50.41 &\cellcolor{myback2}\underline{58.85} &\cellcolor{myback2}52.84 &\cellcolor{myback2}48.68 &\cellcolor{myback2}44.96 &\cellcolor{myback2}41.16 &\cellcolor{myback2}55.77 \\
            ~&~&\cellcolor{myback2}(27.80) &\cellcolor{myback2}(17.28) &\cellcolor{myback2}(23.11) &\cellcolor{myback2}(16.69) &\cellcolor{myback2}(15.70) &\cellcolor{myback2}(27.88) &\cellcolor{myback2}(25.38) &\cellcolor{myback2}(26.69) &\cellcolor{myback2}(25.08) &\cellcolor{myback2}\textbf{(27.45)} &\cellcolor{myback2}(20.34) &\cellcolor{myback2}\underline{(23.09)} &\cellcolor{myback2}(27.04) &\cellcolor{myback2}(23.25) &\cellcolor{myback2}(21.05) &\cellcolor{myback2}(11.22) &\cellcolor{myback2}(13.12) \\

            \cmidrule(lr){1-19}         
            \multirow{4}*{HCI-A} &\multirow{2}*{ACC}  &\cellcolor{myback1}63.61&\cellcolor{myback1}67.42            &\cellcolor{myback1}\textbf{71.82}&\cellcolor{myback1}68.26            &\cellcolor{myback1}66.26&\cellcolor{myback1}67.62            &\cellcolor{myback1}65.72&\cellcolor{myback1}67.29            &\cellcolor{myback1}64.89&\cellcolor{myback1}69.60            &\cellcolor{myback1}\underline{70.96}&\cellcolor{myback1}65.25            &\cellcolor{myback1}68.51&\cellcolor{myback1}61.67            &\cellcolor{myback1}64.54&\cellcolor{myback1}64.57
            &\cellcolor{myback1}67.74\\
            ~&~&\cellcolor{myback1}(23.44)&\cellcolor{myback1}(21.71)&\cellcolor{myback1}\textbf{(21.72)}&\cellcolor{myback1}(23.10)&\cellcolor{myback1}(22.69)&\cellcolor{myback1}(24.58)&\cellcolor{myback1}(25.05)&\cellcolor{myback1}(27.73)&\cellcolor{myback1}(27.12)&\cellcolor{myback1}(22.09)&\cellcolor{myback1}\underline{(19.79)}&\cellcolor{myback1}(26.37)&\cellcolor{myback1}(21.72)&\cellcolor{myback1}(25.08)&\cellcolor{myback1}(19.95)&\cellcolor{myback1}(23.95)&\cellcolor{myback1}(17.22)\\
            \cmidrule(lr){2-19} ~&\multirow{2}*{F1} &\cellcolor{myback2}50.99&\cellcolor{myback2}54.50            &\cellcolor{myback2}\textbf{62.89}&\cellcolor{myback2}56.29            &\cellcolor{myback2}55.17&\cellcolor{myback2}57.77            &\cellcolor{myback2}54.82&\cellcolor{myback2}58.04            &\cellcolor{myback2}57.43&\cellcolor{myback2}57.78            &\cellcolor{myback2}57.66&\cellcolor{myback2}53.40            &\cellcolor{myback2}57.18&\cellcolor{myback2}48.25            &\cellcolor{myback2}50.16&\cellcolor{myback2}50.50
            &\cellcolor{myback2}\underline{58.42}\\
            ~&~&\cellcolor{myback2}(24.89)&\cellcolor{myback2}(20.05)&\cellcolor{myback2}\textbf{(24.99)}&\cellcolor{myback2}(23.79)&\cellcolor{myback2}(23.20)&\cellcolor{myback2}(28.86)&\cellcolor{myback2}(24.52)&\cellcolor{myback2}(31.41)&\cellcolor{myback2}(29.50)&\cellcolor{myback2}(27.18)&\cellcolor{myback2}(25.39)&\cellcolor{myback2}(27.67)&\cellcolor{myback2}(26.63)&\cellcolor{myback2}(27.22)&\cellcolor{myback2}(23.81)&\cellcolor{myback2}(26.39)&\cellcolor{myback2}\underline{(19.50)}\\

            \cmidrule(lr){1-19}         
            \multirow{4}*{HCI-VA} &\multirow{2}*{ACC}  &\cellcolor{myback1}46.29&\cellcolor{myback1}38.32            &\cellcolor{myback1}52.00&\cellcolor{myback1}40.00            &\cellcolor{myback1}41.00&\cellcolor{myback1}51.31            &\cellcolor{myback1}48.41&\cellcolor{myback1}\underline{53.06}            &\cellcolor{myback1}49.15&\cellcolor{myback1}49.24           &\cellcolor{myback1}49.46&\cellcolor{myback1}\textbf{53.35}            &\cellcolor{myback1}44.38&\cellcolor{myback1}41.88            &\cellcolor{myback1}40.68&\cellcolor{myback1}44.27
            &\cellcolor{myback1}46.99\\
            ~&~&\cellcolor{myback1}(28.42)&\cellcolor{myback1}(19.51)&\cellcolor{myback1}(26.05)&\cellcolor{myback1}(20.60)&\cellcolor{myback1}(21.58)&\cellcolor{myback1}(31.24)&\cellcolor{myback1}(25.36)&\cellcolor{myback1}\underline{(24.44)}&\cellcolor{myback1}(26.16)&\cellcolor{myback1}(27.97)&\cellcolor{myback1}(23.20)&\cellcolor{myback1}\textbf{(27.07)}&\cellcolor{myback1}(26.78)&\cellcolor{myback1}(23.17)&\cellcolor{myback1}(22.11)&\cellcolor{myback1}(23.08)&\cellcolor{myback1}(20.76)\\
            \cmidrule(lr){2-19} ~&\multirow{2}*{F1} &\cellcolor{myback2}33.91&\cellcolor{myback2}24.56            &\cellcolor{myback2}38.04&\cellcolor{myback2}27.19            &\cellcolor{myback2}27.10&\cellcolor{myback2}37.62            &\cellcolor{myback2}36.10&\cellcolor{myback2}\textbf{39.75}          &\cellcolor{myback2}36.49&\cellcolor{myback2}36.68           &\cellcolor{myback2}35.97&\cellcolor{myback2}\underline{38.42}            &\cellcolor{myback2}31.96&\cellcolor{myback2}26.62            &\cellcolor{myback2}24.72&\cellcolor{myback2}29.90
            &\cellcolor{myback2}34.76\\
            ~&~&\cellcolor{myback2}(27.65)&\cellcolor{myback2}(13.71)&\cellcolor{myback2}(26.88)&\cellcolor{myback2}(13.65)&\cellcolor{myback2}(15.13)&\cellcolor{myback2}(27.81)&\cellcolor{myback2}(22.71)&\cellcolor{myback2}\textbf{(26.01)}&\cellcolor{myback2}(27.58)&\cellcolor{myback2}(27.25)&\cellcolor{myback2}(23.84)&\cellcolor{myback2}\underline{(27.48)}&\cellcolor{myback2}(27.87)&\cellcolor{myback2}(17.51)&\cellcolor{myback2}(16.79)&\cellcolor{myback2}(20.68)&\cellcolor{myback2}(19.69)\\
            
            \cmidrule(lr){1-19}         
            \multirow{4}*{DEAP-V} &\multirow{2}*{ACC}  &\cellcolor{myback1}54.17&\cellcolor{myback1}51.50            &\cellcolor{myback1}57.71&\cellcolor{myback1}51.52            &\cellcolor{myback1}53.52&\cellcolor{myback1}55.21            &\cellcolor{myback1}55.96&\cellcolor{myback1}56.07            &\cellcolor{myback1}56.49&\cellcolor{myback1}57.02            &\cellcolor{myback1}55.90&\cellcolor{myback1}52.21            &\cellcolor{myback1}56.08&\cellcolor{myback1}\underline{57.89}
            &\cellcolor{myback1}57.70&\cellcolor{myback1}\textbf{59.23}
            &\cellcolor{myback1}56.20\\
            ~&~&\cellcolor{myback1}(18.67)&\cellcolor{myback1}(11.57)&\cellcolor{myback1}(14.72)&\cellcolor{myback1}(9.54)&\cellcolor{myback1}(9.29)&\cellcolor{myback1}(19.70)&\cellcolor{myback1}(18.30)&\cellcolor{myback1}(17.15)&\cellcolor{myback1}(18.17)&\cellcolor{myback1}(15.07)&\cellcolor{myback1}(16.24)&\cellcolor{myback1}(16.19)&\cellcolor{myback1}(17.38)&\cellcolor{myback1}\underline{(19.18)}&\cellcolor{myback1}(19.93)&\cellcolor{myback1}\textbf{(14.69)}&\cellcolor{myback1}(18.14)\\
            \cmidrule(lr){2-19} ~&\multirow{2}*{F1} &\cellcolor{myback2}49.73&\cellcolor{myback2}47.85            &\cellcolor{myback2}\textbf{53.41}&\cellcolor{myback2}47.33            &\cellcolor{myback2}48.31&\cellcolor{myback2}47.23            &\cellcolor{myback2}49.84&\cellcolor{myback2}49.08            &\cellcolor{myback2}50.36&\cellcolor{myback2}\underline{51.40}            &\cellcolor{myback2}47.25&\cellcolor{myback2}48.87            &\cellcolor{myback2}48.61&\cellcolor{myback2}50.97            &\cellcolor{myback2}50.23&\cellcolor{myback2}47.42
            &\cellcolor{myback2}48.21\\
            ~&~&\cellcolor{myback2}(18.92)&\cellcolor{myback2}(11.70)&\cellcolor{myback2}\textbf{(15.50)}&\cellcolor{myback2}(9.58)&\cellcolor{myback2}(7.77)&\cellcolor{myback2}(18.05)&\cellcolor{myback2}(17.77)&\cellcolor{myback2}(17.50)&\cellcolor{myback2}(19.57)&\cellcolor{myback2}\underline{(17.25)}&\cellcolor{myback2}(17.55)&\cellcolor{myback2}(18.34)&\cellcolor{myback2}(19.33)&\cellcolor{myback2}(18.88)&\cellcolor{myback2}(20.97)&\cellcolor{myback2}(17.67)&\cellcolor{myback2}(18.73)\\
            
            \cmidrule(lr){1-19}         
            \multirow{4}*{DEAP-A} &\multirow{2}*{ACC}  &\cellcolor{myback1}63.49&\cellcolor{myback1}61.30            &\cellcolor{myback1}63.37&\cellcolor{myback1}57.49            &\cellcolor{myback1}61.83&\cellcolor{myback1}63.57            &\cellcolor{myback1}63.78&\cellcolor{myback1}62.68            &\cellcolor{myback1}\underline{65.95}&\cellcolor{myback1}61.07        &\cellcolor{myback1}\textbf{66.09}&\cellcolor{myback1}60.31            &\cellcolor{myback1}64.60&\cellcolor{myback1}60.51            &\cellcolor{myback1}62.38&\cellcolor{myback1}62.15
            &\cellcolor{myback1}59.74\\
            ~&~&\cellcolor{myback1}(16.72)&\cellcolor{myback1}(15.88)&\cellcolor{myback1}(14.18)&\cellcolor{myback1}(11.86)&\cellcolor{myback1}(14.32)&\cellcolor{myback1}(20.33)&\cellcolor{myback1}(18.92)&\cellcolor{myback1}(19.66)&\cellcolor{myback1}\underline{(17.61)}&\cellcolor{myback1}(16.56)&\cellcolor{myback1}\textbf{(13.91)}&\cellcolor{myback1}(17.29)&\cellcolor{myback1}(19.42)&\cellcolor{myback1}(17.32)&\cellcolor{myback1}(16.78)&\cellcolor{myback1}(15.90)&\cellcolor{myback1}(18.82)\\
            \cmidrule(lr){2-19} ~&\multirow{2}*{F1} &\cellcolor{myback2}53.31&\cellcolor{myback2}53.26            &\cellcolor{myback2}\underline{53.94}&\cellcolor{myback2}50.75            &\cellcolor{myback2}49.68&\cellcolor{myback2}52.93            &\cellcolor{myback2}48.68&\cellcolor{myback2}53.94            &\cellcolor{myback2}\textbf{55.34}&\cellcolor{myback2}50.43            &\cellcolor{myback2}49.27&\cellcolor{myback2}50.21            &\cellcolor{myback2}52.61&\cellcolor{myback2}50.96            &\cellcolor{myback2}48.57&\cellcolor{myback2}53.22
            &\cellcolor{myback2}50.10\\
            ~&~&\cellcolor{myback2}(14.39)&\cellcolor{myback2}(13.05)&\cellcolor{myback2}\underline{(13.76)}&\cellcolor{myback2}(11.30)&\cellcolor{myback2}(9.20)&\cellcolor{myback2}(20.26)&\cellcolor{myback2}(17.91)&\cellcolor{myback2}(20.10)&\cellcolor{myback2}\textbf{(17.86)}&\cellcolor{myback2}(16.36)&\cellcolor{myback2}(12.86)&\cellcolor{myback2}(18.91)&\cellcolor{myback2}(19.85)&\cellcolor{myback2}(16.53)&\cellcolor{myback2}(16.96)&\cellcolor{myback2}(17.99)&\cellcolor{myback2}(18.08)\\
            
            \cmidrule(lr){1-19}         
            \multirow{4}*{DEAP-VA} &\multirow{2}*{ACC}  &\cellcolor{myback1}37.32&\cellcolor{myback1}39.41            &\cellcolor{myback1}38.08&\cellcolor{myback1}35.68            &\cellcolor{myback1}38.20&\cellcolor{myback1}39.30            &\cellcolor{myback1}40.75&\cellcolor{myback1}41.86            &\cellcolor{myback1}38.80&\cellcolor{myback1}37.51            &\cellcolor{myback1}\textbf{44.53}&\cellcolor{myback1}38.78            &\cellcolor{myback1}39.50&\cellcolor{myback1}38.07            &\cellcolor{myback1}41.71&\cellcolor{myback1}42.58
            &\cellcolor{myback1}\underline{43.34}\\
            ~&~&\cellcolor{myback1}(17.24)&\cellcolor{myback1}(11.53)&\cellcolor{myback1}(15.52)&\cellcolor{myback1}(12.08)&\cellcolor{myback1}(11.34)&\cellcolor{myback1}(15.66)&\cellcolor{myback1}(15.79)&\cellcolor{myback1}(11.57)&\cellcolor{myback1}(16.23)&\cellcolor{myback1}(15.29)&\cellcolor{myback1}\textbf{(14.35)}&\cellcolor{myback1}(15.64)&\cellcolor{myback1}(13.99)&\cellcolor{myback1}(14.12)&\cellcolor{myback1}(15.51)&\cellcolor{myback1}(13.01)&\cellcolor{myback1}\underline{(14.49)}\\
            \cmidrule(lr){2-19} ~&\multirow{2}*{F1} &\cellcolor{myback2}25.55&\cellcolor{myback2}\textbf{29.19}            &\cellcolor{myback2}28.90&\cellcolor{myback2}26.75            &\cellcolor{myback2}21.05&\cellcolor{myback2}27.32            &\cellcolor{myback2}28.63&\cellcolor{myback2}\underline{29.12}            &\cellcolor{myback2}27.89&\cellcolor{myback2}27.00            &\cellcolor{myback2}25.88&\cellcolor{myback2}25.75            &\cellcolor{myback2}24.88&\cellcolor{myback2}25.91            &\cellcolor{myback2}28.88&\cellcolor{myback2}27.63
            &\cellcolor{myback2}23.47\\
            ~&~&\cellcolor{myback2}(14.23)&\cellcolor{myback2}\textbf{(10.21)}&\cellcolor{myback2}(13.15)&\cellcolor{myback2}(8.30)&\cellcolor{myback2}(6.13)&\cellcolor{myback2}(13.23)&\cellcolor{myback2}(13.01)&\cellcolor{myback2}\underline{(10.92)}&\cellcolor{myback2}(15.94)&\cellcolor{myback2}(13.52)&\cellcolor{myback2}(12.08)&\cellcolor{myback2}(11.92)&\cellcolor{myback2}(10.79)&\cellcolor{myback2}(10.98)&\cellcolor{myback2}(15.14)&\cellcolor{myback2}(11.43)&\cellcolor{myback2}(14.21)\\
            
            \cmidrule(lr){1-19} \multirow{4}*{SEED-V} &\multirow{2}*{ACC}  &\cellcolor{myback1}34.88&\cellcolor{myback1}36.02            &\cellcolor{myback1}\textbf{65.48}&\cellcolor{myback1}31.45            &\cellcolor{myback1}34.10&\cellcolor{myback1}49.17            &\cellcolor{myback1}47.21&\cellcolor{myback1}57.59            &\cellcolor{myback1}60.68&\cellcolor{myback1}\underline{63.71}            &\cellcolor{myback1}55.31&\cellcolor{myback1}52.37            &\cellcolor{myback1}28.64&\cellcolor{myback1}22.37            &\cellcolor{myback1}27.77&\cellcolor{myback1}49.75
            &\cellcolor{myback1}47.17\\
            ~&~&\cellcolor{myback1}(18.42)&\cellcolor{myback1}(10.15)&\cellcolor{myback1}\textbf{(11.49)}&\cellcolor{myback1}(5.12)&\cellcolor{myback1}(7.21)&\cellcolor{myback1}(17.86)&\cellcolor{myback1}(19.22)&\cellcolor{myback1}(20.36)&\cellcolor{myback1}(18.42)&\cellcolor{myback1}\underline{(17.03)}&\cellcolor{myback1}(12.88)&\cellcolor{myback1}(19.83)&\cellcolor{myback1}(15.62)&\cellcolor{myback1}(4.31)&\cellcolor{myback1}(17.64)&\cellcolor{myback1}(16.01)&\cellcolor{myback1}(16.82)\\
            \cmidrule(lr){2-19} ~&\multirow{2}*{F1} &\cellcolor{myback2}29.83&\cellcolor{myback2}26.77            &\cellcolor{myback2}49.51&\cellcolor{myback2}20.05            &\cellcolor{myback2}26.29&\cellcolor{myback2}36.82            &\cellcolor{myback2}38.02&\cellcolor{myback2}48.59            &\cellcolor{myback2}\underline{50.13}&\cellcolor{myback2}\textbf{51.31}            &\cellcolor{myback2}46.55&\cellcolor{myback2}46.25            &\cellcolor{myback2}22.42&\cellcolor{myback2}7.54            &\cellcolor{myback2}20.02&\cellcolor{myback2}39.25
            &\cellcolor{myback2}31.61\\
            ~&~&\cellcolor{myback2}(16.75)&\cellcolor{myback2}(8.11)&\cellcolor{myback2}(12.34)&\cellcolor{myback2}(2.59)&\cellcolor{myback2}(4.43)&\cellcolor{myback2}(14.18)&\cellcolor{myback2}(16.74)&\cellcolor{myback2}(20.88)&\cellcolor{myback2}\underline{(19.81)}&\cellcolor{myback2}\textbf{(18.91)}&\cellcolor{myback2}(22.17)&\cellcolor{myback2}(19.34)&\cellcolor{myback2}(15.51)&\cellcolor{myback2}(2.16)&\cellcolor{myback2}(17.58)&\cellcolor{myback2}(15.51)&\cellcolor{myback2}(12.85)\\
            
            \cmidrule(lr){1-19}  \multirow{4}*{MPED} &\multirow{2}*{ACC}  &\cellcolor{myback1}21.49&\cellcolor{myback1}18.43            &\cellcolor{myback1}20.89&\cellcolor{myback1}19.01            &\cellcolor{myback1}18.88&\cellcolor{myback1}\textbf{25.75}        &\cellcolor{myback1}\underline{24.43}&\cellcolor{myback1}22.93            &\cellcolor{myback1}22.83&\cellcolor{myback1}22.65            &\cellcolor{myback1}20.72&\cellcolor{myback1}22.19            &\cellcolor{myback1}22.22&\cellcolor{myback1}19.47            &\cellcolor{myback1}21.24&\cellcolor{myback1}20.40
            &\cellcolor{myback1}18.47\\
            ~&~&\cellcolor{myback1}(7.33)&\cellcolor{myback1}(8.99)&\cellcolor{myback1}(5.48)&\cellcolor{myback1}(6.77)&\cellcolor{myback1}(9.97)&\cellcolor{myback1}\textbf{(7.97)}&\cellcolor{myback1}\underline{(9.28)}&\cellcolor{myback1}(6.98)&\cellcolor{myback1}(8.30)&\cellcolor{myback1}(8.06)&\cellcolor{myback1}(5.34)&\cellcolor{myback1}(8.23)&\cellcolor{myback1}(5.17)&\cellcolor{myback1}(7.38)&\cellcolor{myback1}(7.52)&\cellcolor{myback1}(7.48)&\cellcolor{myback1}(4.28)\\
            \cmidrule(lr){2-19} ~&\multirow{2}*{F1} &\cellcolor{myback2}17.97&\cellcolor{myback2}17.80            &\cellcolor{myback2}17.00&\cellcolor{myback2}12.63            &\cellcolor{myback2}13.27&\cellcolor{myback2}\textbf{21.26}        &\cellcolor{myback2}\underline{21.03}&\cellcolor{myback2}19.10            &\cellcolor{myback2}18.88&\cellcolor{myback2}19.00            &\cellcolor{myback2}19.06&\cellcolor{myback2}18.67            &\cellcolor{myback2}16.76&\cellcolor{myback2}14.83            &\cellcolor{myback2}17.50&\cellcolor{myback2}17.86
            &\cellcolor{myback2}16.54\\
            ~&~&\cellcolor{myback2}(5.97)&\cellcolor{myback2}(6.58)&\cellcolor{myback2}(4.74)&\cellcolor{myback2}(6.80)&\cellcolor{myback2}(6.04)&\cellcolor{myback2}\textbf{(6.81)}&\cellcolor{myback2}\underline{(7.64)}&\cellcolor{myback2}(6.28)&\cellcolor{myback2}(8.00)&\cellcolor{myback2}(6.70)&\cellcolor{myback2}(4.88)&\cellcolor{myback2}(7.45)&\cellcolor{myback2}(4.42)&\cellcolor{myback2}(7.39)&\cellcolor{myback2}(9.52)&\cellcolor{myback2}(6.61)&\cellcolor{myback2}(5.96)\\

            \noalign{\global\arrayrulewidth=1pt}
            \hline
        \end{tabular}
    }}
\end{sidewaystable*}

\begin{sidewaystable*}[htp]
	\caption{The mean accuracies and F1 scores (and standard deviations) using the proposed benchmark for cross-subject EER experiment. The top two methods in each scenario are highlighted using bold and underlined formatting.}
	\label{cross}
	\renewcommand{\arraystretch}{1}
	\centering
	{\footnotesize{
        \begin{tabular}{m{1cm}<{\centering}m{0.5cm}<{\centering}m{0.8cm}<{\centering}m{0.8cm}<{\centering}m{0.8cm}<{\centering}m{0.8cm}<{\centering}m{0.8cm}<{\centering}m{0.8cm}<{\centering}m{0.8cm}<{\centering}m{0.8cm}<{\centering}m{0.8cm}<{\centering}m{0.8cm}<{\centering}m{0.8cm}<{\centering}m{0.8cm}<{\centering}m{0.8cm}<{\centering}m{0.8cm}<{\centering}m{0.8cm}<{\centering}m{0.8cm}<{\centering}m{0.8cm}<{\centering}m{0.8cm}<{\centering}}
            \noalign{\global\arrayrulewidth=2pt}
            \hline
            \noalign{\global\arrayrulewidth=0.5pt}	
           \multicolumn{2}{c}{\multirow{2}{*}{Method}}&\multirow{2}{*}{SVM}&
           \multicolumn{3}{c}{CNN}&\multicolumn{3}{c}{RNN}&\multicolumn{5}{c}{GNN}&\multicolumn{5}{c}{DNN}&Trans.\\
           \cmidrule(lr){4-6}\cmidrule(lr){7-9}\cmidrule(lr){10-14}\cmidrule(lr){15-19}\cmidrule(lr){20-20}& &&\scriptsize{EEGNet}&\scriptsize{CDCN}&\scriptsize{TSception}&\scriptsize{ACRNN}&\scriptsize{BiDANN}&\scriptsize{R2G-STNN}&\scriptsize{DGCNN}&\scriptsize{GCBNet}&\scriptsize{GCBNet\quad\_BLS}&\scriptsize{RGNN}&\scriptsize{NSAL-DGAT}&\scriptsize{DBN}&\scriptsize{DAN}&\scriptsize{DANN}&\scriptsize{MS-MDA}&\scriptsize{PR-PL}&\scriptsize{HSLT}\\ 
            \hline
             
            \multirow{4}*{SEED} &\multirow{2}*{ACC} & \cellcolor{myback1}37.07 & \cellcolor{myback1}38.19 & \cellcolor{myback1}57.72 & \cellcolor{myback1}45.60 & \cellcolor{myback1}45.39 & \cellcolor{myback1}59.49 & \cellcolor{myback1}56.84 & \cellcolor{myback1}60.87 & \cellcolor{myback1}56.32 & \cellcolor{myback1}56.32 & \cellcolor{myback1}59.45 & \cellcolor{myback1}\textbf{64.40} & \cellcolor{myback1}36.16 & \cellcolor{myback1}44.49 & \cellcolor{myback1}63.89 & \cellcolor{myback1}\underline{64.00} & \cellcolor{myback1}56.33 & \cellcolor{myback1}56.00\\
            ~&~ & \cellcolor{myback1}(15.54) & \cellcolor{myback1}(17.25) & \cellcolor{myback1}(16.50) & \cellcolor{myback1}(8.71) & \cellcolor{myback1}(10.09) & \cellcolor{myback1}(15.92) & \cellcolor{myback1}(19.20) & \cellcolor{myback1}(15.62) & \cellcolor{myback1}(14.73) & \cellcolor{myback1}(9.91) & \cellcolor{myback1}(15.28) & \cellcolor{myback1}\textbf{(19.00)} & \cellcolor{myback1}(6.32) & \cellcolor{myback1}(7.68) & \cellcolor{myback1}(1.37) & \cellcolor{myback1}\underline{(18.75)} & \cellcolor{myback1}(4.69) & \cellcolor{myback1}(14.76)\\
            \cmidrule(lr){2-20} ~&\multirow{2}*{F1} & \cellcolor{myback2}33.45 & \cellcolor{myback2}31.83 & \cellcolor{myback2}\textbf{58.66} & \cellcolor{myback2}43.54 & \cellcolor{myback2}42.37 & \cellcolor{myback2}48.18 & \cellcolor{myback2}51.61 & \cellcolor{myback2}57.22 & \cellcolor{myback2}55.12 & \cellcolor{myback2}51.43 & \cellcolor{myback2}56.31 & \cellcolor{myback2}\underline{58.47} & \cellcolor{myback2}22.67 & \cellcolor{myback2}33.79 & \cellcolor{myback2}53.37 & \cellcolor{myback2}57.35 & \cellcolor{myback2}48.02 & \cellcolor{myback2}55.75\\
            ~&~ & \cellcolor{myback2}(10.98) & \cellcolor{myback2}(12.33) & \cellcolor{myback2}\textbf{(13.47)} & \cellcolor{myback2}(9.72) & \cellcolor{myback2}(9.74) & \cellcolor{myback2}(18.23) & \cellcolor{myback2}(13.40) & \cellcolor{myback2}(16.41) & \cellcolor{myback2}(12.88) & \cellcolor{myback2}(6.54) & \cellcolor{myback2}(12.56) & \cellcolor{myback2}\underline{(19.50)} & \cellcolor{myback2}(2.57) & \cellcolor{myback2}(5.60) & \cellcolor{myback2}(2.98) & \cellcolor{myback2}(18.67) & \cellcolor{myback2}(2.36) & \cellcolor{myback2}(14.73)\\

            \cmidrule(lr){1-20}   \multirow{4}*{SEED-IV} &\multirow{2}*{ACC} & \cellcolor{myback1}28.98 & \cellcolor{myback1}28.19 & \cellcolor{myback1}31.03 & \cellcolor{myback1}34.19 & \cellcolor{myback1}31.97 & \cellcolor{myback1}42.28 & \cellcolor{myback1}42.39 & \cellcolor{myback1}42.54 & \cellcolor{myback1}32.27 & \cellcolor{myback1}40.54 & \cellcolor{myback1}48.23 & \cellcolor{myback1}\underline{52.29} & \cellcolor{myback1}36.82 & \cellcolor{myback1}38.24 & \cellcolor{myback1}25.55 & \cellcolor{myback1}\textbf{56.07} & \cellcolor{myback1}43.42 & \cellcolor{myback1}30.33\\
            ~&~ & \cellcolor{myback1}(4.09) & \cellcolor{myback1}(3.82) & \cellcolor{myback1}(9.45) & \cellcolor{myback1}(6.99) & \cellcolor{myback1}(7.02) & \cellcolor{myback1}(15.67) & \cellcolor{myback1}(16.02) & \cellcolor{myback1}(22.09) & \cellcolor{myback1}(11.43) & \cellcolor{myback1}(7.33) & \cellcolor{myback1}(16.14) & \cellcolor{myback1}\underline{(16.89)} & \cellcolor{myback1}(12.81) & \cellcolor{myback1}(11.86) & \cellcolor{myback1}(2.54) & \cellcolor{myback1}\textbf{(15.45)} & \cellcolor{myback1}(5.48) & \cellcolor{myback1}(5.47)\\
            \cmidrule(lr){2-20} ~&\multirow{2}*{F1} & \cellcolor{myback2}24.56 & \cellcolor{myback2}28.35 & \cellcolor{myback2}27.01 & \cellcolor{myback2}26.83 & \cellcolor{myback2}18.82 & \cellcolor{myback2}37.31 & \cellcolor{myback2}42.76 & \cellcolor{myback2}43.10 & \cellcolor{myback2}32.89 & \cellcolor{myback2}42.73 & \cellcolor{myback2}\underline{45.43} & \cellcolor{myback2}45.39 & \cellcolor{myback2}32.60 & \cellcolor{myback2}26.04 & \cellcolor{myback2}11.99 & \cellcolor{myback2}\textbf{48.68} & \cellcolor{myback2}42.37 & \cellcolor{myback2}11.64\\
            ~&~ & \cellcolor{myback2}(5.31) & \cellcolor{myback2}(5.44) & \cellcolor{myback2}(6.63) & \cellcolor{myback2}(6.84) & \cellcolor{myback2}(6.80) & \cellcolor{myback2}(14.14) & \cellcolor{myback2}(17.54) & \cellcolor{myback2}(23.71) & \cellcolor{myback2}(12.53) & \cellcolor{myback2}(5.89) & \cellcolor{myback2}\underline{(14.89)} & \cellcolor{myback2}(13.29) & \cellcolor{myback2}(12.81) & \cellcolor{myback2}(11.74) & \cellcolor{myback2}(1.96) & \cellcolor{myback2}\textbf{(13.78)} & \cellcolor{myback2}(2.39) & \cellcolor{myback2}(2.34)\\

            \cmidrule(lr){1-20} \multirow{4}*{HCI-V} &\multirow{2}*{ACC} & \cellcolor{myback1}\underline{70.33} & \cellcolor{myback1}57.06 & \cellcolor{myback1}67.69 & \cellcolor{myback1}57.36 & \cellcolor{myback1}54.53 & \cellcolor{myback1}73.92 & \cellcolor{myback1}64.76 & \cellcolor{myback1}63.19 & \cellcolor{myback1}65.16 & \cellcolor{myback1}\textbf{71.06} & \cellcolor{myback1}64.94 & \cellcolor{myback1}64.04 & \cellcolor{myback1}69.27 & \cellcolor{myback1}61.50 & \cellcolor{myback1}63.61 & \cellcolor{myback1}67.64 & \cellcolor{myback1}68.79 & \cellcolor{myback1}66.94\\
            ~&~ & \cellcolor{myback1}\underline{(8.74)} & \cellcolor{myback1}(8.14) & \cellcolor{myback1}(8.15) & \cellcolor{myback1}(10.99) & \cellcolor{myback1}(9.87) & \cellcolor{myback1}(10.56) & \cellcolor{myback1}(13.28) & \cellcolor{myback1}(8.13) & \cellcolor{myback1}(6.76) & \cellcolor{myback1}\textbf{(13.33)} & \cellcolor{myback1}(6.72) & \cellcolor{myback1}(6.67) & \cellcolor{myback1}(8.09) & \cellcolor{myback1}(11.49) & \cellcolor{myback1}(6.57) & \cellcolor{myback1}(7.88) & \cellcolor{myback1}(10.43) & \cellcolor{myback1}(5.52)\\
            \cmidrule(lr){2-20} ~&\multirow{2}*{F1} & \cellcolor{myback2}\textbf{66.36} & \cellcolor{myback2}53.83 & \cellcolor{myback2}62.67 & \cellcolor{myback2}54.76 & \cellcolor{myback2}52.58 & \cellcolor{myback2}\underline{66.02} & \cellcolor{myback2}59.54 & \cellcolor{myback2}58.75 & \cellcolor{myback2}63.32 & \cellcolor{myback2}61.83 & \cellcolor{myback2}52.50 & \cellcolor{myback2}38.96 & \cellcolor{myback2}65.51 & \cellcolor{myback2}50.97 & \cellcolor{myback2}38.80 & \cellcolor{myback2}53.70 & \cellcolor{myback2}54.90 & \cellcolor{myback2}62.22\\
            ~&~ & \cellcolor{myback2}\textbf{(5.89)} & \cellcolor{myback2}(5.59) & \cellcolor{myback2}(10.57) & \cellcolor{myback2}(9.28) & \cellcolor{myback2}(4.22) & \cellcolor{myback2}\underline{(14.04)} & \cellcolor{myback2}(13.41) & \cellcolor{myback2}(10.52) & \cellcolor{myback2}(2.33) & \cellcolor{myback2}(9.43) & \cellcolor{myback2}(12.87) & \cellcolor{myback2}(2.26) & \cellcolor{myback2}(12.54) & \cellcolor{myback2}(9.86) & \cellcolor{myback2}(2.22) & \cellcolor{myback2}(3.41) & \cellcolor{myback2}(14.21) & \cellcolor{myback2}(4.53)\\

            \cmidrule(lr){1-20}   \multirow{4}*{HCI-A} &\multirow{2}*{ACC} & \cellcolor{myback1}54.41 & \cellcolor{myback1}54.70 & \cellcolor{myback1}55.93 & \cellcolor{myback1}52.30 & \cellcolor{myback1}51.23 & \cellcolor{myback1}49.98 & \cellcolor{myback1}56.21 & \cellcolor{myback1}59.42 & \cellcolor{myback1}52.21 & \cellcolor{myback1}\textbf{60.85} & \cellcolor{myback1}57.36 & \cellcolor{myback1}51.00 & \cellcolor{myback1}\underline{57.50} & \cellcolor{myback1}56.42 & \cellcolor{myback1}49.06 & \cellcolor{myback1}57.23 & \cellcolor{myback1}55.48 & \cellcolor{myback1}49.48\\
            ~&~ & \cellcolor{myback1}(14.26) & \cellcolor{myback1}(14.46) & \cellcolor{myback1}(19.96) & \cellcolor{myback1}(16.16) & \cellcolor{myback1}(16.66) & \cellcolor{myback1}(22.61) & \cellcolor{myback1}(13.33) & \cellcolor{myback1}(19.93) & \cellcolor{myback1}(10.83) & \cellcolor{myback1}\textbf{(16.25)} & \cellcolor{myback1}(22.72) & \cellcolor{myback1}(23.13) & \cellcolor{myback1}\underline{(9.23)} & \cellcolor{myback1}(18.05) & \cellcolor{myback1}(23.43) & \cellcolor{myback1}(21.98) & \cellcolor{myback1}(16.73) & \cellcolor{myback1}(18.17)\\
            \cmidrule(lr){2-20} ~&\multirow{2}*{F1} & \cellcolor{myback2}52.91 & \cellcolor{myback2}54.02 & \cellcolor{myback2}55.15 & \cellcolor{myback2}50.25 & \cellcolor{myback2}49.45 & \cellcolor{myback2}48.53 & \cellcolor{myback2}55.24 & \cellcolor{myback2}\textbf{57.02} & \cellcolor{myback2}51.84 & \cellcolor{myback2}56.26 & \cellcolor{myback2}\underline{56.90} & \cellcolor{myback2}32.41 & \cellcolor{myback2}56.24 & \cellcolor{myback2}45.35 & \cellcolor{myback2}34.68 & \cellcolor{myback2}55.68 & \cellcolor{myback2}54.45 & \cellcolor{myback2}47.43\\
            ~&~ & \cellcolor{myback2}(11.20) & \cellcolor{myback2}(10.21) & \cellcolor{myback2}(11.60) & \cellcolor{myback2}(11.56) & \cellcolor{myback2}(13.70) & \cellcolor{myback2}(12.59) & \cellcolor{myback2}(10.27) & \cellcolor{myback2}\textbf{(11.56)} & \cellcolor{myback2}(12.32) & \cellcolor{myback2}(12.93) & \cellcolor{myback2}\underline{(11.78)} & \cellcolor{myback2}(10.08) & \cellcolor{myback2}(5.51) & \cellcolor{myback2}(12.41) & \cellcolor{myback2}(10.36) & \cellcolor{myback2}(9.65) & \cellcolor{myback2}(9.92) & \cellcolor{myback2}(13.90)\\
            
            \cmidrule(lr){1-20}  \multirow{4}*{HCI-VA} &\multirow{2}*{ACC} & \cellcolor{myback1}31.56 & \cellcolor{myback1}34.84 & \cellcolor{myback1}30.09 & \cellcolor{myback1}26.99 & \cellcolor{myback1}27.21 & \cellcolor{myback1}36.24 & \cellcolor{myback1}38.25 & \cellcolor{myback1}41.54 & \cellcolor{myback1}43.02 & \cellcolor{myback1}36.92 & \cellcolor{myback1}42.63 & \cellcolor{myback1}38.66 & \cellcolor{myback1}28.30 & \cellcolor{myback1}40.35 & \cellcolor{myback1}28.00 & \cellcolor{myback1}\textbf{44.92} & \cellcolor{myback1}\underline{44.24} & \cellcolor{myback1}35.07\\
            ~&~ & \cellcolor{myback1}(9.12) & \cellcolor{myback1}(9.32) & \cellcolor{myback1}(9.81) & \cellcolor{myback1}(8.81) & \cellcolor{myback1}(12.14) & \cellcolor{myback1}(19.56) & \cellcolor{myback1}(11.10) & \cellcolor{myback1}(9.82) & \cellcolor{myback1}(12.60) & \cellcolor{myback1}(18.54) & \cellcolor{myback1}(17.44) & \cellcolor{myback1}(22.77) & \cellcolor{myback1}(19.84) & \cellcolor{myback1}(22.83) & \cellcolor{myback1}(18.26) & \cellcolor{myback1}\textbf{(20.67)} & \cellcolor{myback1}\underline{(17.24)} & \cellcolor{myback1}(13.21)\\
            \cmidrule(lr){2-20} ~&\multirow{2}*{F1} & \cellcolor{myback2}26.29 & \cellcolor{myback2}27.96 & \cellcolor{myback2}24.71 & \cellcolor{myback2}21.95 & \cellcolor{myback2}20.92 & \cellcolor{myback2}24.42 & \cellcolor{myback2}28.48 & \cellcolor{myback2}\textbf{38.08} & \cellcolor{myback2}\underline{36.07} & \cellcolor{myback2}32.99 & \cellcolor{myback2}25.28 & \cellcolor{myback2}22.13 & \cellcolor{myback2}27.58 & \cellcolor{myback2}20.71 & \cellcolor{myback2}13.12 & \cellcolor{myback2}21.65 & \cellcolor{myback2}22.93 & \cellcolor{myback2}27.19\\
            ~&~ & \cellcolor{myback2}(3.57) & \cellcolor{myback2}(3.12) & \cellcolor{myback2}(2.47) & \cellcolor{myback2}(6.24) & \cellcolor{myback2}(4.61) & \cellcolor{myback2}(5.91) & \cellcolor{myback2}(5.52) & \cellcolor{myback2}\textbf{(2.43)} & \cellcolor{myback2}\underline{(7.58)} & \cellcolor{myback2}(12.41) & \cellcolor{myback2}(6.22) & \cellcolor{myback2}(8.92) & \cellcolor{myback2}(8.97) & \cellcolor{myback2}(7.13) & \cellcolor{myback2}(6.28) & \cellcolor{myback2}(8.94) & \cellcolor{myback2}(2.29) & \cellcolor{myback2}(5.64)\\
            
            \cmidrule(lr){1-20}   \multirow{4}*{DEAP-V} &\multirow{2}*{ACC} & \cellcolor{myback1}49.58 & \cellcolor{myback1}52.36 & \cellcolor{myback1}\textbf{57.78} & \cellcolor{myback1}54.44 & \cellcolor{myback1}51.94 & \cellcolor{myback1}53.49 & \cellcolor{myback1}49.97 & \cellcolor{myback1}49.91 & \cellcolor{myback1}53.58 & \cellcolor{myback1}52.34 & \cellcolor{myback1}52.65 & \cellcolor{myback1}52.22 & \cellcolor{myback1}55.02 & \cellcolor{myback1}51.88 & \cellcolor{myback1}44.32 & \cellcolor{myback1}54.82 & \cellcolor{myback1}51.77 & \cellcolor{myback1}\underline{56.56}\\
            ~&~ & \cellcolor{myback1}(6.77) & \cellcolor{myback1}(6.81) & \cellcolor{myback1}\textbf{(5.00)} & \cellcolor{myback1}(6.33) & \cellcolor{myback1}(7.44) & \cellcolor{myback1}(7.07) & \cellcolor{myback1}(7.19) & \cellcolor{myback1}(7.48) & \cellcolor{myback1}(8.72) & \cellcolor{myback1}(4.51) & \cellcolor{myback1}(8.09) & \cellcolor{myback1}(11.83) & \cellcolor{myback1}(7.19) & \cellcolor{myback1}(7.90) & \cellcolor{myback1}(5.94) & \cellcolor{myback1}(10.42) & \cellcolor{myback1}(10.60) & \cellcolor{myback1}\underline{(5.79)}\\
            \cmidrule(lr){2-20} ~&\multirow{2}*{F1} & \cellcolor{myback2}48.49 & \cellcolor{myback2}49.74 & \cellcolor{myback2}\textbf{57.72} & \cellcolor{myback2}48.94 & \cellcolor{myback2}47.37 & \cellcolor{myback2}45.96 & \cellcolor{myback2}47.75 & \cellcolor{myback2}47.09 & \cellcolor{myback2}50.68 & \cellcolor{myback2}49.03 & \cellcolor{myback2}50.22 & \cellcolor{myback2}45.10 & \cellcolor{myback2}53.14 & \cellcolor{myback2}44.61 & \cellcolor{myback2}30.61 & \cellcolor{myback2}50.60 & \cellcolor{myback2}40.21 & \cellcolor{myback2}\underline{56.56}\\
            ~&~ & \cellcolor{myback2}(6.34) & \cellcolor{myback2}(6.32) & \cellcolor{myback2}\textbf{(6.00)} & \cellcolor{myback2}(7.27) & \cellcolor{myback2}(6.66) & \cellcolor{myback2}(9.86) & \cellcolor{myback2}(7.75) & \cellcolor{myback2}(7.31) & \cellcolor{myback2}(11.20) & \cellcolor{myback2}(5.28) & \cellcolor{myback2}(6.23) & \cellcolor{myback2}(10.30) & \cellcolor{myback2}(7.16) & \cellcolor{myback2}(9.50) & \cellcolor{myback2}(2.58) & \cellcolor{myback2}(9.65) & \cellcolor{myback2}(13.48) & \cellcolor{myback2}\underline{(2.21)}\\
            
            \cmidrule(lr){1-20}    \multirow{4}*{DEAP-A} &\multirow{2}*{ACC} & \cellcolor{myback1}\textbf{51.48} & \cellcolor{myback1}48.94 & \cellcolor{myback1}49.73 & \cellcolor{myback1}45.90 & \cellcolor{myback1}44.09 & \cellcolor{myback1}50.07 & \cellcolor{myback1}49.13 & \cellcolor{myback1}49.91 & \cellcolor{myback1}50.05 & \cellcolor{myback1}50.59 & \cellcolor{myback1}46.85 & \cellcolor{myback1}50.94 & \cellcolor{myback1}\underline{50.99} & \cellcolor{myback1}36.86 & \cellcolor{myback1}49.32 & \cellcolor{myback1}26.48 & \cellcolor{myback1}45.01 & \cellcolor{myback1}41.08\\
            ~&~ & \cellcolor{myback2}\textbf{(7.76)} & \cellcolor{myback2}(7.72) & \cellcolor{myback2}(9.38) & \cellcolor{myback2}(8.16) & \cellcolor{myback2}(16.08) & \cellcolor{myback2}(9.01) & \cellcolor{myback2}(7.49) & \cellcolor{myback2}(9.27) & \cellcolor{myback2}(10.60) & \cellcolor{myback2}(19.22) & \cellcolor{myback2}(16.25) & \cellcolor{myback2}(9.60) & \cellcolor{myback2}\underline{(13.86)} & \cellcolor{myback2}(10.64) & \cellcolor{myback2}(16.56) & \cellcolor{myback2}(13.21) & \cellcolor{myback2}(13.05) & \cellcolor{myback2}(14.32)\\
            \cmidrule(lr){2-20} ~&\multirow{2}*{F1} & \cellcolor{myback2}\textbf{50.95} & \cellcolor{myback2}48.94 & \cellcolor{myback2}49.17 & \cellcolor{myback2}45.56 & \cellcolor{myback2}41.51 & \cellcolor{myback2}46.87 & \cellcolor{myback2}47.18 & \cellcolor{myback2}47.09 & \cellcolor{myback2}47.79 & \cellcolor{myback2}46.38 & \cellcolor{myback2}43.90 & \cellcolor{myback2}\underline{50.75} & \cellcolor{myback2}50.09 & \cellcolor{myback2}32.68 & \cellcolor{myback2}33.26 & \cellcolor{myback2}25.39 & \cellcolor{myback2}37.26 & \cellcolor{myback2}41.05\\
            ~&~ & \cellcolor{myback1}\textbf{(6.88)} & \cellcolor{myback1}(6.75) & \cellcolor{myback1}(9.73) & \cellcolor{myback1}(7.29) & \cellcolor{myback1}(7.69) & \cellcolor{myback1}(8.20) & \cellcolor{myback1}(5.83) & \cellcolor{myback1}(6.12) & \cellcolor{myback1}(6.37) & \cellcolor{myback1}(11.85) & \cellcolor{myback1}(10.58) & \cellcolor{myback1}\underline{(8.46)} & \cellcolor{myback1}(12.74) & \cellcolor{myback1}(9.55) & \cellcolor{myback1}(6.78) & \cellcolor{myback1}(10.91) & \cellcolor{myback1}(10.96) & \cellcolor{myback1}(13.94)\\
            
            \cmidrule(lr){1-20}  \multirow{4}*{DEAP-VA} &\multirow{2}*{ACC} & \cellcolor{myback1}24.58 & \cellcolor{myback1}25.41 & \cellcolor{myback1}23.90 & \cellcolor{myback1}24.64 & \cellcolor{myback1}20.89 & \cellcolor{myback1}\underline{29.30} & \cellcolor{myback1}28.02 & \cellcolor{myback1}25.66 & \cellcolor{myback1}\textbf{30.92} & \cellcolor{myback1}20.49 & \cellcolor{myback1}22.40 & \cellcolor{myback1}26.44 & \cellcolor{myback1}27.84 & \cellcolor{myback1}23.87 & \cellcolor{myback1}26.44 & \cellcolor{myback1}16.06 & \cellcolor{myback1}22.85 & \cellcolor{myback1}17.31\\
            ~&~ & \cellcolor{myback2}(8.90) & \cellcolor{myback2}(8.86) & \cellcolor{myback2}(9.11) & \cellcolor{myback2}(8.69) & \cellcolor{myback2}(14.62) & \cellcolor{myback2}\underline{(8.34)} & \cellcolor{myback2}(9.12) & \cellcolor{myback2}(8.22) & \cellcolor{myback2}\textbf{(12.73)} & \cellcolor{myback2}(11.13) & \cellcolor{myback2}(12.36) & \cellcolor{myback2}(14.74) & \cellcolor{myback2}(13.66) & \cellcolor{myback2}(6.08) & \cellcolor{myback2}(14.74) & \cellcolor{myback2}(14.23) & \cellcolor{myback2}(8.14) & \cellcolor{myback2}(14.77)\\
            \cmidrule(lr){2-20} ~&\multirow{2}*{F1} & \cellcolor{myback2}24.70 & \cellcolor{myback2}24.44 & \cellcolor{myback2}22.58 & \cellcolor{myback2}23.24 & \cellcolor{myback2}15.16 & \cellcolor{myback2}23.36 & \cellcolor{myback2}23.14 & \cellcolor{myback2}\underline{24.95} & \cellcolor{myback2}\textbf{30.98} & \cellcolor{myback2}18.41 & \cellcolor{myback2}16.85 & \cellcolor{myback2}9.97 & \cellcolor{myback2}24.36 & \cellcolor{myback2}22.33 & \cellcolor{myback2}9.97 & \cellcolor{myback2}15.19 & \cellcolor{myback2}15.49 & \cellcolor{myback2}16.87\\
            ~&~ & \cellcolor{myback1}(6.98) & \cellcolor{myback1}(6.49) & \cellcolor{myback1}(6.88) & \cellcolor{myback1}(7.46) & \cellcolor{myback1}(7.56) & \cellcolor{myback1}(4.24) & \cellcolor{myback1}(8.70) & \cellcolor{myback1}\underline{(5.32)} & \cellcolor{myback1}\textbf{(5.27)} & \cellcolor{myback1}(10.33) & \cellcolor{myback1}(6.13) & \cellcolor{myback1}(4.58) & \cellcolor{myback1}(5.53) & \cellcolor{myback1}(5.47) & \cellcolor{myback1}(4.58) & \cellcolor{myback1}(5.02) & \cellcolor{myback1}(6.62) & \cellcolor{myback1}(5.17)\\
            
            \cmidrule(lr){1-20}   \multirow{4}*{SEED-V} &\multirow{2}*{ACC} & \cellcolor{myback1}28.03 & \cellcolor{myback1}32.27 & \cellcolor{myback1}33.84 & \cellcolor{myback1}32.63 & \cellcolor{myback1}27.03 & \cellcolor{myback1}36.58 & \cellcolor{myback1}37.23 & \cellcolor{myback1}36.25 & \cellcolor{myback1}30.84 & \cellcolor{myback1}35.68 & \cellcolor{myback1}36.45 & \cellcolor{myback1}\underline{41.56} & \cellcolor{myback1}32.18 & \cellcolor{myback1}20.82 & \cellcolor{myback1}23.99 & \cellcolor{myback1}\textbf{44.33} & \cellcolor{myback1}33.77 & \cellcolor{myback1}30.59\\
            ~&~ & \cellcolor{myback1}(10.91) & \cellcolor{myback1}(3.57) & \cellcolor{myback1}(9.87) & \cellcolor{myback1}(4.90) & \cellcolor{myback1}(12.76) & \cellcolor{myback1}(12.89) & \cellcolor{myback1}(12.47) & \cellcolor{myback1}(2.61) & \cellcolor{myback1}(6.51) & \cellcolor{myback1}(13.22) & \cellcolor{myback1}(5.65) & \cellcolor{myback1}\underline{(13.19)} & \cellcolor{myback1}(6.56) & \cellcolor{myback1}(5.45) & \cellcolor{myback1}(3.98) & \cellcolor{myback1}\textbf{(12.09)} & \cellcolor{myback1}(2.24) & \cellcolor{myback1}(12.13)\\
            \cmidrule(lr){2-20} ~&\multirow{2}*{F1} & \cellcolor{myback2}19.54 & \cellcolor{myback2}22.22 & \cellcolor{myback2}27.45 & \cellcolor{myback2}25.33 & \cellcolor{myback2}18.27 & \cellcolor{myback2}32.72 & \cellcolor{myback2}31.48 & \cellcolor{myback2}29.77 & \cellcolor{myback2}27.70 & \cellcolor{myback2}30.02 & \cellcolor{myback2}31.93 & \cellcolor{myback2}\underline{35.33} & \cellcolor{myback2}25.29 & \cellcolor{myback2}14.97 & \cellcolor{myback2}17.96 & \cellcolor{myback2}\textbf{38.28} & \cellcolor{myback2}29.64 & \cellcolor{myback2}26.54\\
            ~&~ & \cellcolor{myback2}(11.93) & \cellcolor{myback2}(3.42) & \cellcolor{myback2}(11.00) & \cellcolor{myback2}(5.61) & \cellcolor{myback2}(11.65) & \cellcolor{myback2}(12.71) & \cellcolor{myback2}(13.62) & \cellcolor{myback2}(2.25) & \cellcolor{myback2}(3.64) & \cellcolor{myback2}(16.43) & \cellcolor{myback2}(10.42) & \cellcolor{myback2}\underline{(12.84)} & \cellcolor{myback2}(7.71) & \cellcolor{myback2}(4.54) & \cellcolor{myback2}(4.08) & \cellcolor{myback2}\textbf{(13.48)} & \cellcolor{myback2}(3.00) & \cellcolor{myback2}(5.87)\\

            \cmidrule(lr){1-20}  \multirow{4}*{MPED} &\multirow{2}*{ACC} & \cellcolor{myback1}19.56 & \cellcolor{myback1}20.33 & \cellcolor{myback1}17.83 & \cellcolor{myback1}17.65 & \cellcolor{myback1}17.53 & \cellcolor{myback1}\textbf{25.75} & \cellcolor{myback1}\underline{22.29} & \cellcolor{myback1}19.64 & \cellcolor{myback1}18.82 & \cellcolor{myback1}19.01 & \cellcolor{myback1}18.85 & \cellcolor{myback1}19.07 & \cellcolor{myback1}17.83 & \cellcolor{myback1}19.10 & \cellcolor{myback1}17.77 & \cellcolor{myback1}20.67 & \cellcolor{myback1}18.76 & \cellcolor{myback1}17.94\\
            ~&~ & \cellcolor{myback1}(3.29) & \cellcolor{myback1}(5.44) & \cellcolor{myback1}(1.09) & \cellcolor{myback1}(1.14) & \cellcolor{myback1}(2.41) & \cellcolor{myback1}\textbf{(7.97)} & \cellcolor{myback1}\underline{(5.40)} & \cellcolor{myback1}(5.46) & \cellcolor{myback1}(2.69) & \cellcolor{myback1}(1.63) & \cellcolor{myback1}(3.35) & \cellcolor{myback1}(0.57) & \cellcolor{myback1}(1.09) & \cellcolor{myback1}(1.89) & \cellcolor{myback1}(1.44) & \cellcolor{myback1}(2.15) & \cellcolor{myback1}(0.39) & \cellcolor{myback1}(4.21)\\
            \cmidrule(lr){2-20} ~&\multirow{2}*{F1} & \cellcolor{myback2}16.36 & \cellcolor{myback2}17.27 & \cellcolor{myback2}15.31 & \cellcolor{myback2}15.65 & \cellcolor{myback2}15.28 & \cellcolor{myback2}\textbf{21.26} & \cellcolor{myback2}\underline{18.19} & \cellcolor{myback2}16.25 & \cellcolor{myback2}15.42 & \cellcolor{myback2}15.42 & \cellcolor{myback2}15.22 & \cellcolor{myback2}15.55 & \cellcolor{myback2}15.31 & \cellcolor{myback2}17.17 & \cellcolor{myback2}15.54 & \cellcolor{myback2}16.79 & \cellcolor{myback2}16.10 & \cellcolor{myback2}15.27\\
            ~&~ & \cellcolor{myback2}(3.75) & \cellcolor{myback2}(3.60) & \cellcolor{myback2}(1.19) & \cellcolor{myback2}(1.70) & \cellcolor{myback2}(4.11) & \cellcolor{myback2}\textbf{(6.81)} & \cellcolor{myback2}\underline{(5.44)} & \cellcolor{myback2}(5.56) & \cellcolor{myback2}(2.89) & \cellcolor{myback2}(2.07) & \cellcolor{myback2}(2.37) & \cellcolor{myback2}(1.48) & \cellcolor{myback2}(1.19) & \cellcolor{myback2}(1.89) & \cellcolor{myback2}(1.50) & \cellcolor{myback2}(3.51) & \cellcolor{myback2}(0.60) & \cellcolor{myback2}(3.92)\\    

            \noalign{\global\arrayrulewidth=1pt}
            \hline
        \end{tabular}
    }}
\end{sidewaystable*}

\begin{figure*} [!ht]
	
	\begin{center}    
		\subfloat[subject-dependent] {    
			\includegraphics[width=\columnwidth]{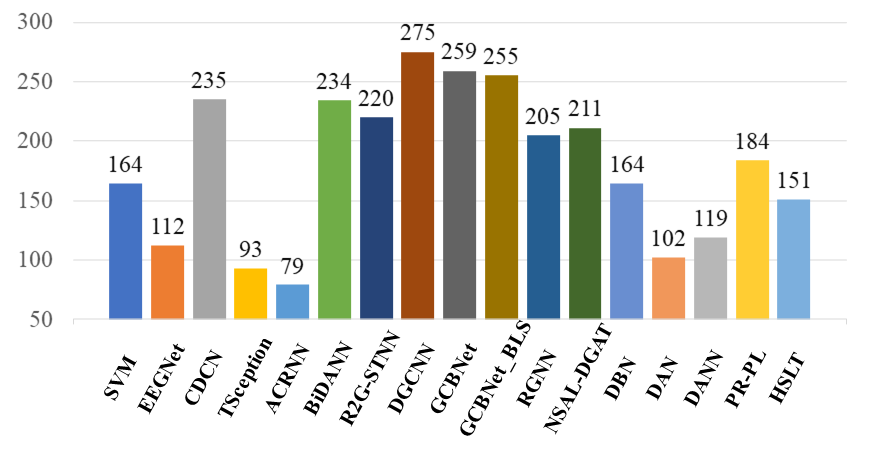}  
		}     
		\subfloat[cross-subject] {     
			\includegraphics[width=\columnwidth]{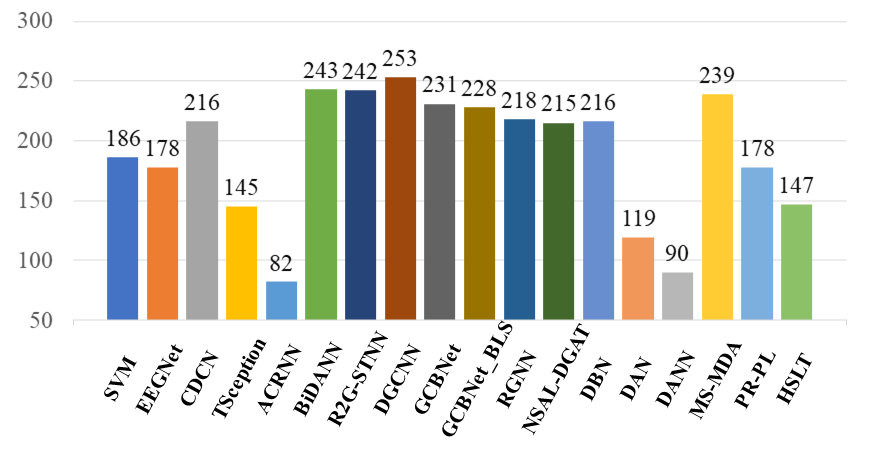}     
		}    
	\end{center}
	\caption{The model performance scoring on subject-dependent and cross-subject tasks based on ACC and F1.}
	\label{score}  
    \vspace{-0.15cm}
\end{figure*}

\subsection{Model Reproduction}
To ensure the reliability of our benchmark, we conducted a systematic reproduction of the representative EER methods by strictly following the experimental settings reported in their original papers, including model configurations and dataset usage. Table~\ref{reproduction} presents both our reproduction results and the original reported results, along with the performance gap. Considering that SEED and DEAP are the two most widely used public datasets, and all included methods have been evaluated on at least one of them, we reproduced the results specifically on SEED and DEAP. If the gap was within 10\%, we regarded the reproduced implementation as sufficiently reliable for subsequent unified benchmark comparisons. In addition, Table IV provides the specific original experimental setups used in the reproduction experiments, clarifying the differences in settings across different models compared to our unified benchmark. All verified models have been integrated into our LibEER algorithm library for direct use.

From the results in the Table~\ref{reproduction} and the replication process, we can draw the following conclusions:

(1) The reproduced model performances are generally slightly lower than the results reported in the original paper, averaging 3.85\% lower, with a maximum difference of less than 10\%, indicating a satisfactory overall replication quality. The reproduction results for ACRNN, R2G-STNN, GCBNet\_BLS, RGNN, DBN, and DAN showed more significant declines compared to the original papers ($>$5\%). The reproduction results for BiDANN, DGCNN, GCBNet, NSAL-DGAT, and PR-PL showed smaller deviations ($<$5\%). Conversely, some scenarios in CDCN, TSception, MS-MDA, and HSLT exhibited improvements in the reproduced results. The primary reason for these discrepancies is the lack of detailed descriptions regarding model specifications and experimental setups in some papers. We determine these parameters and settings through tuning and experience, which could introduce some biases into the results. Consequently, even when attempts are made to replicate the results as closely as possible to the descriptions in the original work, it remains challenging to achieve identical outcomes.

(2) There are substantial inconsistencies in the experimental setups reported in the original papers, including differences in datasets, preprocessing, tasks, and evaluation protocols. Notably, all original settings do not adopt a three-way split (train/validation/test), which increases the risk of overfitting and inflating performance due to the lack of validation-based model selection. This substantial variability in experimental configurations prevents fair comparisons among the models.
Therefore, we established a unified benchmark under a consistent and rational experimental framework, as described in Section III, where all methods are evaluated using standardized preprocessing, dataset usage, a three-way split, and consistent evaluation metrics.

\subsection{Model Comparison}
As shown in Table~\ref{reproduction}, the implementation details of different methods vary significantly, making it difficult to fairly compare their performance and draw reliable conclusions for the advancement of the EER field.
To this end, we conduct a fair comparison of seventeen selected methods based on the unified and reasonable benchmark proposed in Section \uppercase\expandafter{\romannumeral3}. The results for the two experimental tasks are reported in Tables~\ref{dependent} and~\ref{cross}, which clearly present results that differ significantly from those in Table~\ref{reproduction}.
We select SVM~\cite{SVM}, one of the most commonly used machine learning algorithms, as the baseline model. Its results are presented first. In the two tables, the top two methods in each scenario are highlighted using bold and underlined formatting.
From these results, we make the following key observations.

\noindent\textbf{Observation 1. Modeling the spatiotemporal information of EEG signals is crucial for EER tasks. Several GNN-based and RNN-based models achieved superior performance. In addition, integrating transfer learning can effectively alleviate inter-subject variability in EEG data, demonstrating considerable potential.}

\begin{figure*}[t]
    \centering
        \includegraphics[width=0.9\linewidth]{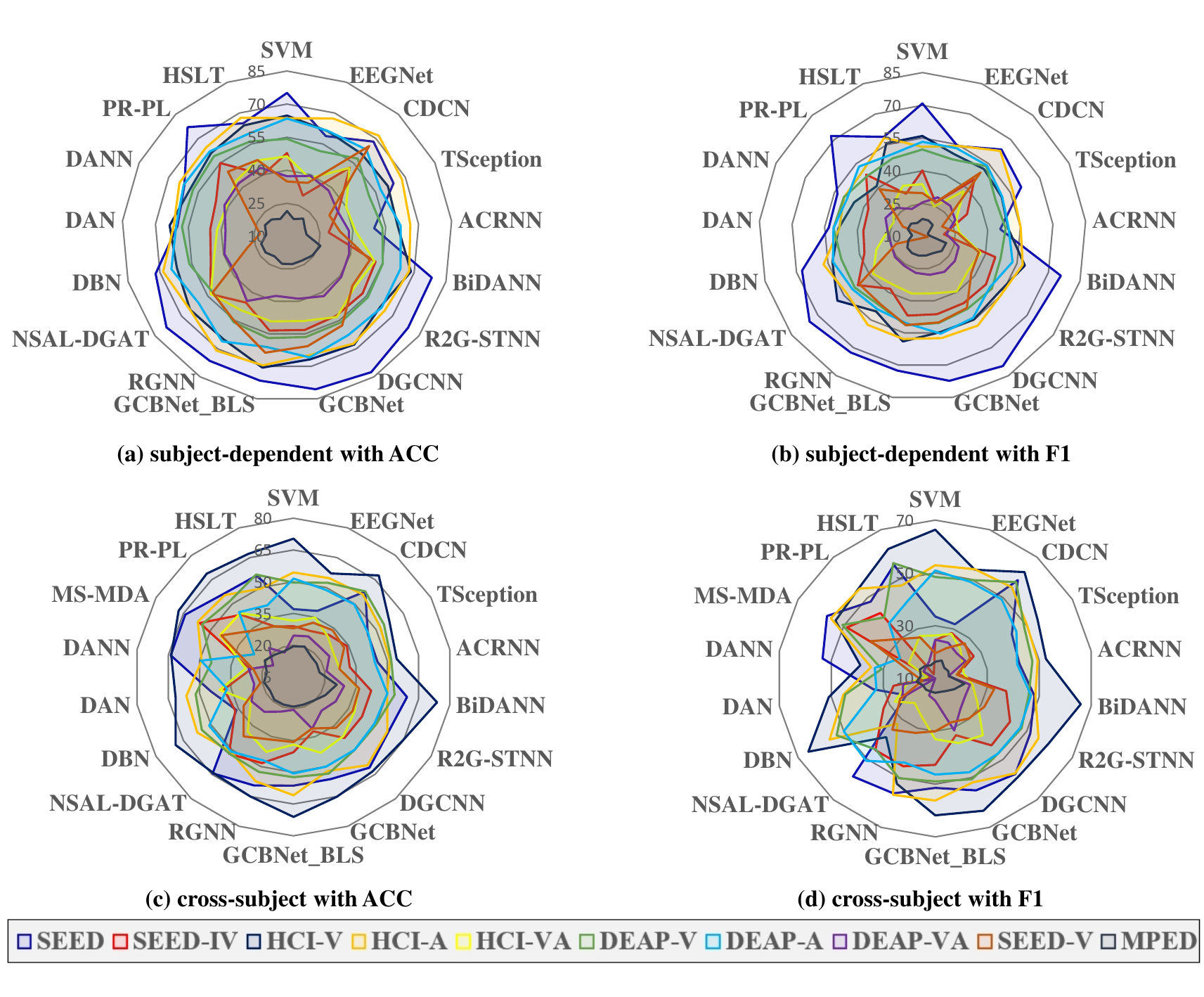}
        \caption{Radar chart of model performance across different datasets on subject-dependent and cross-subject tasks.}
    \label{radar}
    \vspace{-0.15cm}
\end{figure*}
To provide a clearer illustration of the performance of each method under the benchmark, we designed a scoring scheme that simultaneously considers accuracy and F1-score across all datasets. Specifically, for ten datasets (SEED, SEED-IV, HCI-V, HCI-A, HCI-VA, DEAP-V, DEAP-A, DEAP-VA, SEED-V, MPED), all $n$ methods are evaluated based on both accuracy and F1 score, yielding a total of 20 rankings (10 datasets × 2 metrics). In each ranking, the top-performing method receives $n$ points, while the lowest receives 1 point. The final score of each method is computed as the sum of its points across all 20 rankings, as shown in Fig.~\ref{score} (a) and (b).

From Fig.~\ref{score} (a), it can be observed that in the subject-dependent task, DGCNN achieved the best performance with a total score of 275. Following closely are GCBNet and its variant GCBNet\_BLS, scoring 259 and 255 points, respectively. CDCN and BiDANN ranked fourth and fifth with scores of 235 and 234, respectively. In contrast, EEGNet, Tsception, ACRNN, DAN, and DANN performed relatively poorly. 
In the cross-subject task (see Fig.~\ref{score} (b)), the overall performance ranking of each model is similar to that in the subject-dependent scenario. Several GNN-based methods still achieved relatively superior performance. BiDANN and R2G-STNN, two models based on RNN and transfer learning that already performed well in subject-dependent settings, achieved even better performance in the cross-subject scenario, ranking second and third, respectively.
EEGNet and DBN obtained better results on this task compared to their performance in subject-dependent settings. The MS-MDA model, specifically designed for cross-subject scenarios, resulted in a moderate ranking in the overall score.

In terms of model types, the four GNN-based methods demonstrated superior performance on both tasks, providing strong evidence that the graph-structured information between EEG signal channels is crucial for EER tasks. In contrast, the CNN-based methods exhibited varying levels of performance on both tasks. The DNN and Transformer-based methods showed relatively weaker overall performance, though some individual models demonstrated promising results on specific tasks. However, for the Transformer-based methods, as only one model was selected, the results may contain some occasionality. RNN-based models exhibited notable performance divergence, with both BiDANN and R2G-STNN ranking within the top five and even top three across both tasks, while ACRNN consistently ranked last. This may be attributed to the incorporation of transfer learning in BiDANN and R2G-STNN, which effectively models inter-sample variability. However, the two DNN-based transfer learning models, DAN and DANN, performed poorly, indicating that beyond employing transfer learning, it is equally important to capture the intrinsic spatiotemporal characteristics of EEG signals through architectures such as RNNs and GNNs.
Based on the current results, we recommend that researchers in the EER field prioritize models capable of capturing the spatiotemporal dynamics of EEG signals, such as those based on RNN or GNN, and consider incorporating transfer learning techniques to enhance model generalizability across subjects.

\begin{figure*} [!t]
	
	\begin{center}    
		\subfloat[Subject-dependent task on the SEED-IV dataset] {    
		\includegraphics[width=0.49\columnwidth]{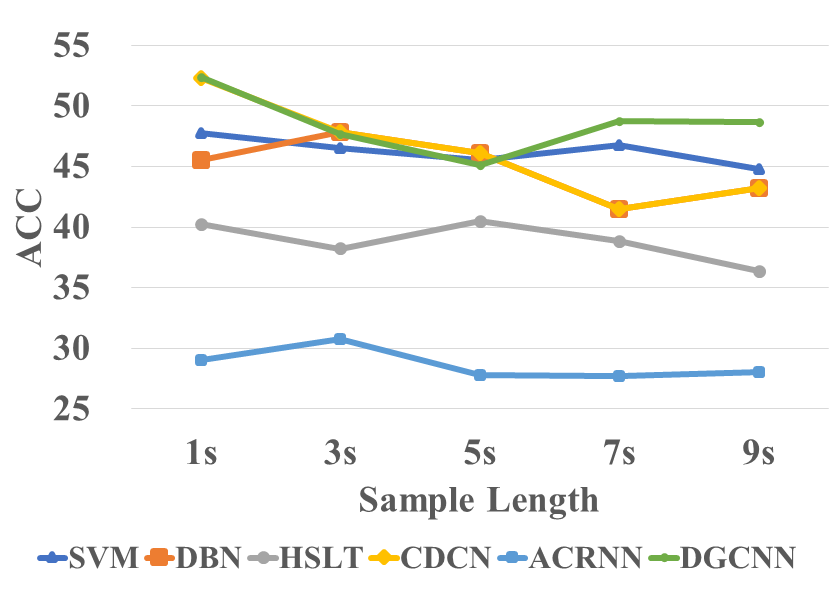} 
        \includegraphics[width=0.49\columnwidth]{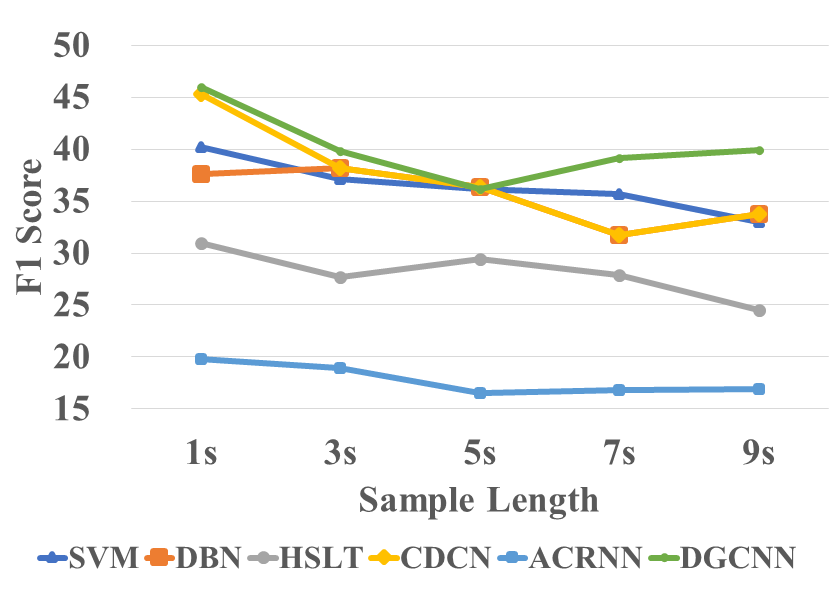}
		}     
		\subfloat[Cross-subject task on the SEED-IV dataset] {     
			\includegraphics[width=0.49\columnwidth]{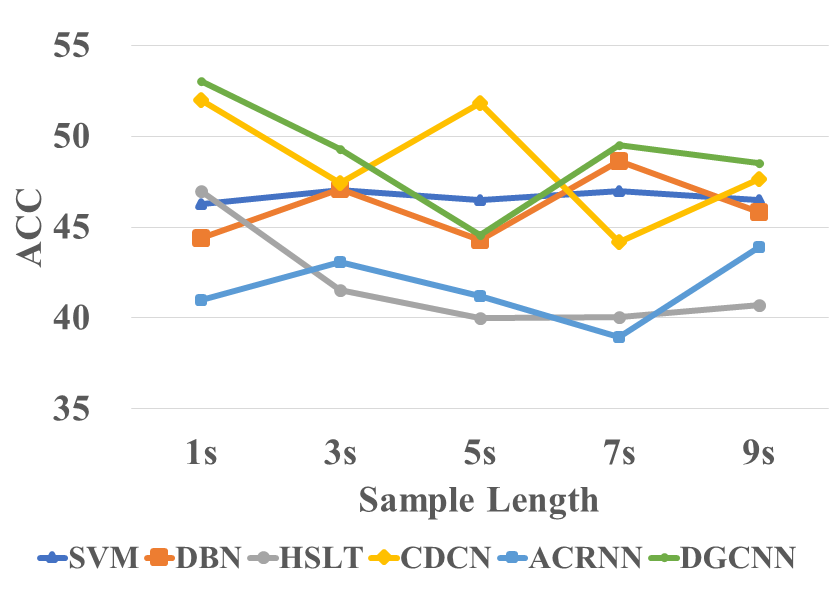} 
        \includegraphics[width=0.49\columnwidth]{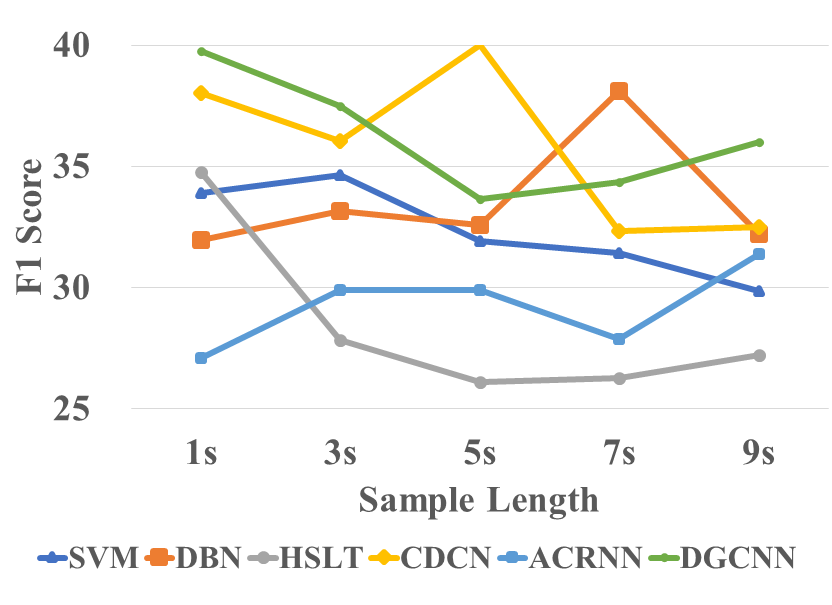}   
		}    \\
  \subfloat[Subject-dependent task on the HCI-VA dataset] {     
			\includegraphics[width=0.49\columnwidth]{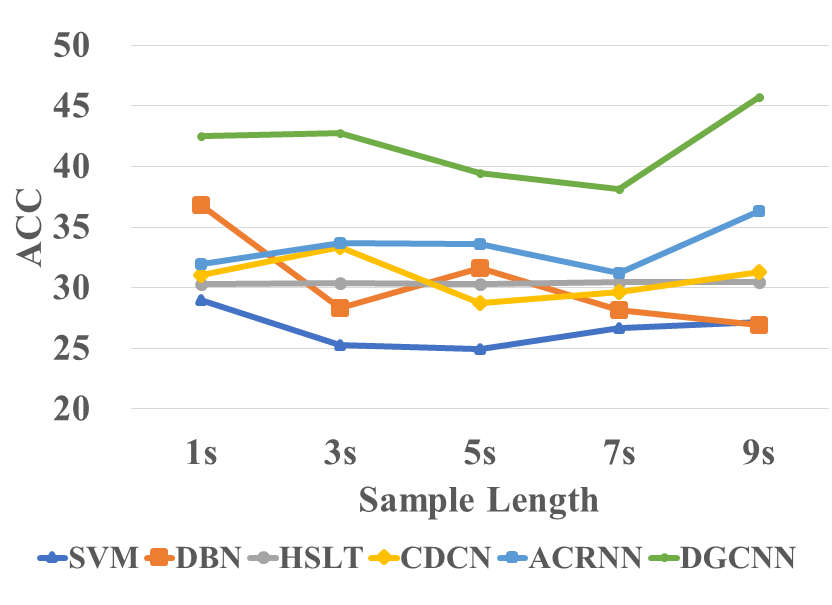} 
        \includegraphics[width=0.49\columnwidth]{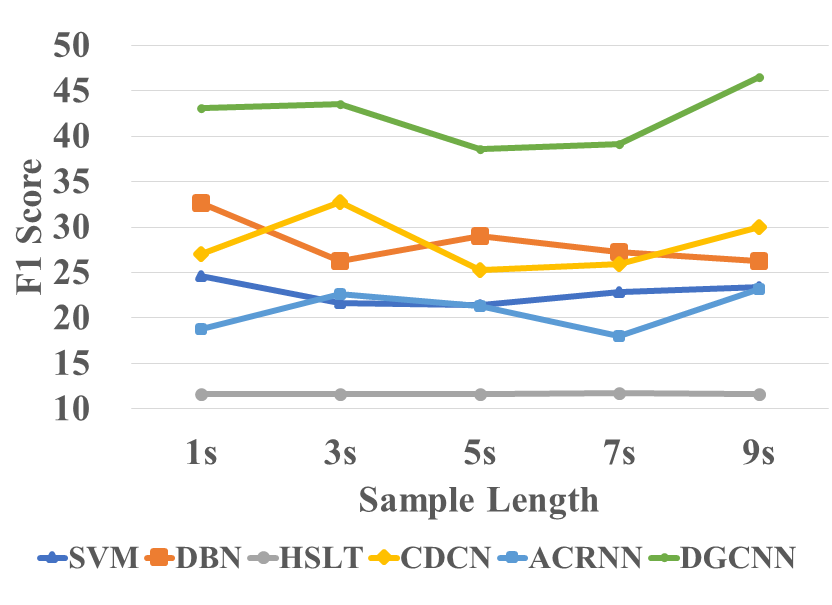}   
		}
  \subfloat[Cross-subject task on the HCI-VA dataset] {     
			\includegraphics[width=0.49\columnwidth]{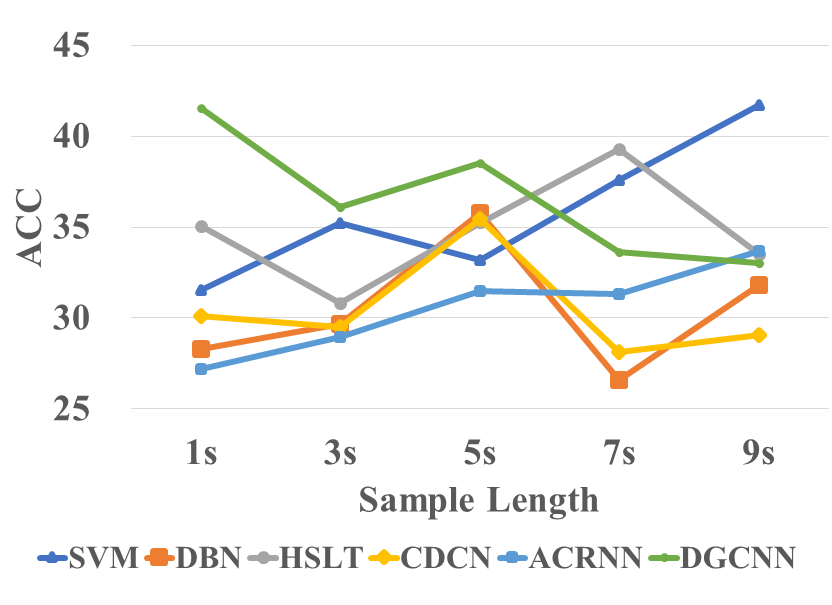} 
        \includegraphics[width=0.49\columnwidth]{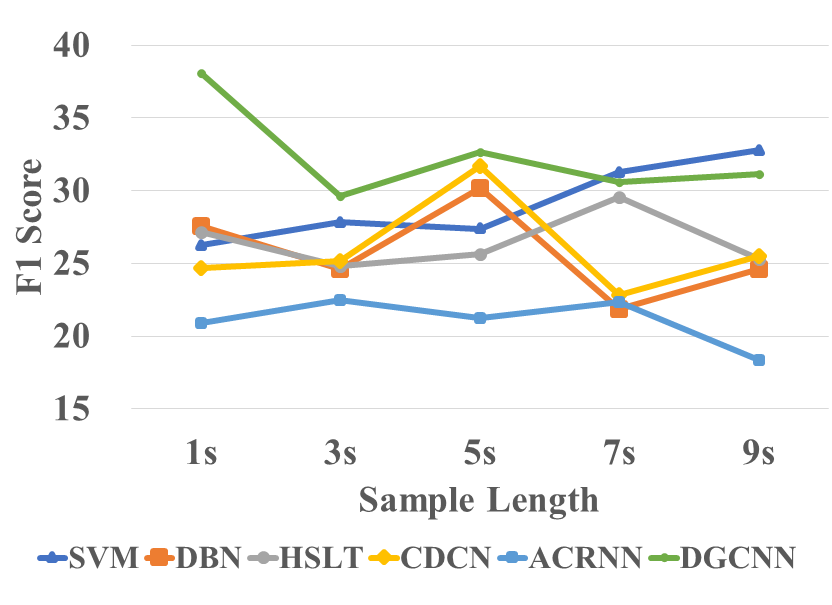}   
		}
	\end{center}
	\caption{Experimental results on sample length for subject-dependent and cross-subject tasks on the SEED-IV and HCI-VA datasets.}
	\label{ablation1}  
    \vspace{-0.3cm}
\end{figure*}

\begin{figure*} [!t]
	
	\begin{center}    
		\subfloat[Subject-dependent task on the SEED-IV dataset] {    
		\includegraphics[width=0.45\columnwidth]{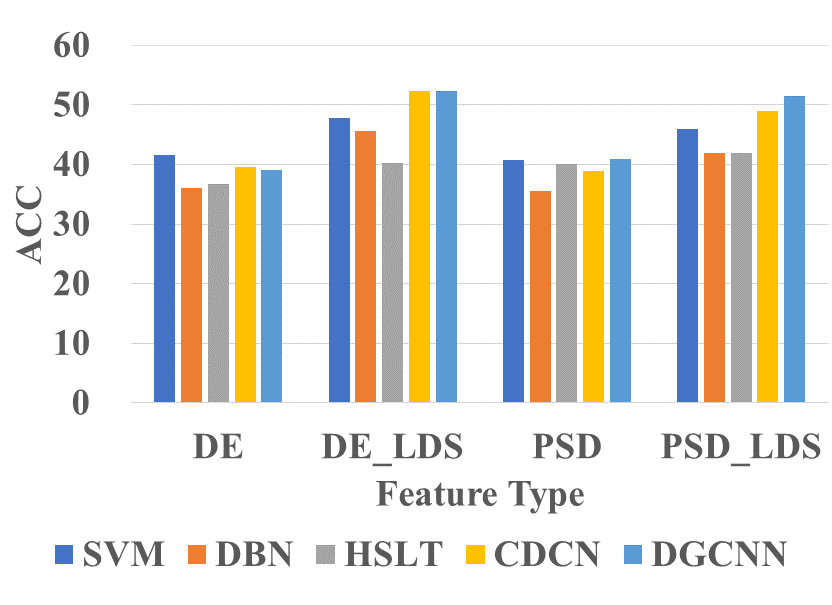} 
        \includegraphics[width=0.45\columnwidth]{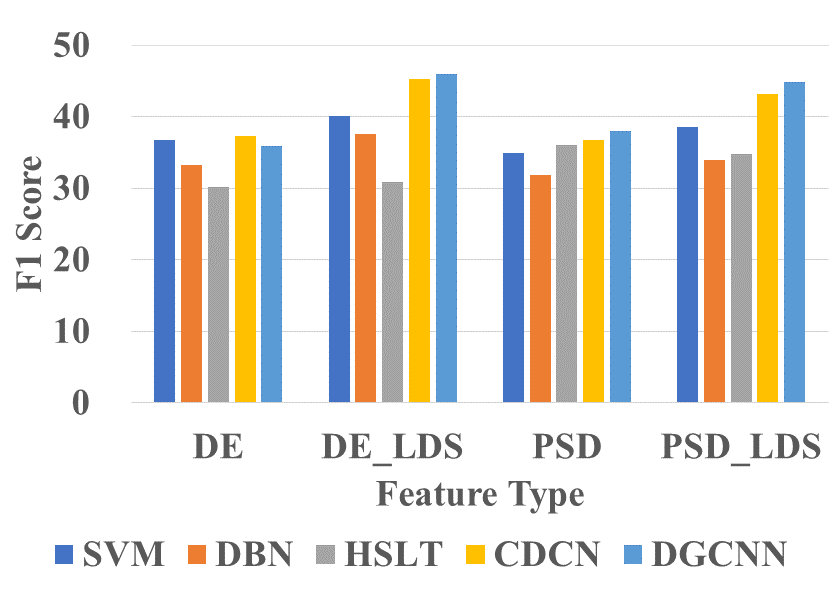}
		}     
		\subfloat[Cross-subject task on the SEED-IV dataset] {     
			\includegraphics[width=0.45\columnwidth]{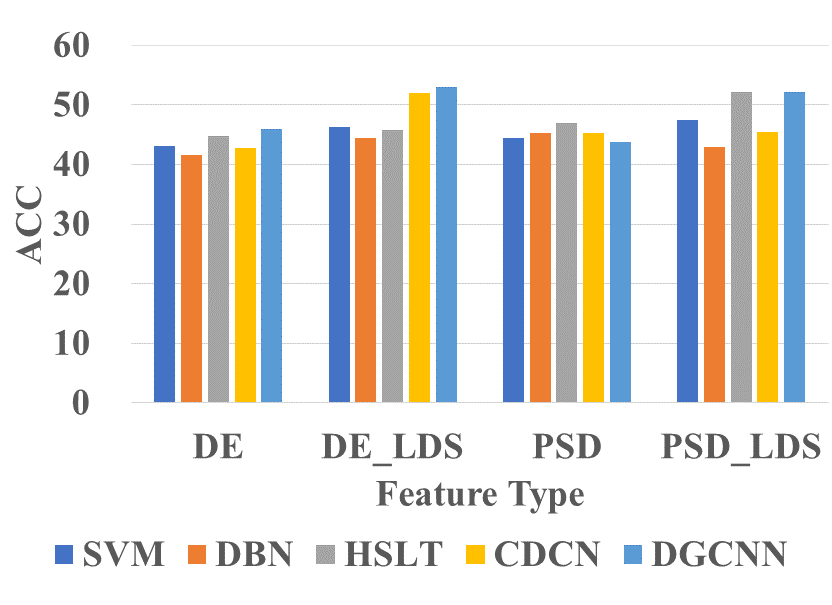} 
        \includegraphics[width=0.45\columnwidth]{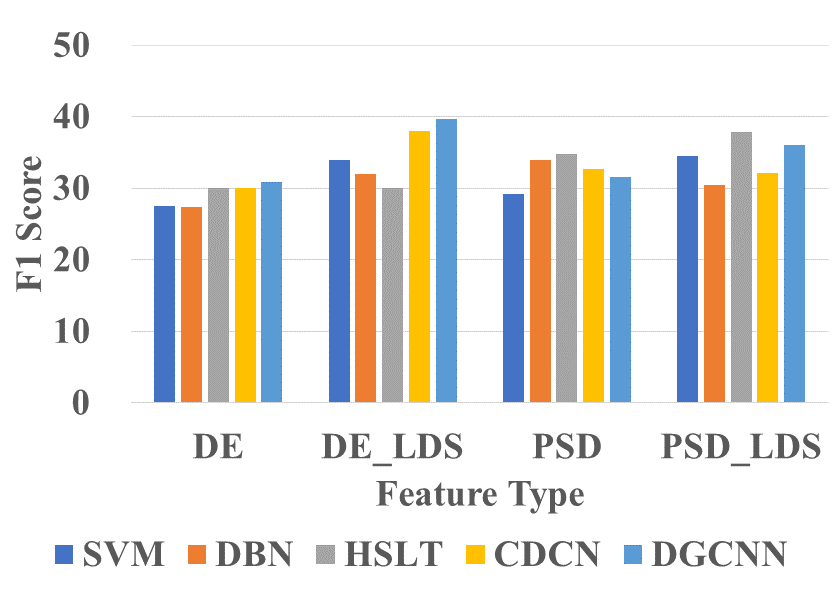}   
		}    \\
  \subfloat[Subject-dependent task on the HCI-VA dataset] {     
			\includegraphics[width=0.45\columnwidth]{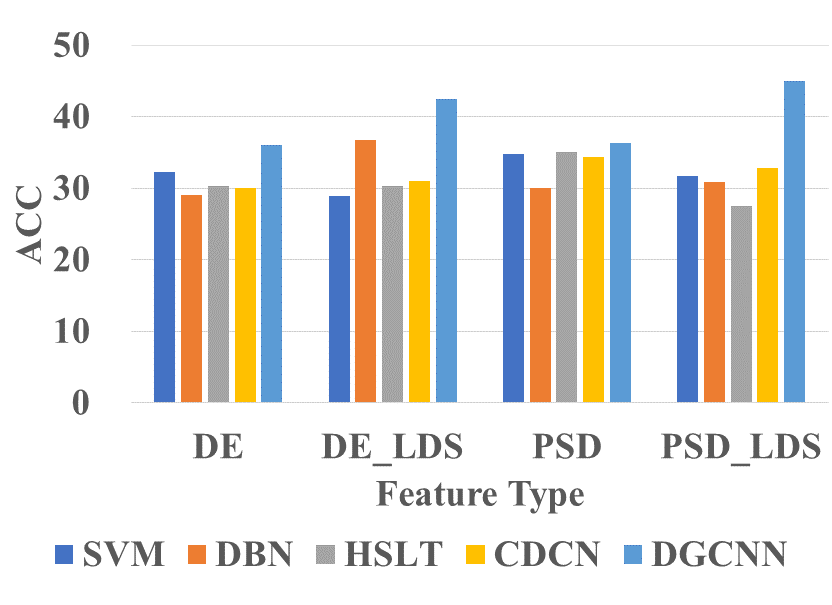} 
        \includegraphics[width=0.45\columnwidth]{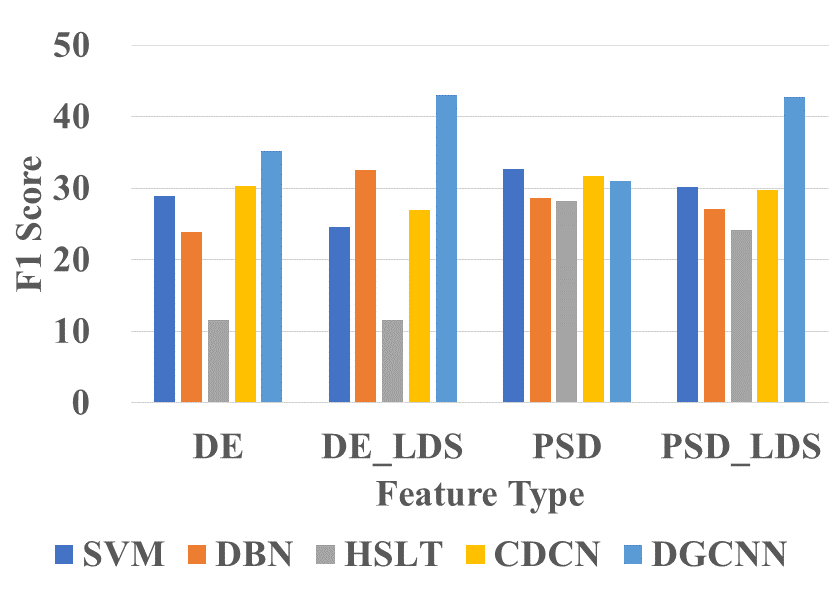}   
		}
  \subfloat[Cross-subject task on the HCI-VA dataset] {     
			\includegraphics[width=0.45\columnwidth]{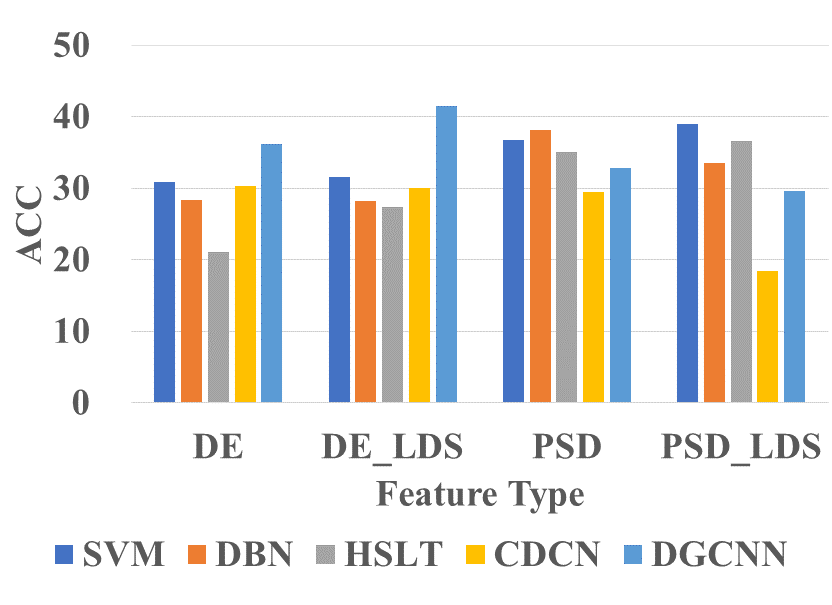} 
        \includegraphics[width=0.45\columnwidth]{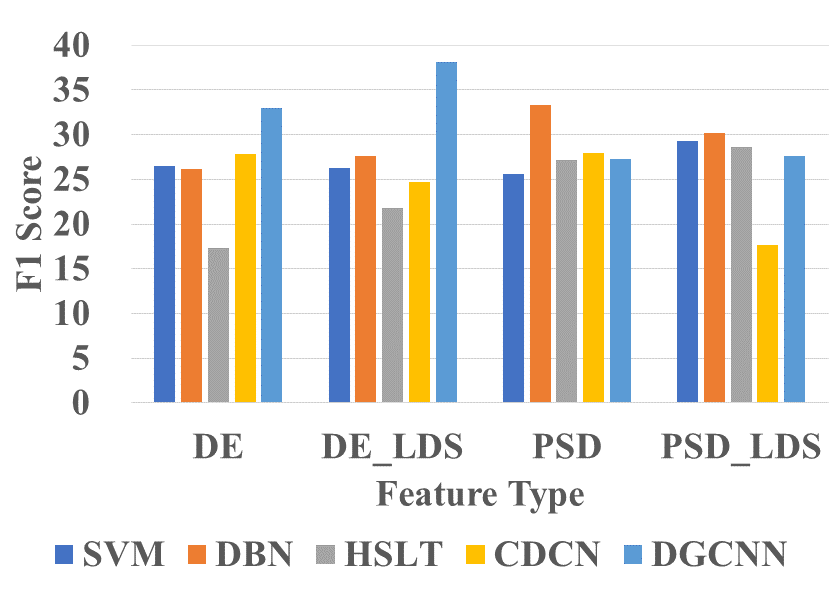}   
		}
	\end{center}
	\caption{Experimental results on feature type for subject-dependent and cross-subject tasks on the SEED-IV and HCI-VA datasets.}
	\label{ablation2}  
    \vspace{-0.3cm}
\end{figure*}

\noindent\textbf{Observation 2. The issue of significant variability in EEG data among different subjects remains unaddressed, with all methods showing large standard deviations in subject-dependent scenarios and relatively low performance in cross-subject scenarios.}


Table~\ref{dependent} demonstrates that the standard deviations for each method across various scenarios are significantly high, averaging approximately 15-20\%. Table~\ref{cross} indicates that the performance of the methods is relatively low, suggesting considerable room for improvement. The pronounced variability in EEG data among subjects has long been a critical issue in the field of EER, despite some methods asserting the inclusion of specific modules to tackle this challenge. However, under our unified and systematic benchmark, the models failed to effectively address the variability in data from different subjects, as reported in the original studies. Consequently, further in-depth investigation is necessary to adequately confront the challenges posed by the substantial differences in EEG data among various subjects.

\begin{table}[!t]
	\caption{Experimental results on training methods for subject-dependent task on the SEED-IV and HCI-VA datasets.}
	\label{ablation3_dependent}
	\renewcommand{\arraystretch}{1.2}
	\centering
	{\small{
        \begin{tabular}{m{0.7cm}<{\centering}m{0.4cm}<{\centering}m{0.75cm}<{\centering}m{0.75cm}<{\centering}m{0.75cm}<{\centering}m{0.75cm}<{\centering}m{0.75cm}<{\centering}m{0.75cm}<{\centering}}
            \noalign{\global\arrayrulewidth=2pt}
            \hline
            \noalign{\global\arrayrulewidth=0.5pt}	
           \multicolumn{2}{c}{\multirow{2}{*}{Method}} & \multicolumn{3}{c}{SEED-IV} & \multicolumn{3}{c}{HCI-VA} \\
           \cmidrule(lr){3-5}\cmidrule(lr){6-8}
            & &Ours&R1&R2&Ours&R1&R2\\
            
            \hline
           
              \multirow{4}*{DBN}
            &\multirow{2}*{ACC}
            &\cellcolor{myback1}45.56&\cellcolor{myback1}\textbf{49.50}&\cellcolor{myback1}43.96&\cellcolor{myback1}44.38&\cellcolor{myback1}\textbf{46.58}&\cellcolor{myback1}42.95\\
            ~&~&\cellcolor{myback1}(21.19)&\cellcolor{myback1}\textbf{(22.02)}&\cellcolor{myback1}(22.47)&\cellcolor{myback1}(26.78)&\cellcolor{myback1}\textbf{(23.59)}&\cellcolor{myback1}(23.24)\\
            \cmidrule(lr){2-8}                 
            ~&\multirow{2}*{F1}
            &\cellcolor{myback2}37.61&\cellcolor{myback2}\textbf{43.34}&\cellcolor{myback2}37.83&\cellcolor{myback2}31.96&\cellcolor{myback2}\textbf{34.87}&\cellcolor{myback2}31.87\\
            ~&~&\cellcolor{myback2}(20.68)&\cellcolor{myback2}\textbf{(22.03)}&\cellcolor{myback2}(21.79)&\cellcolor{myback2}(27.87)&\cellcolor{myback2}\textbf{(23.67)}&\cellcolor{myback2}(21.52) \\
            \cmidrule(lr){1-8}

            \multirow{4}*{HSLT}
            &\multirow{2}*{ACC}
            &\cellcolor{myback1}\textbf{40.28}&\cellcolor{myback1}38.10&\cellcolor{myback1}33.15&\cellcolor{myback1}\textbf{46.99}&\cellcolor{myback1}44.02&\cellcolor{myback1}43.03\\
            ~&~&\cellcolor{myback1}\textbf{(23.80)}&\cellcolor{myback1}(24.28)&\cellcolor{myback1}(20.25)&\cellcolor{myback1}\textbf{(20.76)}&\cellcolor{myback1}(20.09)&\cellcolor{myback1}(19.15) \\
            \cmidrule(lr){2-8}                 
            ~&\multirow{2}*{F1}
            &\cellcolor{myback2}30.92&\cellcolor{myback2}\textbf{31.03}&\cellcolor{myback2}22.03&\cellcolor{myback2}\textbf{34.76}&\cellcolor{myback2}32.46&\cellcolor{myback2}25.19 \\
            ~&~&\cellcolor{myback2}(24.47)&\cellcolor{myback2}\textbf{(24.19)}&\cellcolor{myback2}(19.80)&\cellcolor{myback2}\textbf{(19.69)}&\cellcolor{myback2}(18.85)&\cellcolor{myback2}(13.04) \\
            \cmidrule(lr){1-8}

            \multirow{4}*{CDCN}
            &\multirow{2}*{ACC}
            &\cellcolor{myback1}52.26&\cellcolor{myback1}52.18&\cellcolor{myback1}\textbf{52.83}&\cellcolor{myback1}\textbf{52.00}&\cellcolor{myback1}40.41&\cellcolor{myback1}39.84 \\
            ~&~&\cellcolor{myback1}(21.97)&\cellcolor{myback1}(21.85)&\cellcolor{myback1}\textbf{(22.64)}&\cellcolor{myback1}\textbf{(26.05)}&\cellcolor{myback1}(19.52)&\cellcolor{myback1}(23.59) \\
            \cmidrule(lr){2-8}                 
            ~&\multirow{2}*{F1}
            &\cellcolor{myback2}45.26&\cellcolor{myback2}44.25&\cellcolor{myback2}\textbf{47.17}&\cellcolor{myback2}\textbf{38.04}&\cellcolor{myback2}24.31&\cellcolor{myback2}25.02 \\
            ~&~&\cellcolor{myback2}(23.00)&\cellcolor{myback2}(23.53)&\cellcolor{myback2}\textbf{(23.95)}&\cellcolor{myback2}\textbf{(26.88)}&\cellcolor{myback2}(14.34)&\cellcolor{myback2}(15.63) \\
            \cmidrule(lr){1-8}

            \multirow{4}*{ACRNN}
            &\multirow{2}*{ACC}
            &\cellcolor{myback1}\textbf{29.01}&\cellcolor{myback1}27.54&\cellcolor{myback1}27.45&\cellcolor{myback1}41.00&\cellcolor{myback1}\textbf{41.57}&\cellcolor{myback1}38.58 \\
            ~&~&\cellcolor{myback1}\textbf{(7.10)}&\cellcolor{myback1}(6.86)&\cellcolor{myback1}(3.74)&\cellcolor{myback1}(21.58)&\cellcolor{myback1}\textbf{(23.37)}&\cellcolor{myback1}(22.25) \\
            \cmidrule(lr){2-8}                 
            ~&\multirow{2}*{F1}
            &\cellcolor{myback2}\textbf{19.80}&\cellcolor{myback2}18.79&\cellcolor{myback2}11.74&\cellcolor{myback2}27.10&\cellcolor{myback2}\textbf{31.78}&\cellcolor{myback2}25.34\\
            ~&~&\cellcolor{myback2}\textbf{(5.42)}&\cellcolor{myback2}(6.06)&\cellcolor{myback2}(2.23)&\cellcolor{myback2}(15.13)&\cellcolor{myback2}\textbf{(20.28)}&\cellcolor{myback2}(14.75)  \\
            \cmidrule(lr){1-8}

            \multirow{4}*{DGCNN}
            &\multirow{2}*{ACC}
            &\cellcolor{myback1}\textbf{52.39}&\cellcolor{myback1}47.89&\cellcolor{myback1}50.01&\cellcolor{myback1}\textbf{53.06}&\cellcolor{myback1}50.16&\cellcolor{myback1}44.38 \\
            ~&~&\cellcolor{myback1}\textbf{(24.32)}&\cellcolor{myback1}(23.05)&\cellcolor{myback1}(26.99)&\cellcolor{myback1}\textbf{(24.44)}&\cellcolor{myback1}(25.13)&\cellcolor{myback1}(21.71) \\
            \cmidrule(lr){2-8}                 
            ~&\multirow{2}*{F1}
            &\cellcolor{myback2}\textbf{45.94}&\cellcolor{myback2}41.12&\cellcolor{myback2}43.49&\cellcolor{myback2}\textbf{39.75}&\cellcolor{myback2}37.94&\cellcolor{myback2}31.49 \\
            ~&~&\cellcolor{myback2}\textbf{(24.17)}&\cellcolor{myback2}(22.44)&\cellcolor{myback2}(27.22)&\cellcolor{myback2}\textbf{(26.01)}&\cellcolor{myback2}(26.70)&\cellcolor{myback2}(22.08) \\
            \cmidrule(lr){1-8}

            \multirow{4}*{Average}
            &\multirow{2}*{ACC}
            &\cellcolor{myback1}\textbf{43.90}&\cellcolor{myback1}43.04&\cellcolor{myback1}41.48&\cellcolor{myback1}\textbf{47.49}&\cellcolor{myback1}44.55&\cellcolor{myback1}41.76  \\
            ~&~&\cellcolor{myback1}\textbf{(19.68)}&\cellcolor{myback1}(19.61)&\cellcolor{myback1}(19.22)&\cellcolor{myback1}\textbf{(23.92)}&\cellcolor{myback1}(22.34)&\cellcolor{myback1}(21.99) \\
            \cmidrule(lr){2-8}                 
            ~&\multirow{2}*{F1}
            &\cellcolor{myback2}\textbf{35.91}&\cellcolor{myback2}35.71&\cellcolor{myback2}32.45&\cellcolor{myback2}\textbf{34.32}&\cellcolor{myback2}32.27&\cellcolor{myback2}27.78  \\
            ~&~&\cellcolor{myback2}\textbf{(19.55)}&\cellcolor{myback2}(19.65)&\cellcolor{myback2}(19.00)&\cellcolor{myback2}\textbf{(23.12)}&\cellcolor{myback2}(20.77)&\cellcolor{myback2}(17.40)\\

            \noalign{\global\arrayrulewidth=1pt}
            \hline
        \end{tabular}
    }}
\end{table}

\begin{table}[!t]
	\caption{Experimental results on training methods for cross-subject task on the SEED-IV and HCI-VA datasets.}
	\label{ablation3_cross}
	\renewcommand{\arraystretch}{1.2}
	\centering
	{\small{
        \begin{tabular}{m{0.8cm}<{\centering}m{0.4cm}<{\centering}m{0.75cm}<{\centering}m{0.75cm}<{\centering}m{0.75cm}<{\centering}m{0.75cm}<{\centering}m{0.75cm}<{\centering}m{0.75cm}<{\centering}}
            \noalign{\global\arrayrulewidth=2pt}
            \hline
            \noalign{\global\arrayrulewidth=0.5pt}	
           \multicolumn{2}{c}{\multirow{2}{*}{Method}} & \multicolumn{3}{c}{SEED-IV} & \multicolumn{3}{c}{HCI-VA} \\
           \cmidrule(lr){3-5}\cmidrule(lr){6-8}
            & &Ours&R1&R2&Ours&R1&R2\\
            
            \hline

            \multirow{2}*{DBN}
            &{ACC}
            &\cellcolor{myback1}\textbf{36.82}&\cellcolor{myback1}29.08&\cellcolor{myback1}30.68&\cellcolor{myback1}28.30&\cellcolor{myback1}\textbf{36.22}&\cellcolor{myback1}17.77 \\
            \cmidrule(lr){2-8}                 
            ~&{F1}
            &\cellcolor{myback2}\textbf{32.60}&\cellcolor{myback2}25.52&\cellcolor{myback2}19.79&\cellcolor{myback2}27.58&\cellcolor{myback2}\textbf{29.53}&\cellcolor{myback2}17.56 \\
            \cmidrule(lr){1-8}

            \multirow{2}*{HSLT}
            &{ACC}
            &\cellcolor{myback1}\textbf{30.33}&\cellcolor{myback1}\textbf{30.33}&\cellcolor{myback1}\textbf{30.33}&\cellcolor{myback1}\textbf{35.07}&\cellcolor{myback1}30.77&\cellcolor{myback1}33.12 \\
            \cmidrule(lr){2-8}                 
            ~&{F1}
            &\cellcolor{myback2}\textbf{11.64}&\cellcolor{myback2}\textbf{11.64}&\cellcolor{myback2}\textbf{11.64}&\cellcolor{myback2}\textbf{27.19}&\cellcolor{myback2}24.82&\cellcolor{myback2}18.46 \\
            \cmidrule(lr){1-8}

            \multirow{2}*{CDCN}
            &{ACC}
            &\cellcolor{myback1}31.03&\cellcolor{myback1}29.35&\cellcolor{myback1}\textbf{32.94}&\cellcolor{myback1}\textbf{30.09}&\cellcolor{myback1}24.88&\cellcolor{myback1}25.77 \\
            \cmidrule(lr){2-8}                 
            ~&{F1}
            &\cellcolor{myback2}27.01&\cellcolor{myback2}\textbf{27.62}&\cellcolor{myback2}31.99&\cellcolor{myback2}\textbf{24.71}&\cellcolor{myback2}20.89&\cellcolor{myback2}21.72 \\
            \cmidrule(lr){1-8}

            \multirow{2}*{ACRNN}
            &{ACC}
            &\cellcolor{myback1}31.97&\cellcolor{myback1}\textbf{32.86}&\cellcolor{myback1}30.33&\cellcolor{myback1}\textbf{27.21}&\cellcolor{myback1}23.28&\cellcolor{myback1}24.19 \\
            \cmidrule(lr){2-8}                 
            ~&{F1}
            &\cellcolor{myback2}18.82&\cellcolor{myback2}\textbf{20.60}&\cellcolor{myback2}11.64&\cellcolor{myback2}\textbf{20.92}&\cellcolor{myback2}17.03&\cellcolor{myback2}16.05 \\
            \cmidrule(lr){1-8}

            \multirow{2}*{DGCNN}
            &{ACC}
            &\cellcolor{myback1}42.54&\cellcolor{myback1}42.08&\cellcolor{myback1}\textbf{42.93}&\cellcolor{myback1}\textbf{41.54}&\cellcolor{myback1}34.84&\cellcolor{myback1}26.59 \\
            \cmidrule(lr){2-8}                 
            ~&{F1}
            &\cellcolor{myback2}\textbf{43.10}&\cellcolor{myback2}40.52&\cellcolor{myback2}42.05&\cellcolor{myback2}\textbf{38.08}&\cellcolor{myback2}33.06&\cellcolor{myback2}20.00 \\
            \cmidrule(lr){1-8}

            \multirow{2}*{Average}
            &{ACC}
            &\cellcolor{myback1}\textbf{34.54}&\cellcolor{myback1}32.74&\cellcolor{myback1}33.44&\cellcolor{myback1}\textbf{32.44}&\cellcolor{myback1}30.00&\cellcolor{myback1}25.49 \\
            \cmidrule(lr){2-8}                 
            ~&{F1}
            &\cellcolor{myback2}\textbf{26.63}&\cellcolor{myback2}25.18&\cellcolor{myback2}23.42&\cellcolor{myback2}\textbf{27.70}&\cellcolor{myback2}25.07&\cellcolor{myback2}18.76 \\

            \noalign{\global\arrayrulewidth=1pt}
            \hline
        \end{tabular}
    }}
\end{table}

\noindent\textbf{Observation 3: The scarcity of EEG data limits the effective representation learning capabilities of deep learning methods, leading to some deep learning approaches performing even worse than traditional machine learning methods, such as SVM.}

As demonstrated in Tables~\ref{dependent} and~\ref{cross}, several carefully designed deep learning algorithms did not outperform the baseline SVM, as reported in their original studies. A significant contributing factor to this discrepancy is the scarcity of EEG data, characterized by both limited quantity and scale of datasets. Deep networks typically necessitate substantial data volumes to fully leverage their potential for representation learning. Additionally, many studies lack validation sets in their experimental designs, which can easily result in overfitting and inflated performance metrics in contexts with limited data. In contrast, our rigorous experimental setup has highlighted these shortcomings. Therefore, a prudent strategy would be to utilize various data augmentation techniques~\cite{Luo_2020, Zhang_Zhong_Liu_2024,8512865} to expand the dataset and to implement strategies aimed at reducing model overfitting.

\noindent\textbf{Observation 4. The data collection methods for different EEG datasets vary, resulting in differences in data quality and task difficulty. Several methods performed better on the SEED series datasets, which feature a more reasonable data collection approach.}

To visually illustrate the performance differences of various models across different datasets, we created radar charts for both experimental tasks with ACC and F1 score, as shown in Fig.~\ref{radar}. From the results, we summarize that models performed better on the three-class SEED dataset compared to the binary classification tasks in DEAP and HCI. We believe this is due to the higher label quality in SEED, where discrete emotion labels are easier for self-assessment, while continuous Valence-Arousal-Dominance labels tend to be more subjective. Additionally, SEED provides expectations for the type of emotions elicited by stimuli, whereas DEAP and HCI do not.
As the number of labels increases, various models achieve relatively similar average accuracy on SEED-IV, HCI-VA, and SEED-V, with only DEAP-VA showing notably lower performance. However, the performance differences among models are more pronounced on SEED-IV and SEED-V compared to the other two datasets. Moreover, due to the more balanced label distribution in the SEED series datasets, their F1 scores are significantly higher than those of the other datasets.
The MPED dataset, which contains the largest number of emotion categories, poses significant challenges to most models, indicating that fine-grained recognition of human emotions in the EER field still requires further investigation.


\begin{figure*} [!t]
	
	\begin{center}    
		\subfloat[Subject-dependent task on the SEED-IV dataset] {    
		\includegraphics[width=0.49\columnwidth]{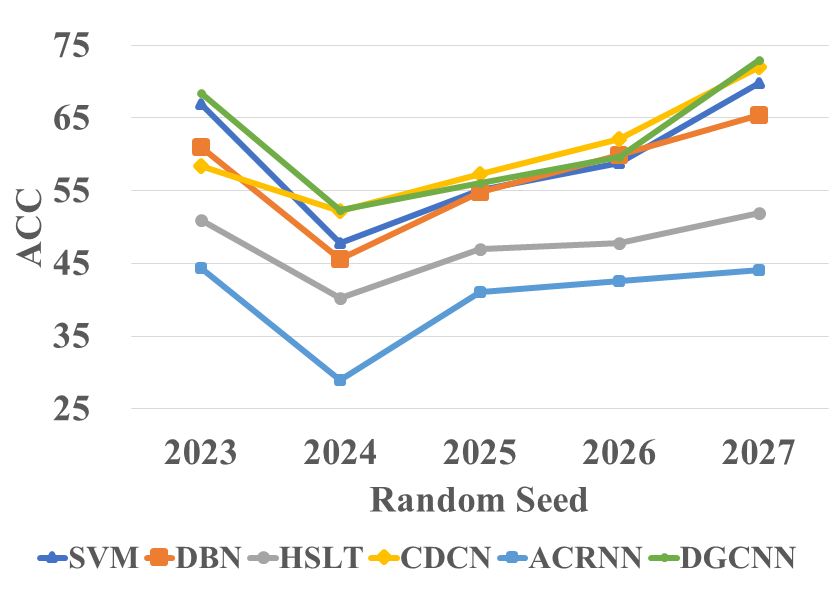} 
        \includegraphics[width=0.49\columnwidth]{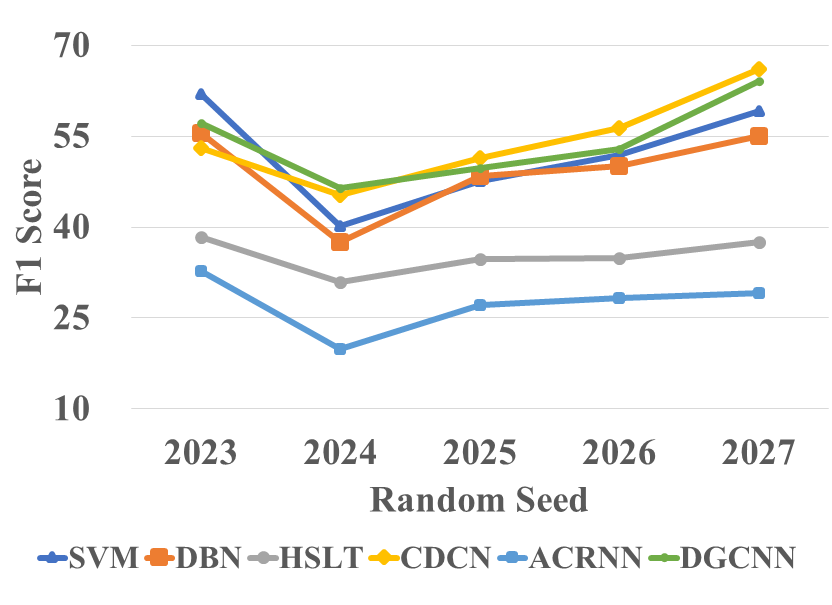}
		}     
		\subfloat[Cross-subject task on the SEED-IV dataset] {     
			\includegraphics[width=0.49\columnwidth]{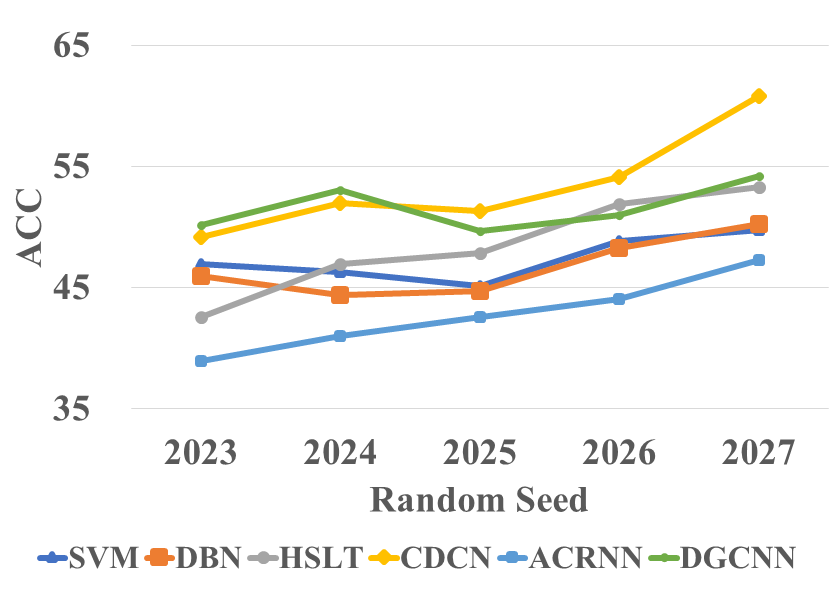} 
        \includegraphics[width=0.49\columnwidth]{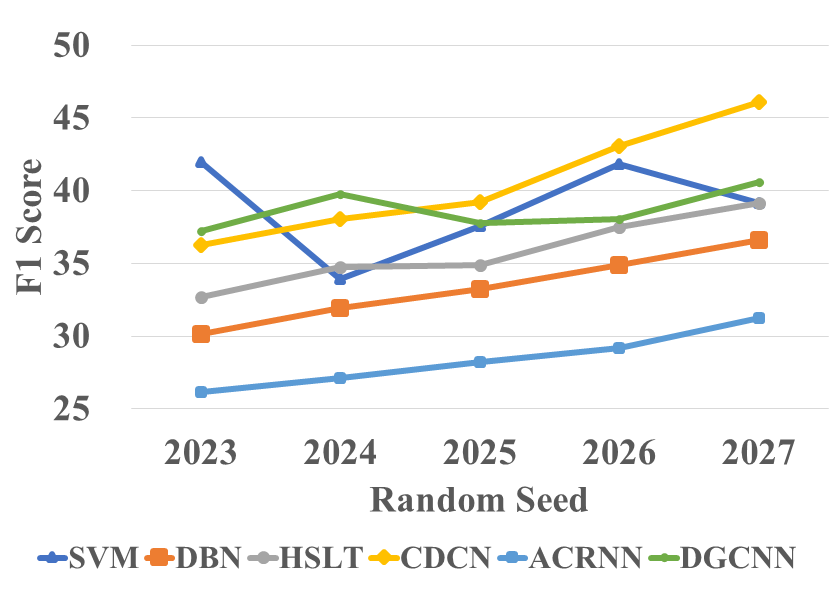}   
		}    \\
  \subfloat[Subject-dependent task on the HCI-VA dataset] {     
			\includegraphics[width=0.49\columnwidth]{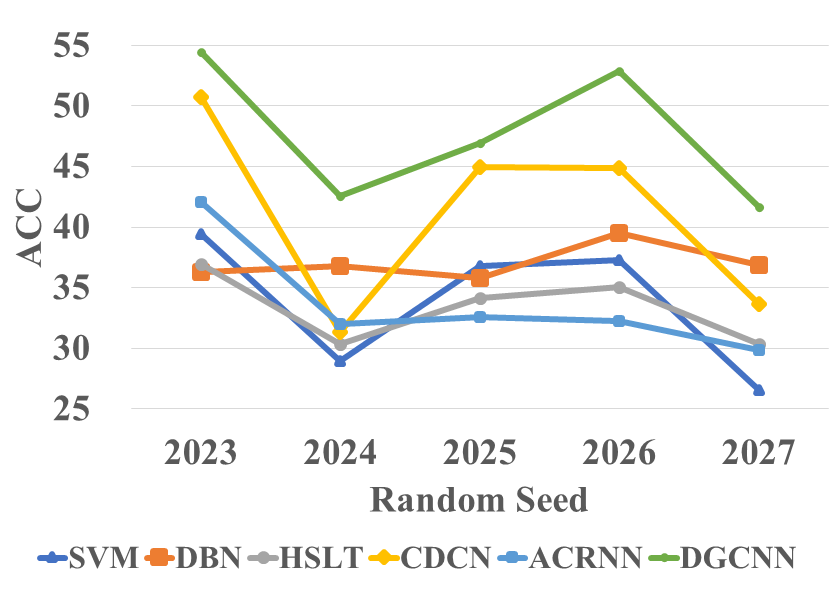} 
        \includegraphics[width=0.49\columnwidth]{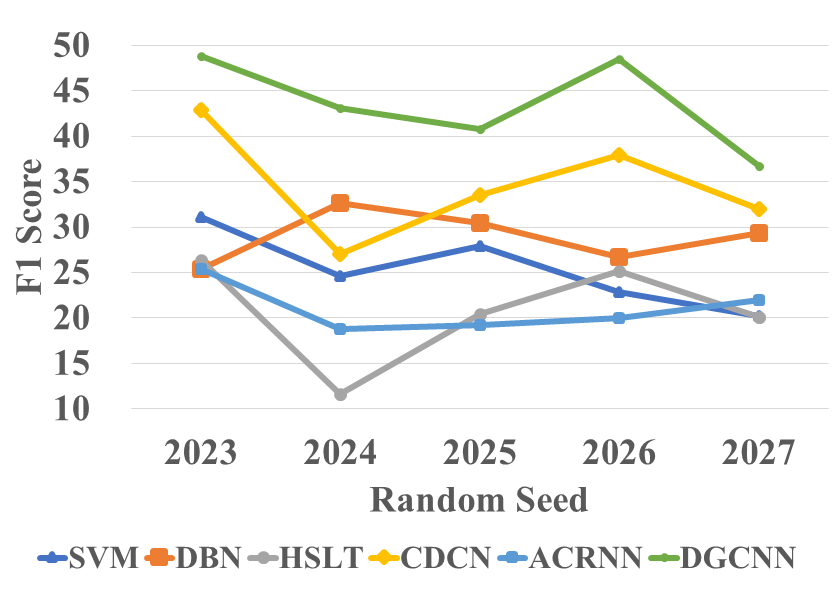}   
		}
  \subfloat[Cross-subject task on the HCI-VA dataset] {     
			\includegraphics[width=0.49\columnwidth]{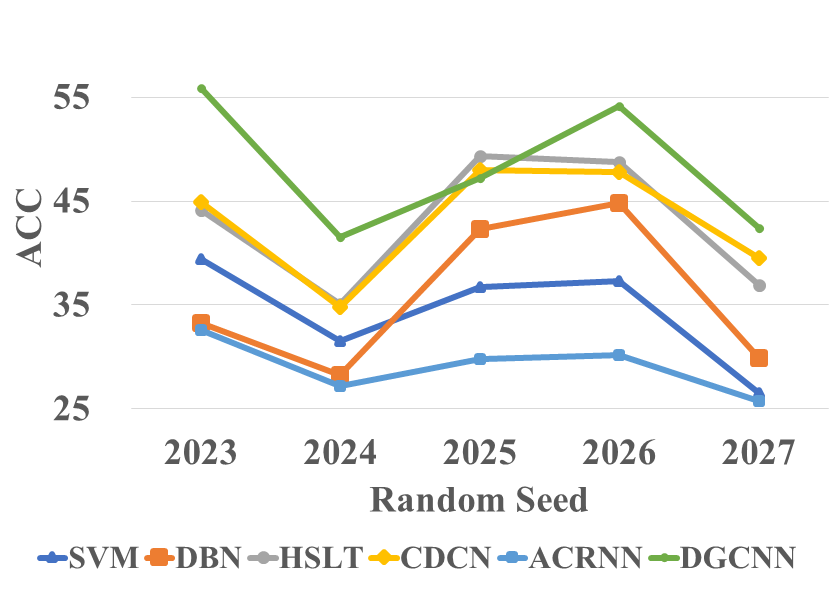} 
        \includegraphics[width=0.49\columnwidth]{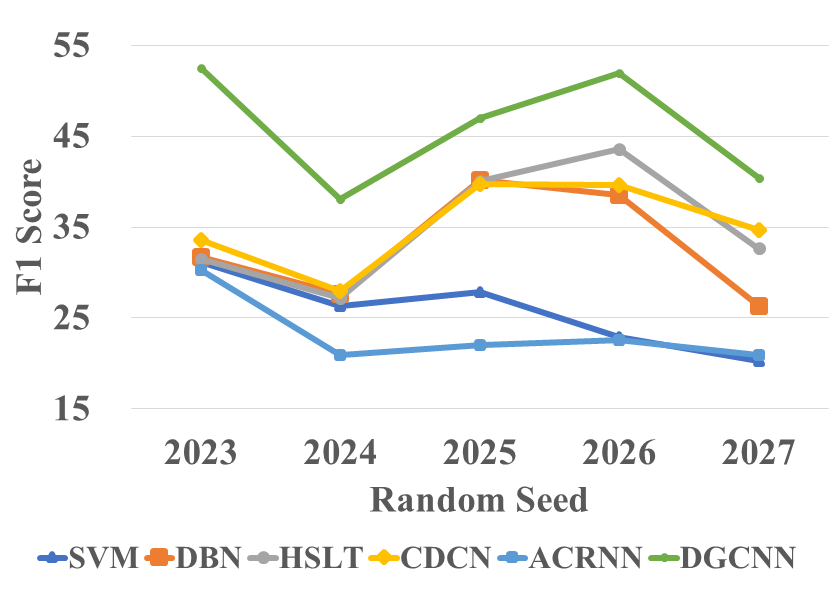}   
		}
	\end{center}
	\caption{Experimental results on random seed for subject-dependent and cross-subject tasks on the SEED-IV and HCI-VA datasets.}
	\label{ablation4}  
    \vspace{-0.3cm}
\end{figure*}

\subsection{Discussion of Experiment Settings}

In this section, we select multiple representative methods from each kind of EER method and conduct comparative experiments on four crucial factors, i.e., sample length, feature type, evaluation methods and random seeds. To be specific, SEED-IV and MAHNOB-HCI with four classification labels are taken as the dataset and SVM, DBN, HSLT, CDCN, ACRNN, and DGCNN will be selected as the representative methods. The ACC and F1 scores for both experimental scenarios are reported.    

\subsubsection{Data Splitting Length}
Existing studies have chosen to divide trials into samples of 1 to 10 seconds in length~\cite{dgcnn, eegnet, acrnn}. To explore the effects and impact of different sample lengths, we examined lengths of 1s, 3s, 5s, 7s, and 9s, with the results presented in Fig.~\ref{ablation1}. As illustrated in Fig.~\ref{ablation1}, in the subject-dependent scenario, the performance of most methods tends to decline as sample length increases. We attribute this trend to the sparsity of data in the subject-dependent context; shorter sample lengths yield a greater total number of samples, which enhances the learning capacity of deep models.

In the cross-subject scenario, it can be observed from the figure that the performance of various methods with different sample lengths across the two scenarios does not show a clear pattern. Different methods may require different sample lengths in different scenarios, and choosing the appropriate length has always been a crucial and challenging task in the EER field. This suggests to researchers that the choice of sample length must balance the trade-off between the quantity of sample augmentation and the quality of the samples.


\subsubsection{Data preprocessing Feature}
In the comparative experiments on feature types, we utilized both DE and PSD features, along with their respective variants processed by Locality Differential Sparse (LDS), as inputs. The results are presented in Fig.~\ref{ablation2}. As illustrated in Fig.~\ref{ablation2}, in the subject-dependent scenario, the DE and PSD features demonstrate comparable performance, while both feature types processed by LDS show significant improvements. In the cross-subject scenario, the DE feature exhibited the lowest performance, whereas the other three features displayed similar efficacy. This finding highlights LDS as a valuable feature processing technique, suggesting that the selection of specific features should be adapted flexibly according to the experimental context and the unique characteristics of the model.

\subsubsection{Evaluation Methods}
In the comparative experiments on model training methods, we explored how different training approaches impact model performance. We compared three training methods: (1) the method used in our benchmark, which selects the epoch with the best performance on the validation set (as ``Ours'' in Tables~\ref{ablation3_dependent} and \ref{ablation3_cross}); (2) selecting the model's result from the last epoch (as ``R1'' in Tables~\ref{ablation3_dependent} and \ref{ablation3_cross}), where, for a fair comparison, the model's training set includes the validation set; and (3) selecting the epoch when the loss on the validation set decreases by less than 1\% within the first 10 epochs (as ``R2'' in Tables~\ref{ablation3_dependent} and \ref{ablation3_cross}).

Tables~\ref{ablation3_dependent} and \ref{ablation3_cross} show that the training method used in our benchmark achieved the best performance in the majority of cases and obtained the highest average performance. Although ``R2" also utilized the validation set like ``Ours", its performance was significantly lower than that of ``Ours". On the other hand, ``R1", despite using more data and labels in the training process, lacked a validation set, making it prone to overfitting and occasionality. Therefore, although it outperformed ``Ours" in some cases, its average performance was still lower. These findings highlight the importance of using a reasonable and appropriate training method for the effective training of EER models.

\subsubsection{Random Seeds}
The random seed determines the reproducibility of experimental results, yet it is often unspecified in existing studies. To investigate the impact of random seed settings on model performance, we conducted comparisons using seeds set to 2023, 2024, 2025, 2026, and 2027. The results are shown in Figure~\ref{ablation4}. Two key conclusions can be drawn from the results. First, the random seed has a significant impact on experimental outcomes, with different models exhibiting considerable performance variation under different seeds. This is because changing the random seed affects both the random data splitting process and the stochastic behaviors during model training, such as parameter initialization, mini-batch ordering, and dropout, all of which can lead to variations in the final results. Among all the methods, CDCN is most affected by the random seed, with performance differences reaching approximately 20\%. In contrast, HSLT and ACRNN are relatively less affected, though their variations still reach around 10\%.
Second, in the majority of cases, these models exhibit similar variation patterns when the random seed changes. This suggests that using a unified random seed across all models can help ensure the fairness of comparisons to some extent.
This finding underscores the importance of adopting the same random seed for all methods in the EER field and explicitly reporting the seed value in publications to ensure fairness and reproducibility of results.

\section{Future Direction}\label{sec:future}
While LibEER is able to standardize and accelerate research in EER, we will further improve and enhance the following two aspects in our future work.

\textbf{Model Reproduction.} 
In replicating representative models in the EER field, we faced several challenges despite closely following the papers. Some methods lacked detailed descriptions of model or training parameters\cite{DBN, HSLT}. In other cases~\cite{dgcnn, rgnn, acrnn}, the available code did not fully align with the reported paper details. 
We also found that in DGCNN\cite{dgcnn} minor adjustments during replication improved performance, suggesting the paper's content may not have been updated. When the above issues arose, we used our experience to replicate the models as accurately as possible, adhering to the papers. The replication results (Table~\ref{reproduction}) show discrepancies compared to the original reports. In the future, we aim to collaborate with original authors and the broader EER community to improve and expand this algorithm library.

\textbf{Experimental Settings.} 
LibEER includes several key experimental setups to facilitate research, including six widely recognized public datasets, common data preprocessing techniques, two main tasks, common data partitioning methods, and reasonable performance evaluation methods.
However, LibEER does not cover all experimental configurations used in the EER field. Some studies use other datasets like DREAMER\cite{katsigiannis2017dreamer} and Amigos\cite{miranda2018amigos}. Others develop unique preprocessing methods, including customized filtering, artifact removal\cite{eegnet, Shen2021ContrastiveLO, 10197212}, denoising\cite{li2021multi, HE2022105048}, normalization techniques\cite{HE2022105048, eegnet}, and extracting other types of features~\cite{dgcnn, li2020novel, gcbnet}.
Additionally, some tasks, such as cross-session\cite{9904850,9857562,10160130}, subject-independent\cite{9857562,li2023effective}, and cross-dataset tasks\cite{9373917, HE2022105048}, are designed to assess model performance under different conditions, such as data from different sessions or populations. Furthermore, alternative evaluation strategies, including early stopping\cite{3350871, 7112127, 6890301}, are employed in some studies for more robust model training and assessment.
Future versions of LibEER will aim to incorporate more diverse experimental setups from the EER field.

\section{Conclusion}\label{sec:concl}
In this work, we propose LibEER, a comprehensive benchmark and algorithm library specifically designed for EEG-based emotion recognition. By standardizing the experimental settings, including dataset usage, data preprocessing methods, experimental task selection, data splitting strategies, and evaluation methods, LibEER addresses the prevalent issues of inconsistency and reproducibility in the EER field. Through its unified PyTorch codebase and the comprehensive implementation of seventeen representative models, our benchmark ensures fair comparisons and accurate evaluations of models in both subject-dependent and cross-subject scenarios. Additionally, through fair comparison experiments and investigation of key experimental settings, we present findings and discussions that aim to inspire researchers in the field. We believe that LibEER will not only facilitate the development and benchmarking of new EER models but also promote standardization and transparency, thereby advancing research and applications in this domain. In the future, we plan to further enhance LibEER by incorporating additional experimental settings and more advanced models.

\section*{Acknowledgments}
This work was supported by National Natural Science Foundation of China (62202367), Project of China Knowledge Centre for Engineering Science and Technology.

\bibliographystyle{IEEEtran}
\bibliography{reference}

\begin{thebibliography}{10}
\providecommand{\url}[1]{#1}
\csname url@samestyle\endcsname
\providecommand{\newblock}{\relax}
\providecommand{\bibinfo}[2]{#2}
\providecommand{\BIBentrySTDinterwordspacing}{\spaceskip=0pt\relax}
\providecommand{\BIBentryALTinterwordstretchfactor}{4}
\providecommand{\BIBentryALTinterwordspacing}{\spaceskip=\fontdimen2\font plus
\BIBentryALTinterwordstretchfactor\fontdimen3\font minus \fontdimen4\font\relax}
\providecommand{\BIBforeignlanguage}[2]{{%
\expandafter\ifx\csname l@#1\endcsname\relax
\typeout{** WARNING: IEEEtran.bst: No hyphenation pattern has been}%
\typeout{** loaded for the language `#1'. Using the pattern for}%
\typeout{** the default language instead.}%
\else
\language=\csname l@#1\endcsname
\fi
#2}}
\providecommand{\BIBdecl}{\relax}
\BIBdecl

\bibitem{emontiondef}
R.~J. Dolan, ``Emotion, cognition, and behavior,'' \emph{Science}, vol. 298, no. 5596, pp. 1191--1194, 2002.

\bibitem{liu2024eeg}
H.~Liu, T.~Lou, Y.~Zhang, Y.~Wu, Y.~Xiao, C.~S. Jensen, and D.~Zhang, ``Eeg-based multimodal emotion recognition: a machine learning perspective,'' \emph{IEEE Transactions on Instrumentation and Measurement}, 2024.

\bibitem{zhang2024torcheegemo}
Z.~Zhang, S.-h. Zhong, and Y.~Liu, ``Torcheegemo: A deep learning toolbox towards eeg-based emotion recognition,'' \emph{Expert Systems with Applications}, vol. 249, p. 123550, 2024.

\bibitem{alarcao2017emotions}
S.~M. Alarcao and M.~J. Fonseca, ``Emotions recognition using eeg signals: A survey,'' \emph{IEEE transactions on affective computing}, vol.~10, no.~3, pp. 374--393, 2017.

\bibitem{CAMKD}
H.~Zhang, D.~Chen, and C.~Wang, ``Confidence-aware multi-teacher knowledge distillation,'' in \emph{ICASSP}.\hskip 1em plus 0.5em minus 0.4em\relax IEEE, 2022, pp. 4498--4502.

\bibitem{DBN}
W.-L. Zheng, J.-Y. Zhu, Y.~Peng, and B.-L. Lu, ``Eeg-based emotion classification using deep belief networks,'' in \emph{ICME}.\hskip 1em plus 0.5em minus 0.4em\relax IEEE, 2014, pp. 1--6.

\bibitem{ms-mda}
H.~Chen, M.~Jin, Z.~Li, C.~Fan, J.~Li, and H.~He, ``Ms-mda: Multisource marginal distribution adaptation for cross-subject and cross-session eeg emotion recognition,'' \emph{Frontiers in Neuroscience}, vol.~15, p. 778488, 2021.

\bibitem{DAN}
H.~Li, Y.-M. Jin, W.-L. Zheng, and B.-L. Lu, ``Cross-subject emotion recognition using deep adaptation networks,'' in \emph{ICONIP}.\hskip 1em plus 0.5em minus 0.4em\relax Springer, 2018, pp. 403--413.

\bibitem{PR-PL}
R.~Zhou, Z.~Zhang, H.~Fu, L.~Zhang, L.~Li, G.~Huang, F.~Li, X.~Yang, Y.~Dong, Y.-T. Zhang \emph{et~al.}, ``Pr-pl: A novel prototypical representation based pairwise learning framework for emotion recognition using eeg signals,'' \emph{IEEE Transactions on Affective Computing}, vol.~15, no.~2, pp. 657--670, 2023.

\bibitem{HSLT}
Z.~Wang, Y.~Wang, C.~Hu, Z.~Yin, and Y.~Song, ``Transformers for eeg-based emotion recognition: A hierarchical spatial information learning model,'' \emph{IEEE Sensors Journal}, vol.~22, no.~5, pp. 4359--4368, 2022.

\bibitem{eegnet}
V.~J. Lawhern, A.~J. Solon, N.~R. Waytowich, S.~M. Gordon, C.~P. Hung, and B.~J. Lance, ``Eegnet: a compact convolutional neural network for eeg-based brain--computer interfaces,'' \emph{Journal of neural engineering}, vol.~15, no.~5, p. 056013, 2018.

\bibitem{cdcn}
Z.~Gao, X.~Wang, Y.~Yang, Y.~Li, K.~Ma, and G.~Chen, ``A channel-fused dense convolutional network for eeg-based emotion recognition,'' \emph{IEEE Transactions on Cognitive and Developmental Systems}, vol.~13, no.~4, pp. 945--954, 2020.

\bibitem{tsception}
Y.~Ding, N.~Robinson, S.~Zhang, Q.~Zeng, and C.~Guan, ``Tsception: Capturing temporal dynamics and spatial asymmetry from eeg for emotion recognition,'' \emph{IEEE Transactions on Affective Computing}, vol.~14, no.~3, pp. 2238--2250, 2022.

\bibitem{acrnn}
W.~Tao, C.~Li, R.~Song, J.~Cheng, Y.~Liu, F.~Wan, and X.~Chen, ``Eeg-based emotion recognition via channel-wise attention and self attention,'' \emph{IEEE Transactions on Affective Computing}, vol.~14, no.~1, pp. 382--393, 2020.

\bibitem{dgcnn}
T.~Song, W.~Zheng, P.~Song, and Z.~Cui, ``Eeg emotion recognition using dynamical graph convolutional neural networks,'' \emph{IEEE Transactions on Affective Computing}, vol.~11, no.~3, pp. 532--541, 2018.

\bibitem{rgnn}
P.~Zhong, D.~Wang, and C.~Miao, ``Eeg-based emotion recognition using regularized graph neural networks,'' \emph{IEEE Transactions on Affective Computing}, vol.~13, no.~3, pp. 1290--1301, 2020.

\bibitem{R2G-STNN}
Y.~Li, W.~Zheng, L.~Wang, Y.~Zong, and Z.~Cui, ``From regional to global brain: A novel hierarchical spatial-temporal neural network model for eeg emotion recognition,'' \emph{IEEE Transactions on Affective Computing}, vol.~13, no.~2, pp. 568--578, 2022.

\bibitem{gcbnet}
T.~Zhang, X.~Wang, X.~Xu, and C.~P. Chen, ``Gcb-net: Graph convolutional broad network and its application in emotion recognition,'' \emph{IEEE Transactions on Affective Computing}, vol.~13, no.~1, pp. 379--388, 2019.

\bibitem{PGCN}
M.~Jin, C.~Du, H.~He, T.~Cai, and J.~Li, ``Pgcn: Pyramidal graph convolutional network for eeg emotion recognition,'' \emph{IEEE Transactions on Multimedia}, 2024.

\bibitem{NSAL-DGAT}
Y.~Yang, Z.~Wang, Y.~Song, Z.~Jia, B.~Wang, T.-P. Jung, and F.~Wan, ``Exploiting the intrinsic neighborhood semantic structure for domain adaptation in eeg-based emotion recognition,'' \emph{IEEE Transactions on Affective Computing}, 2025.

\bibitem{DE}
R.-N. Duan, J.-Y. Zhu, and B.-L. Lu, ``Differential entropy feature for eeg-based emotion classification,'' in \emph{NER}.\hskip 1em plus 0.5em minus 0.4em\relax IEEE, 2013, pp. 81--84.

\bibitem{zhang2022eeg}
Y.~Zhang, H.~Liu, D.~Zhang, X.~Chen, T.~Qin, and Q.~Zheng, ``Eeg-based emotion recognition with emotion localization via hierarchical self-attention,'' \emph{IEEE Transactions on Affective Computing}, vol.~14, no.~3, pp. 2458--2469, 2022.

\bibitem{wu2023autoeer}
Y.~Wu, H.~Liu, D.~Zhang, Y.~Zhang, T.~Lou, and Q.~Zheng, ``Autoeer: automatic eeg-based emotion recognition with neural architecture search,'' \emph{Journal of Neural Engineering}, vol.~20, no.~4, p. 046029, 2023.

\bibitem{zhang2024cross}
Y.~Zhang, H.~Liu, D.~Wang, D.~Zhang, T.~Lou, Q.~Zheng, and C.~Quek, ``Cross-modal credibility modelling for eeg-based multimodal emotion recognition,'' \emph{Journal of Neural Engineering}, vol.~21, no.~2, p. 026040, 2024.

\bibitem{pytorch}
A.~Paszke, S.~Gross, F.~Massa, A.~Lerer, J.~Bradbury, G.~Chanan, T.~Killeen, Z.~Lin, N.~Gimelshein, L.~Antiga \emph{et~al.}, ``Pytorch: An imperative style, high-performance deep learning library,'' \emph{Advances in neural information processing systems}, vol.~32, 2019.

\bibitem{tensorflow}
M.~Abadi, P.~Barham, J.~Chen, Z.~Chen, A.~Davis, J.~Dean, M.~Devin, S.~Ghemawat, G.~Irving, M.~Isard \emph{et~al.}, ``$\{$TensorFlow$\}$: a system for $\{$Large-Scale$\}$ machine learning,'' in \emph{OSDI}, 2016, pp. 265--283.

\bibitem{DEAP}
S.~Koelstra, C.~Muhl, M.~Soleymani, J.-S. Lee, A.~Yazdani, T.~Ebrahimi, T.~Pun, A.~Nijholt, and I.~Patras, ``Deap: A database for emotion analysis; using physiological signals,'' \emph{IEEE transactions on affective computing}, vol.~3, no.~1, pp. 18--31, 2011.

\bibitem{SEED}
W.-L. Zheng and B.-L. Lu, ``Investigating critical frequency bands and channels for eeg-based emotion recognition with deep neural networks,'' \emph{IEEE Transactions on autonomous mental development}, vol.~7, no.~3, pp. 162--175, 2015.

\bibitem{HCI}
M.~Soleymani, J.~Lichtenauer, T.~Pun, and M.~Pantic, ``A multimodal database for affect recognition and implicit tagging,'' \emph{IEEE transactions on affective computing}, vol.~3, no.~1, pp. 42--55, 2011.

\bibitem{SEED-IV}
W.-L. Zheng, W.~Liu, Y.~Lu, B.-L. Lu, and A.~Cichocki, ``Emotionmeter: A multimodal framework for recognizing human emotions,'' \emph{IEEE transactions on cybernetics}, vol.~49, no.~3, pp. 1110--1122, 2018.

\bibitem{SEED-V}
W.~Liu, J.-L. Qiu, W.-L. Zheng, and B.-L. Lu, ``Comparing recognition performance and robustness of multimodal deep learning models for multimodal emotion recognition,'' \emph{IEEE Transactions on Cognitive and Developmental Systems}, vol.~14, no.~2, pp. 715--729, 2021.

\bibitem{MPED}
T.~Song, W.~Zheng, C.~Lu, Y.~Zong, X.~Zhang, and Z.~Cui, ``Mped: A multi-modal physiological emotion database for discrete emotion recognition,'' \emph{IEEE Access}, vol.~7, pp. 12\,177--12\,191, 2019.

\bibitem{jianglarge}
W.~Jiang, L.~Zhao, and B.-l. Lu, ``Large brain model for learning generic representations with tremendous eeg data in bci,'' in \emph{ICLR}, 2024.

\bibitem{matlab}
C.~Brunner, A.~Delorme, and S.~Makeig, ``Eeglab--an open source matlab toolbox for electrophysiological research,'' \emph{Biomedical Engineering/Biomedizinische Technik}, vol.~58, no. SI-1-Track-G, p. 000010151520134182, 2013.

\bibitem{liu2023emotionkd}
Y.~Liu, Z.~Jia, and H.~Wang, ``Emotionkd: a cross-modal knowledge distillation framework for emotion recognition based on physiological signals,'' in \emph{ACM MM}, 2023, pp. 6122--6131.

\bibitem{SVM}
J.~A. Suykens and J.~Vandewalle, ``Least squares support vector machine classifiers,'' \emph{Neural Processing Letters}, vol.~9, no.~3, pp. 293--300, 1999.

\bibitem{ding2023lggnet}
Y.~Ding, N.~Robinson, C.~Tong, Q.~Zeng, and C.~Guan, ``Lggnet: Learning from local-global-graph representations for brain--computer interface,'' \emph{IEEE Transactions on Neural Networks and Learning Systems}, 2023.

\bibitem{khare2020time}
S.~K. Khare and V.~Bajaj, ``Time--frequency representation and convolutional neural network-based emotion recognition,'' \emph{IEEE transactions on neural networks and learning systems}, vol.~32, no.~7, pp. 2901--2909, 2020.

\bibitem{hasnul2021electrocardiogram}
M.~A. Hasnul, N.~A.~A. Aziz, S.~Alelyani, M.~Mohana, and A.~A. Aziz, ``Electrocardiogram-based emotion recognition systems and their applications in healthcare—a review,'' \emph{Sensors}, vol.~21, no.~15, p. 5015, 2021.

\bibitem{gauba2017prediction}
H.~Gauba, P.~Kumar, P.~P. Roy, P.~Singh, D.~P. Dogra, and B.~Raman, ``Prediction of advertisement preference by fusing eeg response and sentiment analysis,'' \emph{Neural Networks}, vol.~92, pp. 77--88, 2017.

\bibitem{nandi2021real}
A.~Nandi, F.~Xhafa, L.~Subirats, and S.~Fort, ``Real-time emotion classification using eeg data stream in e-learning contexts,'' \emph{Sensors}, vol.~21, no.~5, p. 1589, 2021.

\bibitem{cheng2020emotion}
J.~Cheng, M.~Chen, C.~Li, Y.~Liu, R.~Song, A.~Liu, and X.~Chen, ``Emotion recognition from multi-channel eeg via deep forest,'' \emph{IEEE Journal of Biomedical and Health Informatics}, vol.~25, no.~2, pp. 453--464, 2020.

\bibitem{BiDANN}
Y.~Li, W.~Zheng, Y.~Zong, Z.~Cui, T.~Zhang, and X.~Zhou, ``A bi-hemisphere domain adversarial neural network model for eeg emotion recognition,'' \emph{IEEE Transactions on Affective Computing}, vol.~12, no.~2, pp. 494--504, 2018.

\bibitem{li2023brain}
D.~Li, L.~Xie, Z.~Wang, and H.~Yang, ``Brain emotion perception inspired eeg emotion recognition with deep reinforcement learning,'' \emph{IEEE Transactions on Neural Networks and Learning Systems}, 2023.

\bibitem{li2023effective}
C.~Li, P.~Li, Y.~Zhang, N.~Li, Y.~Si, F.~Li, Z.~Cao, H.~Chen, B.~Chen, D.~Yao \emph{et~al.}, ``Effective emotion recognition by learning discriminative graph topologies in eeg brain networks,'' \emph{IEEE Transactions on Neural Networks and Learning Systems}, 2023.

\bibitem{khosrowabadi2013ernn}
R.~Khosrowabadi, C.~Quek, K.~K. Ang, and A.~Wahab, ``Ernn: A biologically inspired feedforward neural network to discriminate emotion from eeg signal,'' \emph{IEEE transactions on neural networks and learning systems}, vol.~25, no.~3, pp. 609--620, 2013.

\bibitem{alsolamy2016emotion}
M.~Alsolamy and A.~Fattouh, ``Emotion estimation from eeg signals during listening to quran using psd features,'' in \emph{CSIT}.\hskip 1em plus 0.5em minus 0.4em\relax IEEE, 2016, pp. 1--5.

\bibitem{10697478}
Y.~Zhang, H.~Liu, Y.~Xiao, M.~Amoon, D.~Zhang, D.~Wang, S.~Yang, and C.~Quek, ``Llm-enhanced multi-teacher knowledge distillation for modality-incomplete emotion recognition in daily healthcare,'' \emph{IEEE Journal of Biomedical and Health Informatics}, pp. 1--11, 2024.

\bibitem{8512865}
Y.~Luo and B.-L. Lu, ``Eeg data augmentation for emotion recognition using a conditional wasserstein gan,'' in \emph{EMBC}, 2018, pp. 2535--2538.

\bibitem{Zhang_Zhong_Liu_2024}
\BIBentryALTinterwordspacing
Z.~Zhang, S.~Zhong, and Y.~Liu, ``Beyond mimicking under-represented emotions: Deep data augmentation with emotional subspace constraints for eeg-based emotion recognition,'' \emph{Proceedings of the AAAI Conference on Artificial Intelligence}, vol.~38, no.~9, pp. 10\,252--10\,260, Mar. 2024. [Online]. Available: \url{https://ojs.aaai.org/index.php/AAAI/article/view/28891}
\BIBentrySTDinterwordspacing

\bibitem{Luo_2020}
\BIBentryALTinterwordspacing
Y.~Luo, L.-Z. Zhu, Z.-Y. Wan, and B.-L. Lu, ``Data augmentation for enhancing eeg-based emotion recognition with deep generative models,'' \emph{Journal of Neural Engineering}, vol.~17, no.~5, p. 056021, oct 2020. [Online]. Available: \url{https://dx.doi.org/10.1088/1741-2552/abb580}
\BIBentrySTDinterwordspacing

\bibitem{li2020novel}
Y.~Li, L.~Wang, W.~Zheng, Y.~Zong, L.~Qi, Z.~Cui, T.~Zhang, and T.~Song, ``A novel bi-hemispheric discrepancy model for eeg emotion recognition,'' \emph{IEEE Transactions on Cognitive and Developmental Systems}, vol.~13, no.~2, pp. 354--367, 2020.

\bibitem{9373917}
T.~Song, S.~Liu, W.~Zheng, Y.~Zong, Z.~Cui, Y.~Li, and X.~Zhou, ``Variational instance-adaptive graph for eeg emotion recognition,'' \emph{IEEE Transactions on Affective Computing}, vol.~14, no.~1, pp. 343--356, 2023.

\bibitem{katsigiannis2017dreamer}
S.~Katsigiannis and N.~Ramzan, ``Dreamer: A database for emotion recognition through eeg and ecg signals from wireless low-cost off-the-shelf devices,'' \emph{IEEE journal of biomedical and health informatics}, vol.~22, no.~1, pp. 98--107, 2017.

\bibitem{miranda2018amigos}
J.~A. Miranda-Correa, M.~K. Abadi, N.~Sebe, and I.~Patras, ``Amigos: A dataset for affect, personality and mood research on individuals and groups,'' \emph{IEEE transactions on affective computing}, vol.~12, no.~2, pp. 479--493, 2018.

\bibitem{li2021multi}
R.~Li, Y.~Wang, and B.-L. Lu, ``A multi-domain adaptive graph convolutional network for eeg-based emotion recognition,'' in \emph{ACM MM}, 2021, pp. 5565--5573.

\bibitem{HE2022105048}
\BIBentryALTinterwordspacing
Z.~He, Y.~Zhong, and J.~Pan, ``An adversarial discriminative temporal convolutional network for eeg-based cross-domain emotion recognition,'' \emph{Computers in Biology and Medicine}, vol. 141, p. 105048, 2022. [Online]. Available: \url{https://www.sciencedirect.com/science/article/pii/S0010482521008428}
\BIBentrySTDinterwordspacing

\bibitem{Shen2021ContrastiveLO}
\BIBentryALTinterwordspacing
X.~Shen, X.~Liu, X.~Hu, D.~Zhang, and S.~Song, ``Contrastive learning of subject-invariant eeg representations for cross-subject emotion recognition,'' \emph{IEEE Transactions on Affective Computing}, vol.~14, pp. 2496--2511, 2021. [Online]. Available: \url{https://api.semanticscholar.org/CorpusID:237572066}
\BIBentrySTDinterwordspacing

\bibitem{10197212}
W.~Wang, F.~Qi, D.~P. Wipf, C.~Cai, T.~Yu, Y.~Li, Y.~Zhang, Z.~Yu, and W.~Wu, ``Sparse bayesian learning for end-to-end eeg decoding,'' \emph{IEEE Transactions on Pattern Analysis and Machine Intelligence}, vol.~45, no.~12, pp. 15\,632--15\,649, 2023.

\bibitem{9904850}
Z.~Li, E.~Zhu, M.~Jin, C.~Fan, H.~He, T.~Cai, and J.~Li, ``Dynamic domain adaptation for class-aware cross-subject and cross-session eeg emotion recognition,'' \emph{IEEE Journal of Biomedical and Health Informatics}, vol.~26, no.~12, pp. 5964--5973, 2022.

\bibitem{9857562}
L.~Feng, C.~Cheng, M.~Zhao, H.~Deng, and Y.~Zhang, ``Eeg-based emotion recognition using spatial-temporal graph convolutional lstm with attention mechanism,'' \emph{IEEE Journal of Biomedical and Health Informatics}, vol.~26, no.~11, pp. 5406--5417, 2022.

\bibitem{10160130}
R.~Zhou, Z.~Zhang, H.~Fu, L.~Zhang, L.~Li, G.~Huang, F.~Li, X.~Yang, Y.~Dong, Y.-T. Zhang, and Z.~Liang, ``Pr-pl: A novel prototypical representation based pairwise learning framework for emotion recognition using eeg signals,'' \emph{IEEE Transactions on Affective Computing}, vol.~15, no.~2, pp. 657--670, 2024.

\bibitem{3350871}
J.~Ma, H.~Tang, W.-L. Zheng, and B.-L. Lu, ``Emotion recognition using multimodal residual lstm network,'' in \emph{ACM MM}, 2019, pp. 176--183.

\bibitem{7112127}
M.~Soleymani, S.~Asghari-Esfeden, Y.~Fu, and M.~Pantic, ``Analysis of eeg signals and facial expressions for continuous emotion detection,'' \emph{IEEE Transactions on Affective Computing}, vol.~7, no.~1, pp. 17--28, 2016.

\bibitem{6890301}
M.~Soleymani, S.~Asghari-Esfeden, M.~Pantic, and Y.~Fu, ``Continuous emotion detection using eeg signals and facial expressions,'' in \emph{ICME}.\hskip 1em plus 0.5em minus 0.4em\relax IEEE, 2014, pp. 1--6.

\bibitem{yang2022survey}
Y.~Yang, X.~Xia, D.~Lo, and J.~Grundy, ``A survey on deep learning for software engineering,'' \emph{ACM Computing Surveys (CSUR)}, vol.~54, no. 10s, pp. 1--73, 2022.

\bibitem{alom2019state}
M.~Z. Alom, T.~M. Taha, C.~Yakopcic, S.~Westberg, P.~Sidike, M.~S. Nasrin, M.~Hasan, B.~C. Van~Essen, A.~A. Awwal, and V.~K. Asari, ``A state-of-the-art survey on deep learning theory and architectures,'' \emph{electronics}, vol.~8, no.~3, p. 292, 2019.

\end{thebibliography}

\vspace{-0.6cm}
\begin{IEEEbiography}[{\includegraphics[width=1in,height=1.25in,clip,keepaspectratio]{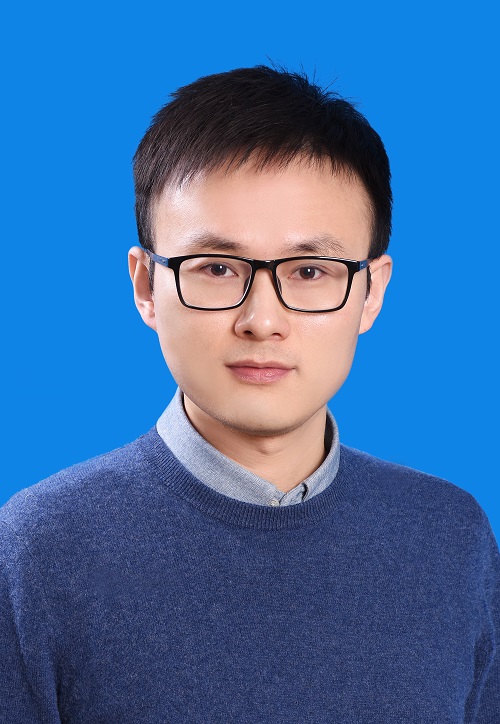}}]{Huan Liu}
received his B.S. degree and Ph.D. degree from Xi'an Jiaotong University in 2013 and 2020, respectively. He is currently an Associate Professor at the Department of Computer Science and Technology, Xi'an Jiaotong University, Xi'an 710049, China. His research interests include affective computing, machine learning, and deep learning.
\end{IEEEbiography}
\vspace{-0.6cm}

\begin{IEEEbiography}
[{\includegraphics[width=1in,height=1.25in,clip,keepaspectratio]{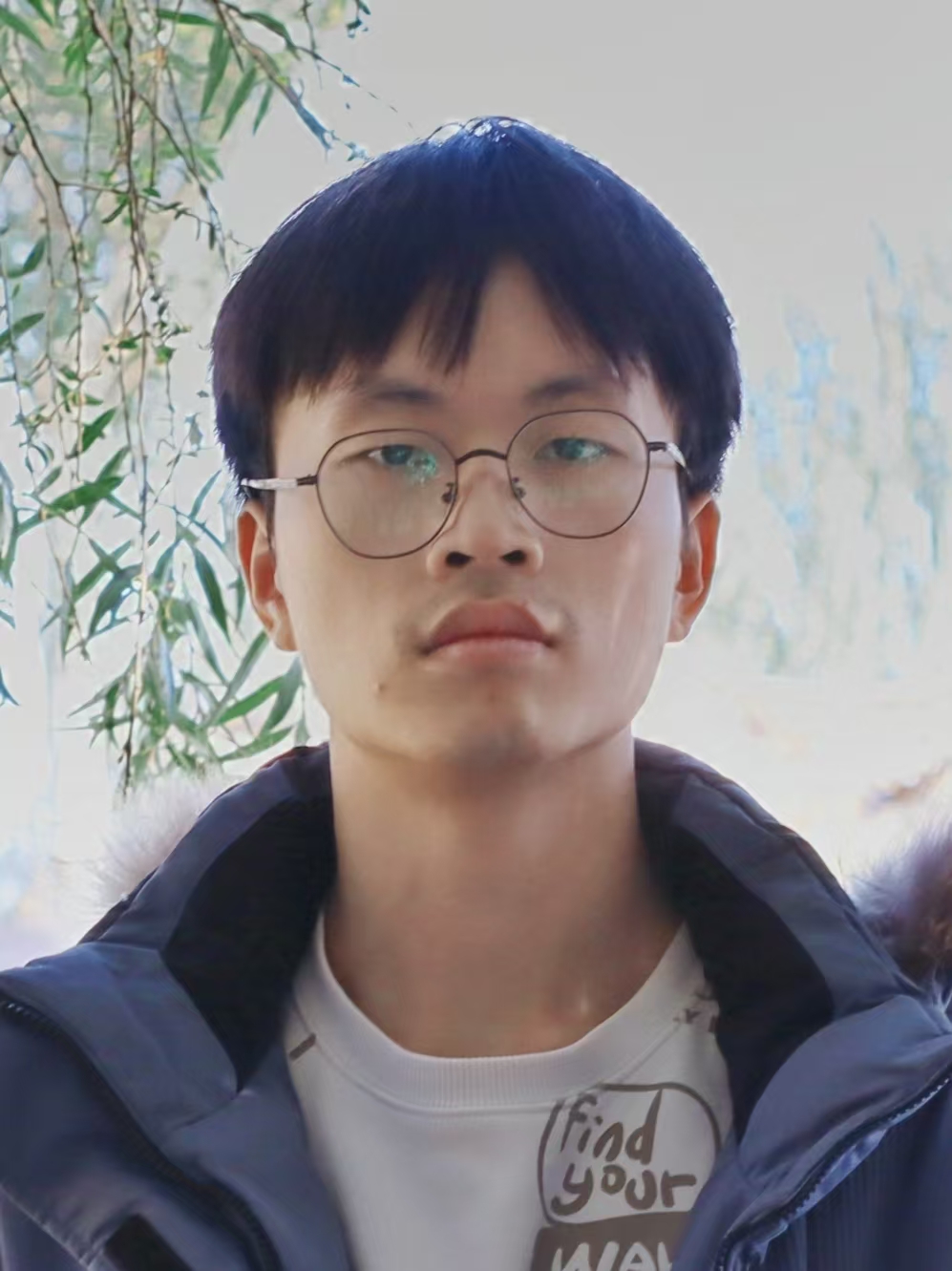}}]{Shusen Yang}
received his B.S. degree from Xi'an Jiaotong University in 2023. He is currently pursuing the M.S. degree with the MOEKLINNS Laboratory, Department of Computer Science and Technology, Xi’an Jiaotong University.
His research interests include EEG-based emotion recognition.
\end{IEEEbiography}
\vspace{-0.6cm}

\begin{IEEEbiography}[{\includegraphics[width=1in,height=1.25in,clip,keepaspectratio]{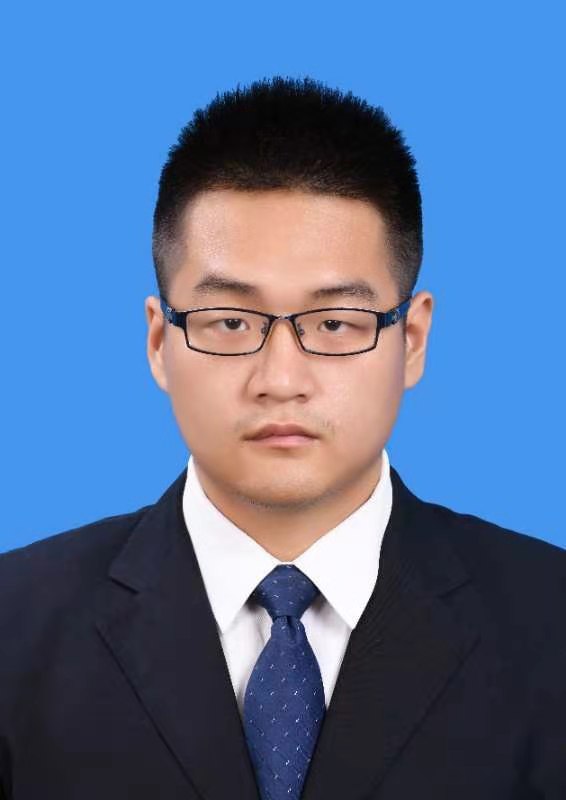}}]{Yuzhe Zhang}
received his B.S. degree from Xi'an Jiaotong University in 2018. He is currently pursuing the Ph.D. degree with the MOEKLINNS Lab, Department of Computer Science and Technology, Xi'an Jiaotong University, Xi'an 710049, China. His research interest includes affective computing and brain–computer interface.
\end{IEEEbiography}
\vspace{-0.6cm}

\begin{IEEEbiography}[{\includegraphics[width=1in,height=1.25in,clip,keepaspectratio]{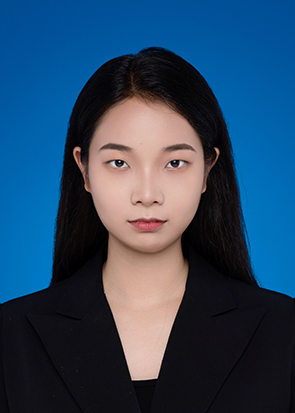}}]{Mengze Wang}
received her B.E. degree from Xi'an Jiaotong University in 2024. She is currently pursuing the Master's degree with the MOEKLINNS Lab, Department of Computer Science and Technology, Xi'an Jiaotong University, Xi'an 710049, China. Her research interest includes retrieval-augmented generation and smart education.
\end{IEEEbiography}
\vspace{-0.6cm}

\begin{IEEEbiography}[{\includegraphics[width=1in,height=1.25in,clip,keepaspectratio]{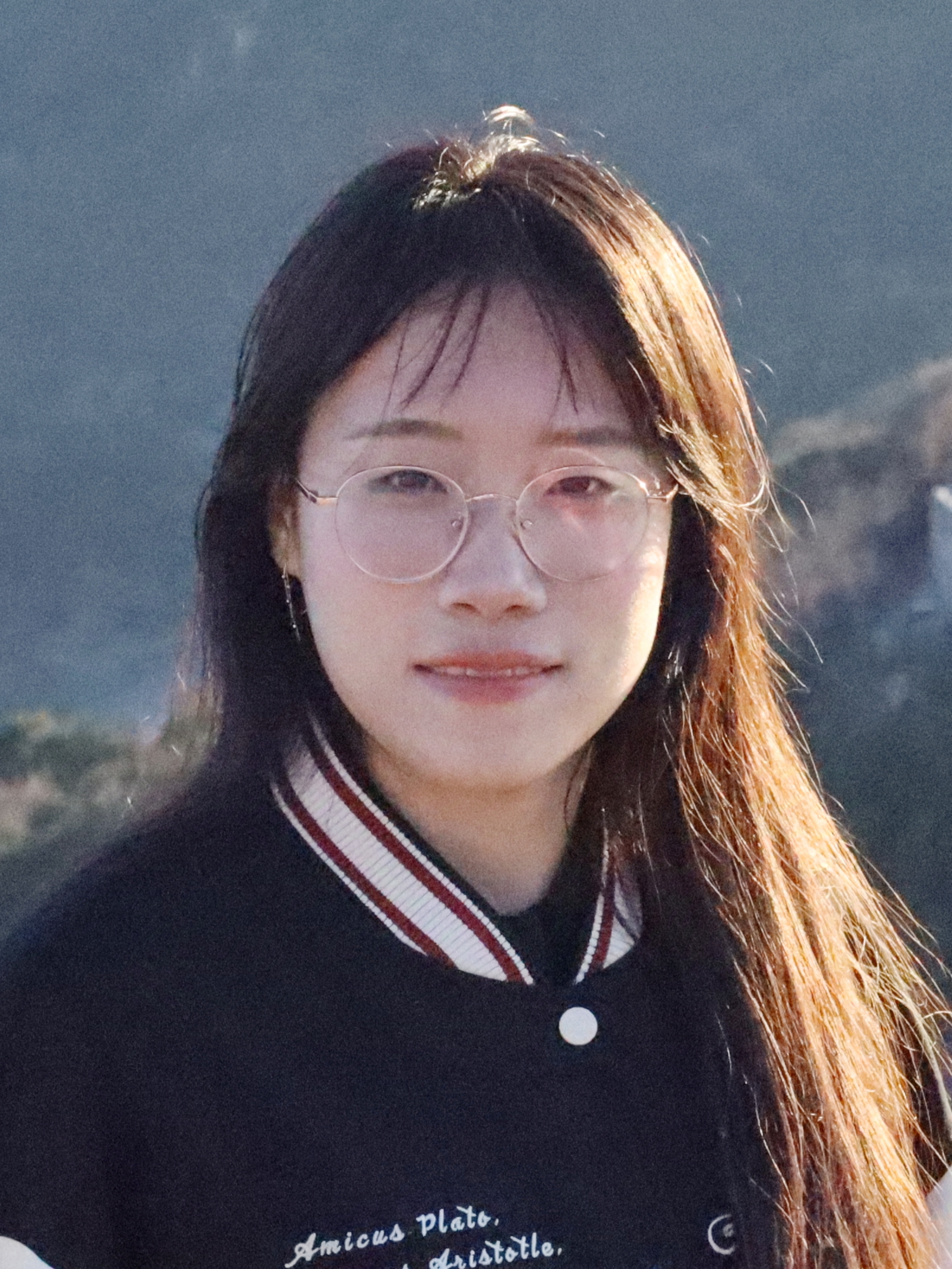}}]{Fanyu Gong}
received her B.S. degree from Xi'an Jiaotong University in 2023. She is currently pursuing the M.S. degree with the MOEKLINNS Laboratory, Department of Computer Science and Technology, Xi’an Jiaotong University.
Her research interests include EEG-based emotion recognition.
\end{IEEEbiography}
\vspace{-0.6cm}

\begin{IEEEbiography}[{\includegraphics[width=1in,height=1.25in,clip,keepaspectratio]{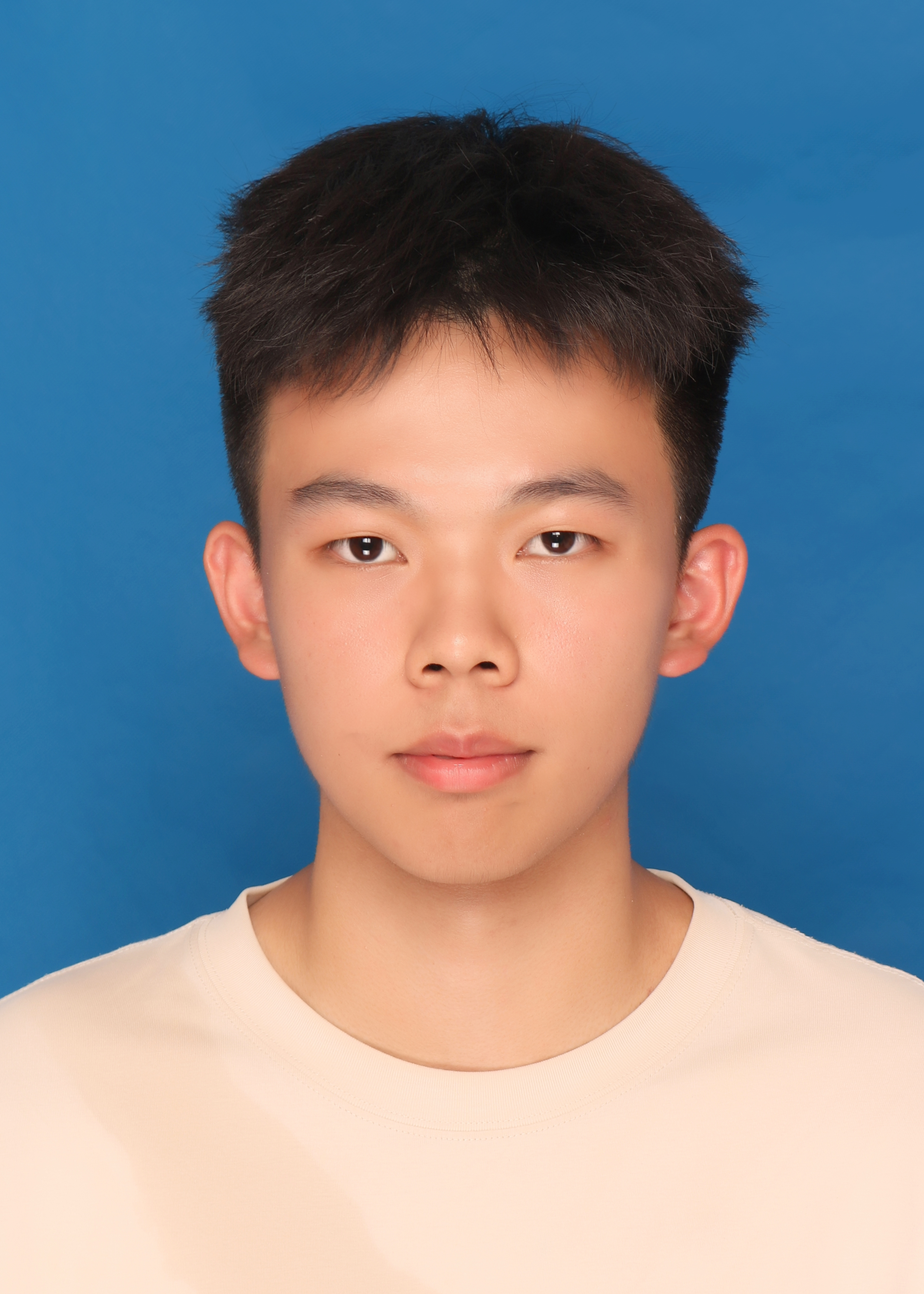}}]{Chengxi Xie}
is currently pursuing the B.S. degree in Joint School 
of Design and Innovation, Xi'an Jiaotong University, Xi'an 710049, China. His research interests include emotion recognition and machine learning.
\end{IEEEbiography}
\vspace{-0.65cm}

\begin{IEEEbiography}[{\includegraphics[width=1in,height=1.25in,clip,keepaspectratio]{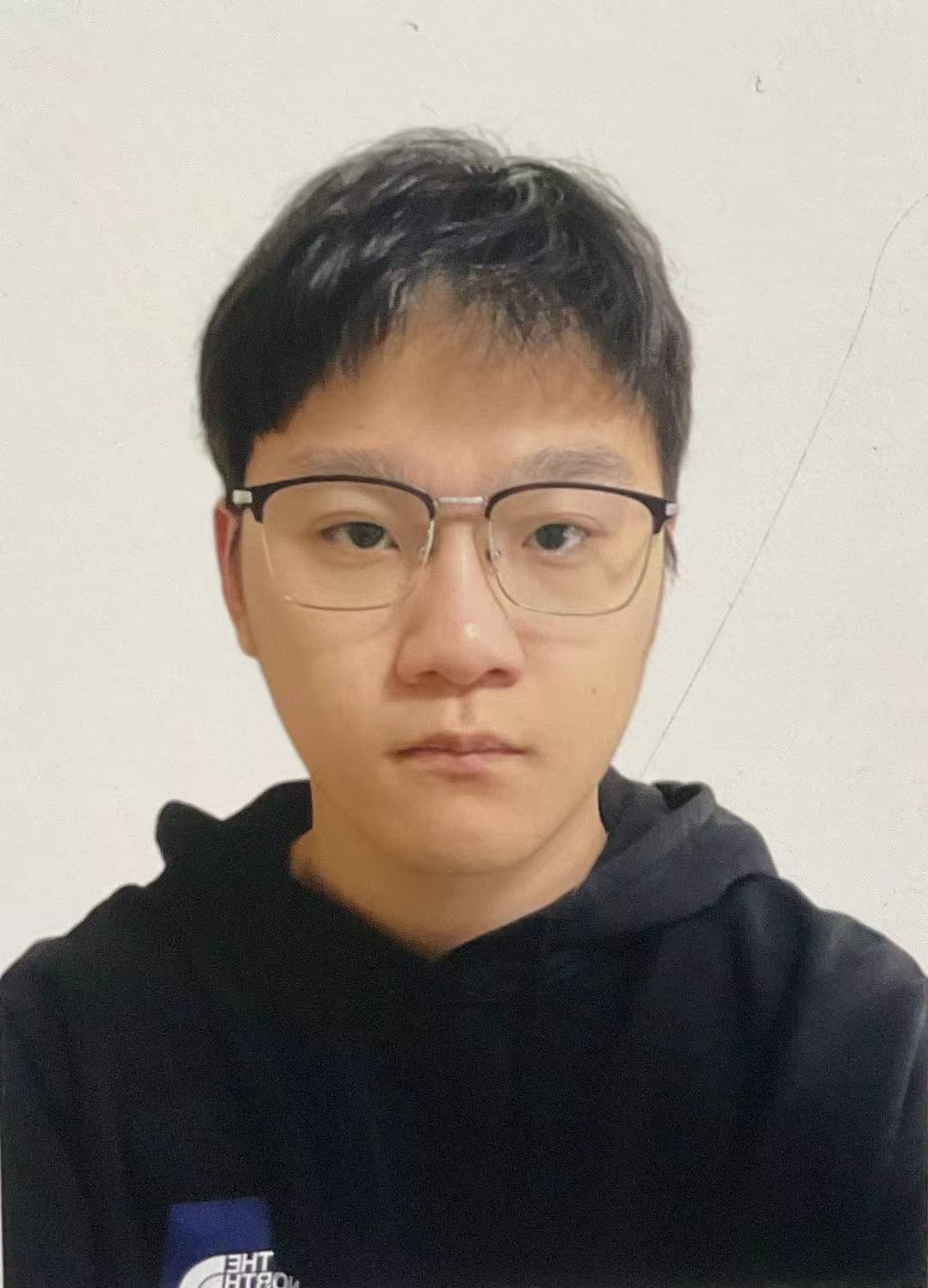}}]{Guanjian Liu} received his B.E. degree from Xi’an Jiaotong University in 2024. He is currently pursuing the master’s degree with the MOEKLINNS Lab, Department of Computer Science and Technology, Xi'an Jiaotong University, Xi'an 710049, China. His research interest includes brain-computer interface.
\end{IEEEbiography}
\vspace{-0.65cm}

\begin{IEEEbiography}[{\includegraphics[width=1in,height=1.25in,clip,keepaspectratio]{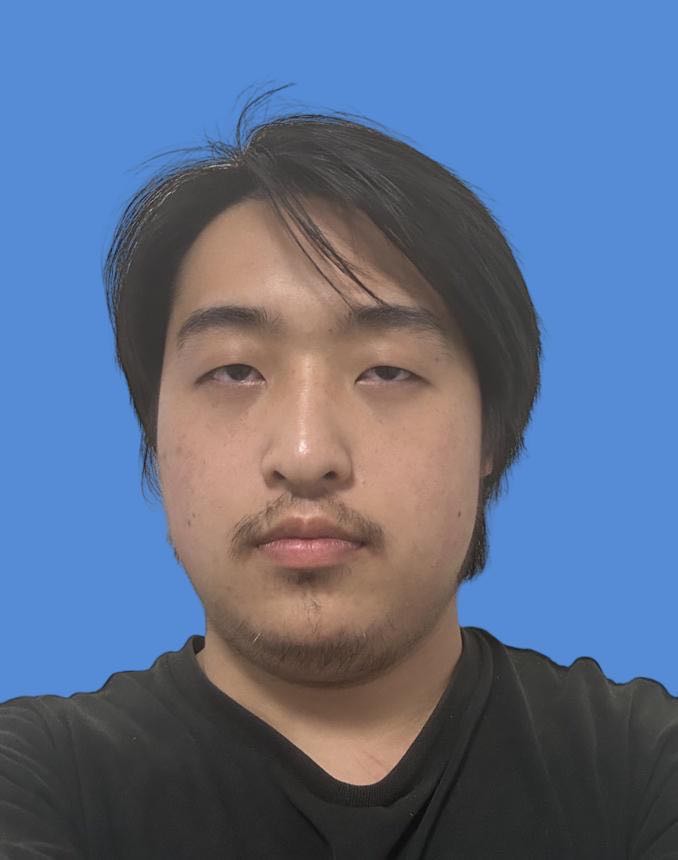}}]{Zejun Liu}
is currently pursuing the B.S. degree in Joint School of Design and Innovation, Xi'an Jiao-tong University, Xi'an 710049, China. His research interest includes emotion recognition and deep learning.
\end{IEEEbiography}
\vspace{-0.65cm}

\begin{IEEEbiography}[{\includegraphics[width=1in,height=1.25in, clip,keepaspectratio]{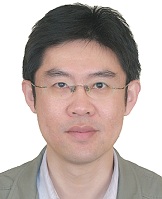}}]{Yong-Jin Liu} (Senior Member, IEEE) received the B.Eng. degree from Tianjin University, China, in 1998, and the Ph.D. degree from The Hong Kong University of Science and Technology, Hong Kong, China, in 2004.
He is currently a Full Professor with the Department of Computer Science and Technology, Tsinghua University, China. 
His research interests include machine learning, cognitive computation, computer graphics, and computer-aided design.
\end{IEEEbiography}
\vspace{-0.65cm}

\begin{IEEEbiography}[{\includegraphics[width=1in,height=1.25in,clip,keepaspectratio]{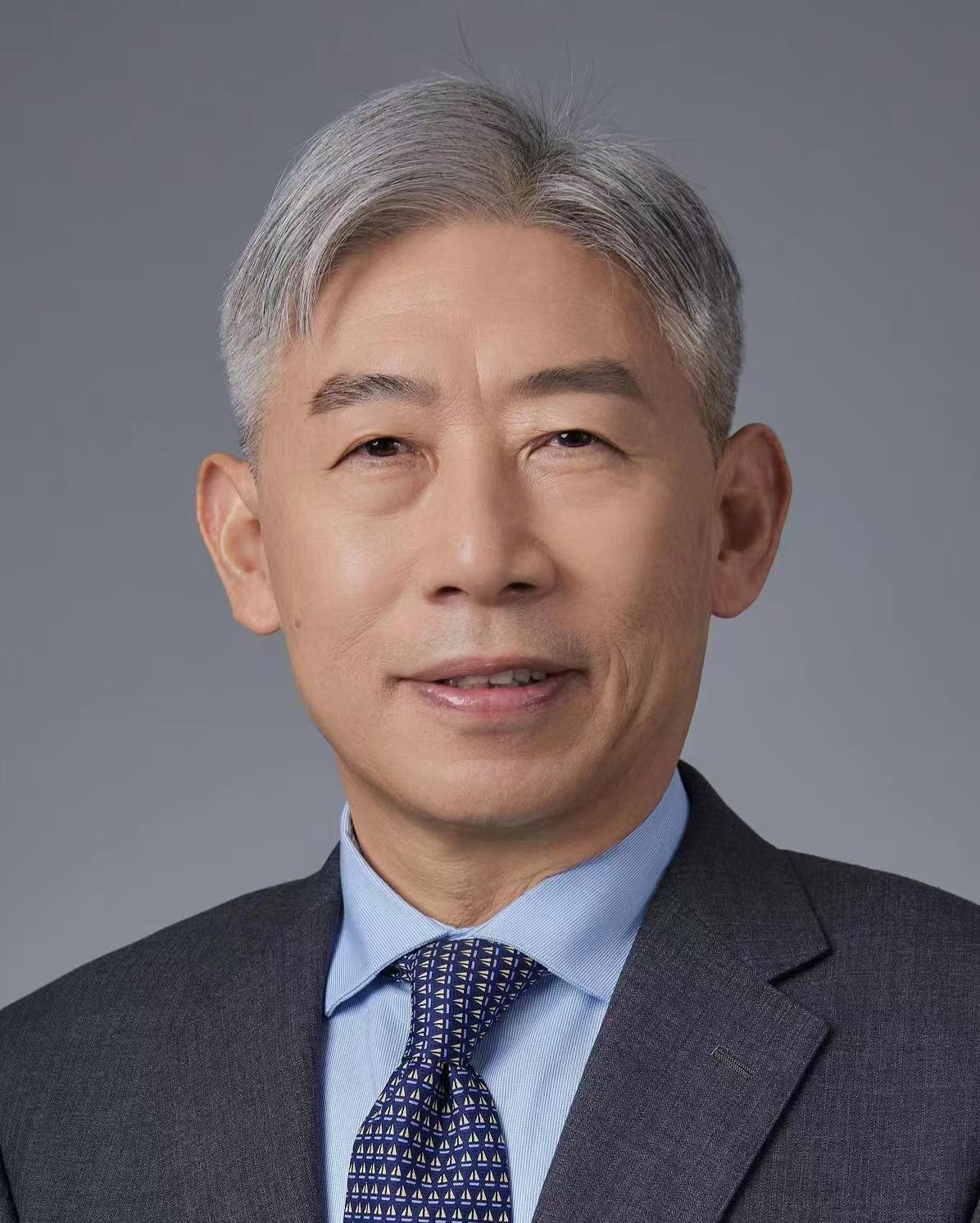}}]{Bao-Liang Lu}  (M’94-SM’01-F’21) received the B.S. degree in instrument and control engineering from Qingdao University of Science and Technology, Qingdao, China, in 1982, the M.S. degree in computer science and technology from Northwestern Polytechnical University, Xi’an, China, in 1989, and the Dr. Eng. degree in electrical engineering from Kyoto University, Kyoto, Japan, in 1994. Since 2002, he has been a Full Professor with the Department of Computer Science and Engineering, Shanghai Jiao Tong University, Shanghai, China. He is currently the Associate Editors of IEEE Transactions on Affective Computing, IEEE Transactions on Cognitive and Development Systems, and Journal of Neural Engineering. His current research interests include brain-like computing, deep learning, affective computing, and brain-computer interface. 
\end{IEEEbiography}
\vspace{-0.65cm}

\begin{IEEEbiography}[{\includegraphics[width=1in,height=1.25in,clip,keepaspectratio]{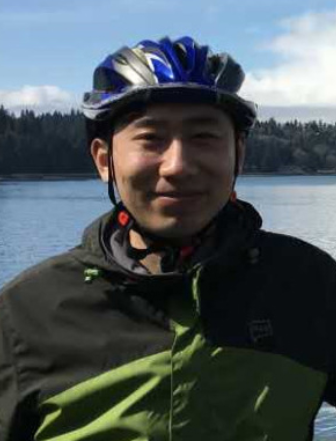}}]{Dalin Zhang} (Senior Member, IEEE) received his B.S. degree from Jilin University in 2012, M.S. degree from University of Chinese Academy of Sciences in 2015,
Ph.D. degree from the University of New South Wales Sydney in 2020. He is currently a Professor at the Hangzhou Dianzi University, China. His research interest includes the brain-computer interface, human activity recognition and the Internet of Things.
\end{IEEEbiography}

\end{document}